# Crystalline silicates and dust processing in the protoplanetary disks of the Taurus young cluster

Dan M. Watson[1], Jarron M. Leisenring[1,2], Elise Furlan[3,4], C.J. Bohac[1], B. Sargent[1], W.J. Forrest[1], Nuria Calvet[5], Lee Hartmann[5], Jason T. Nordhaus[1], Joel D. Green[1], K.H. Kim[1], G.C. Sloan[6], C.H. Chen[7,8], L.D. Keller[9], Paola d'Alessio[10], J. Najita[7], Keven I. Uchida[6], and J.R. Houck[6]

## Abstract

We characterize the crystalline silicate content and spatial distribution of small dust grains in a large sample of protoplanetary disks in the Taurus-Auriga young cluster, using Spitzer Space Telescope mid-infrared spectra. In turn we use the results to analyze the evolution of structure and composition of these 1-2 Myr-old disks around Solar- and later-type young stars, and test the standard models of dust processing which result in the conversion of originally amorphous dust into minerals. We find strong evidence of evolution of the dust crystalline mass fraction in parallel with that of the structure of the disks, in the sense that increasing crystalline mass fraction is strongly linked to dust settling to the disk midplane. We also confirm that the crystalline silicates are confined to small radii, $r \lesssim 10$ AU. However, we see no significant correlation of crystalline mass fraction with stellar mass or luminosity, stellar accretion rate, disk mass, or disk/star mass ratio, as would be expected in the standard models of dust processing based upon photo-evaporation and condensation close to the central star, accretion-heating-driven annealing at $r \lesssim 1$ AU, or spiral-shock heating at $r \lesssim 10$ AU, with or without effective large-scale radial mixing mechanisms. Either another grain-crystallizing mechanism dominates over these, or another process must be at work within the disks to erase the correlations they produce. We propose one of each sort that seem to be worth further investigation, namely X-ray heating and annealing of dust grains, and modulation of disk structure by giant-planetary formation and migration.

[1] Department of Physics and Astronomy, University of Rochester, Rochester, NY 14627, USA; dmw@pas.rochester.edu.

[2] Department of Astronomy, University of Virginia, Charlottesville, VA 22904 USA.

[3] NASA Astrobiology Institute, and Department of Physics and Astronomy, UCLA, Los Angeles, CA 90095, USA.

[4] Current address: Jet Propulsion Laboratory, Caltech, Mail Stop 264-767, 4800 Oak Grove Drive, Pasadena, CA 91109.

[5] Department of Astronomy, University of Michigan, Ann Arbor, MI 48109, USA.

[6] Center for Radiophysics and Space Research, Cornell University, Ithaca, NY 14853, USA.

[7] NOAO, Tucson, AZ 85719, USA.

[8] *Spitzer* Fellow.

[9] Department of Physics, Ithaca College, Ithaca, NY 14850, USA.

[10] Centro de Radioastronomía y Astrofísica, UNAM, 58089 Morelia, Michoacán, México.

## 1 Introduction

The solid denizens of the Solar system are thought to have formed during the first 10 Myr or so of the Solar system's life, from within the circum-Solar, protoplanetary disk.

Precisely when, and how, are major unanswered questions within the domains of astrophysics and planetary science. Clues to the identity of the relevant processes have long been sought in observations of the mid-infrared ($\lambda = 5 - 50 \ \mu m$) emission features of small silicate dust grains in protoplanetary disks in young stellar objects (YSOs), as these contain evidence of the conversion of originally amorphous, submicron interstellar grains into much larger, partially-crystalline rocks. Crystallization may also trace other processes that are harder to detect unambiguously, such as dust-grain growth.

The processes usually thought to be most important in the conversion of pristine, interstellar dust grains to minerals are two (see, e.g., Wooden et al. 2005):

1. evaporation of the original grains, and re-condensation under conditions of high temperature and density, such that the products resemble those of chemical equilibrium (Grossman 1972, Gail 2001, 2004). This in turn favors magnesium-silicate minerals and metallic iron. Suitable conditions ($T > 1200$ K) are found only in close proximity to the central star, and are produced by stellar and stellar-accretion luminosity.

2. annealing of the grains at temperatures somewhat below their sublimation points (800-1200 K), such as can be found at $r \lesssim 1$ AU when viscous heating of the disk by accretion processes is substantial (Nuth et al. 2000, Gail 2001, Bockelée-Morvan et al. 2002, Wehrstedt & Gail 2002); or as can be produced over the central 10 AU by spiral density-wave shocks which grow from fluid-mechanical instabilities in the disk (Harker & Desch 2002).

If the source of crystalline silicates is in the inner disk $(r \lesssim 1 \text{ AU})$, a mechanism for large-scale radial mixing of the grains is also required, to match the distribution of crystalline silicate grains in the Solar system's remaining primitive material, and the profiles of silicate features in protoplanetary disks. Leading candidates for the radial grain-transport method include turbulent diffusion within the disk, with (Boss 2004) or without (Gail 2001, Bockelée-Morvan et al. 2002, Wehrstedt & Gail 2002) convection; large-scale meridional flows in an α-disk (Keller & Gail 2004); and reflux of material ejected from the inner disk by an X-wind (Shu et al. 1996, 2001). Turbulent diffusion and meridional flows would work as long as the disk is not broken by gaps cleared by planets; the X-wind method works as long as the object drives a vigorous outflow. Spiral-shock heating does not require an accompanying radial transport mechanism. There is currently no direct observational confirmation of any of these formation and distribution mechanisms, and there may be viable alternatives.

The great sensitivity of the Spitzer Space Telescope (Werner et al. 2004) permits observation of large samples of spectra of young stellar objects that are complete at small stellar masses. Thus we can explore the degree of dust-grain crystallization over a wide range of other properties of the systems, to search thereby for the relation of crystallization to the structure of the disk, and the relative importance of the mechanisms of crystallization and mixing. Here we report such a search for trends involving dust-grain emission features, in a large, coeval sample of protoplanetary disks around stars of Solar mass and smaller. We use the results of our *Spitzer* Infrared Spectrograph[11] (IRS; Houck et al. 2004) survey of members of the Taurus-Auriga YSO cluster (Forrest et al. 2004; d'Alessio et al. 2005; Calvet et al. 2005; Furlan et al. 2005, 2006; Sargent et al. 2006).

## 2   Observations and data reduction

The observations are described in more detail by Furlan et al. (2005, 2006). The present targets are 84 classical T Tauri stars with Class II infrared spectral energy distributions; all lie within the Taurus cloud, to which we adopt a distance of 140 pc (Kenyon & Hartmann 1995). All of the spectra were obtained in 2004 February, March and August, using *Spitzer*-IRS. We covered the full 5.3-38 $\mu$m IRS spectral range for all targets. Both orders of the IRS short-wavelength, low-spectral-resolution (SL) spectrograph were always used, in combination with either both orders of the long-wavelength, low-spectral-resolution spectrograph (LL), or, for targets brighter than 600 mJy at $\lambda = 12$ $\mu$m, both the short-wavelength (SH) and long-wavelength (LH) high-resolution spectrographs. The observations were designed to achieve a signal-to-noise ratio $(S/N)$ of at least 50 per spectral resolution element throughout the full wavelength range – that is, $S/N$ equal to the IRS specification for flat-field accuracy.

---

[11] The IRS was a collaborative venture between Cornell University and Ball Aerospace Corporation funded by NASA through the Jet Propulsion Laboratory and the Ames Research Center.

All data reduction was carried out with the SMART software package developed by the IRS instrument team (Higdon et al. 2004), along with software automation of our own devise. We began with the *Spitzer* Science Center IRS data pipeline, v. S13.2: the Basic Calibrated Data (BCD) product for low-resolution spectra, and the non-flatfielded "droop" product for high resolution. Permanently-bad detector pixels, and "rogue" pixels for which responsivity and noise are strongly illumination-dependent, were identified in *Spitzer* facility dark-current measurements made during the first two years of observatory operation, and were repaired by interpolation of signals from neighboring, undamaged pixels displaced in the dispersion direction within the two-dimensional spectral images. Sky emission in the off-target orders was subtracted from the on-target SL and LL observations. In our SH and LH observations, sky emission is negligible compared to object emission, as we verified by inspection of small raster maps (Furlan et al. 2006), so no sky emission was subtracted in these cases. We extracted point-source spectra from the SL and LL spectral images, using a variable-width window matched to the IRS point-spread function (Sargent et al. 2006; henceforth S06). Full-slit extraction sufficed for the SH and LH spectra. Identically-prepared spectra were made for two standard stars, $\alpha$ Lac (A1V) for SL and LL, and $\xi$ Dra (K2III) for SH and LH. Relative spectral response functions (RSRFs) were made by dividing these spectra into the template spectra corresponding to the stars (Cohen et al. 2003; Cohen 2004, private communication). Multiplication of RSRFs by our target spectra, nod position by nod position, and then averaging of the nods, completed the calibration process. Finally, as unresolved spectral lines were rarely present, we convolved and rebinned all SH and LH spectra to the same resolution and sampling as SL and LL. The resulting spectrophotometric uncertainty is 2-5%, as we estimate by comparison of the propagated noise-based uncertainties, the differences between nodded observations, and comparison to mid-infrared photometric measurements in the literature.

# 3 Analysis: extraction of emission features, and derivation of crystalline indices

## 3.1 Identification of the major emission features

The spectra exhibit a wide variety of solid-state features from small dust grains, superposed on the smooth continuum produced by each object's optically thick, flared, protoplanetary disk. A good example, the spectrum of IS Tau, appears in Figure 1. Several other examples, chosen to illustrate the full range of the emission features, appear in Figure 2; the complete collection is shown in the **online-only** supplementary material, and is also shown in somewhat different form by Furlan et al. (2006). Besides the broad signatures of amorphous silicates, which peak at 9.7 and 18 $\mu$m, we see relatively sharp features associated with several silicon-bearing minerals. Especially and frequently prominent are features characteristic of the simplest olivine minerals $\left(\mathrm{Mg}_{2x}\mathrm{Fe}_{2[1-x]}\mathrm{SiO}_4\right)$, centered at wavelengths 10.0, 10.8, 11.1, 12.0, 16.5, 19.0, 23.0, 23.7, 28.1 and 33.6 $\mu$m. Usually the peaks line up very well with those of forsterite, $\mathrm{Mg}_2\mathrm{SiO}_4$, as shown in Figure 1. So do the similar features seen in spectra of other kinds of primitive sub-planetary bodies such as that of Comet Hale-Bopp, also shown in Figure 1. However, and in contrast to this cometary spectrum, our sample objects typically also exhibit features of two other mineral families: the simplest pyroxenes $\left(Mg_xFe_{[1-x]}SiO_3\right)$, especially *via* a peak at 9.3 $\mu$m, and silica $\left(SiO_2\right)$, which is most conspicuous as a narrow feature at 12.5 $\mu$m, and as a shoulder on the short-wavelength side of the 10 $\mu$m complex, at about 8.6 $\mu$m. As shown in Figure 1, the positions of the strongest features of these species agree rather well with those of orthoenstatite $\left(MgSiO_3\right)$ and α-quartz.

The presence of these minerals is nothing new in itself. Olivines and pyroxenes in the dust of massive, deeply embedded YSOs have been studied for quite some time (e.g. Knacke et al. 1993, Malfait et al. 1998, Bouwman et al. 2001), and have been seen more recently in many low-mass YSOs like those in the present sample (e.g. Alexander et al. 2003, Przygodda et al. 2003, Meeus et al. 2003, Kessler-Silacci et al. 2005, 2006, Honda et al. 2006). Silicas are not as frequently studied, but are far from unknown (e.g. Honda et al. 2003, Uchida et al. 2004, S06, Bouwman et al. 2008).

Two examples of more-unusual spectral features are also shown in Figure 2. In the survey targets with the strongest 12.5 $\mu$m silica features, as in ZZ Tau (Figure 2), the relative strength of the three major silica features at 9.2, 12.5 and 20 $\mu$m are not often well explained by α-quartz. This might indicate a different polymorph of $SiO_2$, like tridymite or cristobalite (e.g. Speck 1998), or the effect of grain size or shape; we explore this possibility in a separate paper (Sargent et al. 2008a). In UX Tau A (Figure 2), the 10 $\mu$m silicate complex is joined by sharp but spectrally-resolved features at 11.3, 12.0 and 12.7 $\mu$m, evidently the family of C-H out-of-plane bending modes of polycyclic aromatic hydrocarbons (PAHs). This is strange in two respects. First, PAH features are commonly seen in intermediate-mass YSOs such as Herbig Ae stars (van Boekel et al. 2005, Sloan et al. 2005, Keller et al. 2008), that can provide abundant ultraviolet excitation for the molecules; with spectral type K5, UX Tau A is much cooler than is common for young stars with PAH emission spectra. Second, PAH features at 6.2, 7.7 and 8.6 $\mu$m appear in the spectra of more massive YSOs (e.g. in V892 Tau, AB Aur and SU Aur; Furlan et al. 2006), and are usually much brighter than the 11-13 $\mu$m features, but are weak or absent in the spectrum of UX Tau A.

Spectra of members of our sample have been produced by many other observers. The largest current collection of overlapping measurements is that by Honda et al. (2006; henceforth H06), whose ground-based spectroscopic observations in the $8-13$ $\mu$m atmospheric window covered twenty-four of the 84 objects on our list. There are, unfortunately, major differences between the present spectra and those by H06. Leaving aside three multiple-star systems which H06 could resolve spatially and we could not, there are 18 objects to compare, 12 of which were observed by H06 within two months of our observations. Of these, only two of the H06 spectra (those of CZ Tau and IQ Tau) agree with ours within the uncertainties, and even in these there are significant differences in the silicate profiles. The other sixteen H06 spectra are quite different in flux-density level and silicate profile from ours. Unresolved multiplicity (along with the larger size of the IRS entrance slit) cannot account for the differences, as the vast majority of these objects have been imaged at infrared wavelengths with 0.05-0.1 arcsec resolution, and all companions so discovered are accounted for in our comparison. Nor can intrinsic variability plausibly explain the differences. There is no tendency for the twelve H06 observations within two months of ours to agree better than the six taken two years earlier. The median magnitude difference between the spectra at a wavelength of 8 $\mu$m is 0.24 mag, and variation at this wavelength and on this scale is seen in only 1-2% of Class II objects in the similar Cha I cloud (see, e.g., Luhman et al. 2008). Also, eleven of the overlapping Taurus objects have been observed with *Spitzer*-IRS in five epochs over 1.5 years starting with the observations we discuss here. With the possible exceptions of the famously-variable DG Tau and RY Tau, none of the other vary within the IRS band by enough to account for the difference between the present spectra and those in H06 (J. Bary, private communication, 2008; Leisenring et al. 2008, in preparation).

## 3.2 Objects with "pristine," interstellar-like, dust emission features

Only five of the survey targets lack any evidence of minerals, presenting silicate-feature profiles indistinguishable from those of interstellar dust. Three of the five are the transitional-disk objects CoKu Tau/4, DM Tau, and GM Aur, in which extremely dust-free, several-AU-scale, central clearings or radial gaps are inferred from absence of infrared excess at the shorter IRS wavelengths. As we have discussed elsewhere (Forrest et al. 2004, d'Alessio et al. 2005, Calvet et al. 2005), the interstellar-like silicate profile is explained naturally in a leading scenario for the formation of the clearings: the gravitational action of a giant-planetary (or, in the case of CoKu Tau/4, stellar) companion, which rapidly drives the warmest (and perhaps most thoroughly processed) parts of the disk toward the star to be accreted or ejected, and leaves behind the distant, outer disk, in which no significant dust-grain processing has yet taken place. We have modeled the composition of dust in these systems (S06), and obtained upper limits on the mass fraction of crystalline silicates below the best upper limits for the interstellar medium, 1-1.5% (Min et al. 2007).

The other two objects with essentially interstellar-like silicate profiles are LkCa 15 and UY Aur (Figure 2). These objects have infrared excesses at all IRS wavelengths, and therefore have optically-thick disks with inner edges very close to their central stars. As they are bright and the quality of their spectra is high, we have adopted LkCa 15 and UY Aur as exemplars of emission from amorphous, submicron, interstellar-like (henceforth "pristine") dust grains. A few other members of the sample (e.g. FM Tau) have silicate features nearly as pristine.

The existence of objects like LkCa 15 and UY Aur – disks a million years old with inner edges near their central stars,[12] that yet have essentially no small crystalline-silicate

________________________

[12] The LkCa 15 disk has a large gap (Piétu et al. 2006) – that is, it is a transitional disk – but the gap is separated from the star by an optically-thick inner disk, which in turn extends to the dust-sublimation point (Espaillat et al. 2007; Espaillat et al. 2008); thus the disk still has an inner edge near the central star.

grains – is perhaps the most interesting finding of this study, and these objects will figure prominently in the following discussion.

### 3.3 Crystalline dust emission features and crystalline mass-fraction indices

We detect silicate and silica emission features from an optically-thin layer of dust on the nearer surface of the disk, heated by starlight to temperatures higher than the opaque disk beneath. That the layer is optically thin, and the emission thermal in origin, means that an emission-feature flux is proportional to the column density of the species that gives it rise, and thus that the fluxes can be used to determine the relative abundances or mass fractions of the optically-thin dust's components. We explore in the following the dependence of crystalline abundances on other properties of the present Class II YSO systems, in search of the mechanism of the processing that has led to the partial crystallization.

In principle, the crystalline mass fractions can be determined by fitting the spectra with physical models of the disks and their optically-thin "atmospheres" in combination with laboratory optical constants of the dust components. With various simplifying assumptions, this technique has been applied to silicate emission in the spectra of Class II objects and Herbig Ae/Be stars by many previous workers (e.g. Przygodda et al. 2003, Meeus et al. 2003, van Boekel et al. 2005, Kessler-Silacci et al. 2005, Bouwman et al. 2008). We have used it in models of the $10 \mu$m complex for ten of the present objects (S06). Such models are difficult to construct, as there are many details of grain shape, size and porosity, of composition (e.g. Mg/Fe relative abundance), and of radial and vertical structure of the disks, for which we must account.

However, we have enough information at present to construct good *proxies* for the crystalline mass fractions, in the form of indices based upon the contrast of the various dust emission features. These indices are relatively easy to calculate, and can be used reliably to search for trends. This alternative has been fruitful before in the analysis of Class II YSO spectra (e.g. Kessler-Silacci et al. 2005, 2006) and Herbig Ae/Be stars (e.g. Bouwman et al. 2001, van Boekel et al. 2003).

In Figure 3 we illustrate the calculation of the crystalline silicate indices that we have found useful in this case. First we fit a smooth curve – a fifth-order polynomial in wavelength – to the ranges in the observed flux density $(F_\nu)$ spectrum within which dust-feature emission is small: $5.61 - 7.94$ $\mu$m, $13.02 - 13.50$ $\mu$m, $14.32 - 14.83$ $\mu$m, $30.16 - 32.19$ $\mu$m, and $35.07 - 35.92$ $\mu$m, thus producing a continuum flux-density spectrum $F_{\nu,C}$. From this we construct a quantity that we may call the equivalent width per channel, $W_\nu$:

$$W_\nu(\nu) = \frac{F_\nu - F_{\nu,C}}{F_{\nu,C}} \quad . \tag{1}$$

In the limit of small dust-feature optical depth $\tau_\nu$, $W_\nu \rightarrow \tau_\nu B_\nu(T_{\text{feature}})/B_\nu(T_{\text{continuum}})$ $\approx \tau_\nu$. The integral over frequency is the standard definition of equivalent width, $W = \int W_\nu d\nu$. We construct a "pristine-dust" reference spectrum, $W_{\nu,0}$, by averaging $W_\nu$ for the two bright objects with interstellar-like silicate features, LkCa 15 and UY Aur. The object spectrum $F_\nu$, continuum $F_{\nu,C}$, equivalent width per channel $W_\nu$, and pristine reference $W_{\nu,0}$ are all plotted for each of the examples shown in Figures 2-3.

Useful indices of crystalline-species relative abundance can be constructed by comparison of the contrast of a feature in a spectrum to the same ratio in the pristine reference spectrum. We define indices for the strongest features of pyroxenes, olivines and silica in the 10 $\mu$m complex, lying respectively at wavelengths $\lambda_P, \lambda_O, \lambda_S = 9.21$, 11.08 and 12.46 $\mu$m, by using ratios to the values of $W_\nu$ near the peak of the pristine silicate feature, $\lambda_R = 9.94$ $\mu$m, integrated over a frequency range equivalent to $2\Delta\lambda = 0.545$ $\mu$m, and divided by the same ratios for the "pristine-dust" reference spectrum, $W_{\nu,0}$, as shown in Figure 3:

$$X_{10} = \frac{\int_{\nu_X - \Delta\nu}^{\nu_X + \Delta\nu} W_\nu d\nu}{\int_{\nu_R - \Delta\nu}^{\nu_R + \Delta\nu} W_\nu d\nu} \frac{\int_{\nu_R - \Delta\nu}^{\nu_R + \Delta\nu} W_{\nu,0} d\nu}{\int_{\nu_X - \Delta\nu}^{\nu_X + \Delta\nu} W_{\nu,0} d\nu} \quad , \quad X = P, O, S. \tag{2}$$

An index thus approaches unity if the emission feature approaches the pristine profile, and increases above unity with increasing prominence of a crystalline signature above the pristine profile. We will refer henceforth to the indices $P_{10}$, $O_{10}$, and $S_{10}$ as the ten-micron pyroxene, olivine and silica indices.

In Figure 4 we plot the orthoenstatite, forsterite, and large (5 μm) grain abundances determined from the 10 $\mu$m silicate complex in twelve Class II objects (ten in Taurus) by S06, against the pyroxene and olivine indices, $P_{10}$ and $O_{10}$, calculated from the same spectra. The correlations evident in this plot demonstrate that the indices $P_{10}$ and $O_{10}$ are good tracers of the fraction of silicates of pyroxene and olivine composition that subsist in crystalline form, and are *not* good tracers of the fraction of solid material present as large grains. We expect that the same would prove true of the silica index $S_{10}$, but too few of the objects modeled by S06 have silica features to make a comparison useful.

A rather stronger correlation between the $O_{10}$ index and the large-grain mass fraction would be obtained if we use the opacity-modelling results by H06. These authors obtain systematically larger mass fractions in large grains, modeling their own observations, than do S06, modeling *Spitzer*-IRS observations of the same population of objects. This seems mostly to be a result of the substantial differences between the H06 observations and ours, as discussed above (section 3.1), rather than any difference in the modeling methods of H06 and S06. It also seems likely that some of the difference between the dust composition derived in H06 and S06 arises from the restriction of the H06 observations and models to a smaller wavelength range than our spectra, and in particular from H06 not having access to the $13-15\,\mu$m range where the silicate-emission minimum lies. Thus any correlation of our $O_{10}$ index with large-grain mass fractions inferred from the H06 results would need to be regarded with caution.

Similar indices can be defined for the crystalline-silicate features superposed on the 18 $\mu$m complex. However, in this case we also have recourse to the 33.6 $\mu$m emission feature of forsterite, the only crystalline-silicate feature in our spectra that is distinct from the 10- and 18-$\mu$m complexes, and this turns out to render indices constructed from the $16-28$ $\mu$m features somewhat redundant. We define an index $O_{33}$ for this feature based upon a ratio to one of the stretches of the 18-$\mu$m complex that has no strong crystalline features evident, the 1.53 $\mu$m wide band centered on $\lambda = 21.7$ $\mu$m. $O_{33}$ is calculated in analogy with Equation 2, using $\lambda_R = 21.7$ $\mu$m. We do not, however, normalize $O_{33}$ to the pristine profiles, since this would involve dividing by a number approaching zero: the pristine exemplars have essentially zero silicate-feature emission at 33.6 $\mu$m. Unlike the other indices, this long-wavelength olivine index thus approaches *zero* as the silicate profiles approach the pristine ones, and increases with increasing prominence of the 33 $\mu$m olivine feature relative to the underlying pristine component of the 18 $\mu$m complex.

Table 1 is a list of the crystalline indices $P_{10}, O_{10}, S_{10}$, and $O_{33}$, the equivalent widths of the 10 and 18 $\mu$m complexes and the 33.6 $\mu$m feature, $W_{10}, W_{20}$, and $W_{33}$, and the full width at half maximum of the 10 $\mu$m complex, $\Delta\lambda_{10}$, together with the uncertainty in each quantity as propagated from the uncertainties in each spectrum, for each of the systems in our study. The rather large range of crystalline indices is illustrated with histograms in Figure 5.

### 3.4 Continuum spectral indices

Furlan et al. (2005, 2006) have defined continuum spectral indices, $n_{6-13}$ and $n_{13-25}$, for this collection of objects, from their emission at wavelengths near $\lambda = 6, 13,$ and $25\ \mu m$ :

$$n_{a-b} = \log\left(\lambda_b F_\lambda\left[\lambda_b\right]/\lambda_a F_\lambda\left[\lambda_a\right]\right)/\log\left(\lambda_b/\lambda_a\right) \quad . \tag{3}$$

for which there are no strong spectral features, and for which the slope of the spectrum indicates that the emission is heavily dominated by the optically-thick disk (see, e.g., Hartmann 1998). The relationship between these continuum indices has been shown by Furlan et al. (see also d'Alessio et al. 2006) to reveal the depletion of dust from the higher-elevation portions of the disk, or, in other words, to indicate the degree of settling of dust to the midplane. In particular, Furlan et al. show by comparison to grids of detailed models that smaller values of $n_{13-25}$ among the sample correspond to the "most settled" disks, with dust-to-gas ratios at the highest elevations smaller by 2-3 orders of magnitude than in the disks with the largest values of $n_{13-25}$. We present here another, analogous, index, $n_{13-31}$. This index involves a longer-wavelength band, at $30.16 - 32.19\ \mu m$, chosen to avoid all significant amorphous or crystalline silicate emission features over the continuum from the optically-thick disk. Like $n_{13-25}$, $n_{13-31}$ tracks the scale height of dust, and varies inversely with the degree of sedimentation of the disk; as we will see, it behaves the same as $n_{13-25}$ in most other respects as well. All three continuum spectral indices are presented in Table 2.

# 4  Discussion: trends among the spectral features, and the processing of dust

## 4.1  Properties of the stars and disks in the present sample

The 84 YSOs in the present sample are T Tauri stars that have infrared excesses with Class II spectral indices, most of which are well-studied at many wavelengths. They are described in more detail by Furlan et al. (2006). About half of them are known to belong to multiple-star systems, and the other half are currently thought to be single stars. Reliable spectral types have been determined for most of them. All but a few have strong emission lines (i.e. are classical T Tauri stars) within significantly veiled visible and near-infrared spectra, from which mass accretion rates have been derived. Recent submillimeter continuum observations have been used to estimate the masses of most of their protoplanetary disks (Andrews & Williams 2005).

Values for luminosity and mass can be found for many of the stars in numerous publications. In order to ensure that the masses and luminosities that we use in the following have been obtained in a consistent fashion, we have by standard means re-derived these quantities from the spectral types and from visible and near-infrared (2MASS) broadband photometry. We have first estimated the extinction toward the stars from their *V-I* or *V-J* colors, assuming this extinction to be due to diffuse interstellar dust with total-to-selective-extinction ratio $R_V = 3.1$ (Cardelli et al. 1989, Mathis 1990). Then we used each star's corrected visible and near-infrared magnitudes and spectral type to estimate its luminosity and effective temperature. From comparison of these results to theoretical stellar-evolutionary tracks and isochrones, and an assumed uncertainty of a quarter of a spectral class, we estimated the age of the Taurus population, and interpolated the mass and an associated uncertainty for each star. This last step is illustrated in the H-R diagram shown in Figure 6, in which the stellar absolute magnitudes and effective temperatures are plotted along with pre-main-sequence isochrones and tracks by Siess et al. (2000). As many have done before us, we interpret this diagram as that of a stellar cluster formed over a time interval smaller than the typical stellar age, with scatter in the luminosity direction due to variation in uncorrected selective extinction and accretion power, and to undetected multiplicity. We take the typical age from the center of the distribution in Figure 6 to be 2 Myr, and use the corresponding isochrone to interpolate masses for the stars. As a check we also compare in Table 3 stellar masses computed as above with the masses of six single stars – BP Tau, CY Tau, DL Tau, DM Tau, GM Aur and LkCa 15 – determined from the kinematics of the disks around these stars (Simon, Dutrey and Guilloteau 2001). Good agreement is found thereby; the average ratio of the masses is 0.97±0.25.

All of the parameters of the stars and disks we seek to compare to the dust properties are listed in Table 3. The sample covers a considerable range of luminosity, accretion rate, estimated mass of disk relative to star, and stellar mass in the range $0.3-2M_{\odot}$, with high statistical significance.

## 4.2 Location of the crystalline silicates, and the large range of inner-disk crystalline mass fraction in the sample

The transitional disks around CoKu Tau/4 and DM Tau are free of small dust grains in their central regions (radii 10 AU and 3 AU respectively), and also have relatively weak and pristine-looking $10\,\mu$m silicate features. S06 have noted this to be an indication that the crystalline silicates in the other disks are mostly confined to $r \lesssim 10\,\mathrm{AU}$, as has been observed directly in Herbig Ae/Be disks by van Boekel et al. (2004). Within the present sample, this suggestion finds support from the statistics of the long-wavelength (20-38 $\mu$m) crystalline silicate features. All of the objects with strong $10\,\mu$m silicate features have strong $18\,\mu$m features. However, although more than 90% of these objects also have strong $10\,\mu$m crystalline silicate features, only about 50% have detectable 20-38 $\mu$m crystalline silicate features. Thus the domain of crystalline silicates is significantly smaller than that of the amorphous silicates, which in turn requires $T \sim 200\,\mathrm{K}$, and radii within a few AU of a star, for the crystalline grains.

Though the range of ages within the sample is small compared to the average age (Kenyon & Hartmann 1995; see also Figure 6), we see a very large range of the $10\,\mu$m crystalline indices (Figure 5), and infer a correspondingly large range in crystalline-silicate mass fraction in the inner disks. Objects such as IS Tau have inferred inner-disk crystalline silicate mass fractions in the 80-100% range, while the "pristine exemplars" have less than 0.5%; yet in most respects besides crystallinity, these systems are similar today. Such a large range of crystalline mass fraction is hard to understand in terms of *steady* application of any of the dust thermal processing and mixing mechanisms that have been mentioned above (section 1). Particularly vexing are the systems in which there are very few crystalline-silicate grains, like LkCa 15, UY Aur, and FM Tau, as this would apparently require peculiarly low levels of crystallization and mixing activity through 1-2 Myr despite luminosity and accretion rate in the normal range.

### 4.3 Trends among the dust emission features: crystallinity and sedimentation

We have searched for correlations among the crystalline indices and widths derived from the spectra (Table 1), the continuum spectral indices (Table 2), and the global properties of the star-disk systems (Table 3). Table 4 is list of linear correlation coefficients (Pearson's $r$) for all pairs $(x, y)$ of these quantities:

$$r = \sum_i (x_i - \bar{x})(y_i - \bar{y}) \left[ \sum_i (x_i - \bar{x})^2 \sum_i (y_i - \bar{y})^2 \right]^{-1/2} \quad , \tag{4}$$

and the probability $p_{\text{rand}}(r, N)$ that the correlation could have been produced by a random distribution in a sample of the same size, $N$:

$$p_{\text{rand}}(N, r) = \frac{2\Gamma\left(\frac{N-1}{2}\right)}{\sqrt{\pi}\,\Gamma\left(\frac{N-2}{2}\right)} \int_{|r|}^{1} \left(1 - u^2\right)^{\frac{N-4}{2}} du \tag{5}$$

(see, for instance, Taylor 1997). Pairs of quantities linked with small $p_{\text{rand}}$ ($\leq 1\%$) are significantly correlated; pairs linked with particularly large $p_{\text{rand}}$ are extremely unlikely to be related. The strongest correlations listed in Table 4 have $|r| = 0.6 - 0.7$, and are observed among pairs like mass-luminosity and $W_{10} - W_{20}$ that are known from other evidence to be correlated.

In Table 4 one can see many significant trends between the indices and equivalent widths we have extracted from the mid-infrared spectra. Some of these trends are illustrated in Figure 7, which contains several plots with the 10 $\mu$m olivine index, $O_{10}$, as the independent variable. In general we find that $O_{10}$ is strongly and positively correlated with the other 10 $\mu$m crystalline indices, and with the width of the 10 $\mu$m silicate feature, $\Delta\lambda_{10}$. The crystalline indices are all negatively correlated with the equivalent widths, $W_{10}$ and $W_{20}$, and with the continuum spectral indices, $n_{6-13}$, $n_{13-25}$, and $n_{13-31}$. The longer-wavelength crystalline indices, here represented by $O_{33}$, also are strongly and positively correlated with the shorter-wavelength ones; curiously, $O_{33}$ is even more strongly correlated with $P_{10}$ than with $O_{10}$.

From these trends several firm conclusions emerge. First, the positive correlation among all of the crystalline indices indicates that the agent responsible for the crystallization of the initially amorphous material does not favor one mineral family over another, which would have given weaker correlation. The crystallization must also not involve much transformation or incorporation of one mineral into another, which would have given negative correlations among the indices. Second, the positive correlation between $\Delta\lambda_{10}$ and the crystalline indices, taken along with the typically high contrast of the narrow crystalline features, is consistent with the width of the 10-μm feature growing because of the appearance of the mineral features in small dust grains, along with whatever broadening of the profiles arises from growth of grains to larger diameters ($\gtrsim 1\ \mu$m). That the width of the 10-μm feature can grow substantially without large dust grains can easily be demonstrated from lab-opacity-based grain models, as can be seen in the recent work by Voshchinnikov & Henning (2008) and Min et al. (2008). Third, the negative correlations between the crystalline indices and the continuum spectral indices indicate that crystalline dust mass fractions are generally larger, the more the dust has settled to the disk midplane. The latter is an encouraging sign that two important markers of protoplanetary-disk evolution – dust mineralization and sedimentation – track each other, but it begs the question of the process by which the dust is partially crystallized.

## 4.4   Disk-structure evolution, or grain growth?

Large particles settle to the midplane, within the optically-thick disk, much faster than small ones (Goldreich & Ward 1973). Thus observations of the mid-infrared emission features, which probe the optically-thin disk "atmospheres", are strongly biased toward small grains. Nevertheless, such emission has often been searched for signatures of dust-grain growth, in efforts to find constraints on the timescale for planetesimal development, and some of the trends that appear in Table 3 and Figure 7 have previously been searched for, or noted, in this context.

In particular, the negative correlation we observe between the 10 $\mu$m olivine index, $O_{10}$, and the 10 $\mu$m silicate-feature equivalent width, $W_{10}$, is the same as that reported by Kessler-Silacci et al. (2006) for a sample of protoplanetary disks covering several young clusters. By comparison with the behavior of opacity in dust over a range of grain sizes, Kessler-Silacci et al. show that this trend is consistent with growth of grains to several-micron diameter, and conclude that grain growth is what they observe. But this conclusion involves a tacit assumption that all the disks have the same shape (degree of flaring and sedimentation), so that any variation in $W_{10}$ can be attributed to a change in grain properties[13]. In fact, protoplanetary disks of a given age and type have a range of sedimentation, measured by the continuum spectral indices $n_{13-25}$ and $n_{13-31}$ (Furlan et al. 2005, 2006; d'Alessio et al. 2006), so they will have a corresponding variation of brightness of the dust component "atmospheres," measured by $W_{10}$. As we have seen above, $O_{10}$ is also negatively correlated, and $W_{10}$ positively correlated, with $n_{13-25}$ and $n_{13-31}$, *so the trend between crystallinity and silicate-feature equivalent width can also be explained by small grains in disks with a range of dust-settling to the midplane.*

To illustrate this point, we compare our $n_{13-31}$ and $W_{10}$ data with values calculated from model protoplanetary disks in which the degree of sedimentation can be adjusted

---

[13] This also involves an assumption that other aspects of grain structure, like porosity (Voshchinnikov & Henning 2008), do not significantly affect $W_{10}$ or $\Delta\lambda_{10}$.

(d'Alessio et al. 2006). In these models we include only small (radius $\leq 0.25\,\mu m$) dust grains with interstellar-like opacity (Weingartner & Draine 2001) in the upper disk layers, taken to be that above a tenth of the local gas-pressure scale height. We vary several other system parameters over ranges appropriate for our Taurus sample: stellar mass $M$, luminosity $L$, and radius $R$; disk accretion rate $dM/dt$ (assumed equal to the stellar accretion rate $dM_{\text{star}}/dt$); viscosity parameter $\alpha$; disk axis orientation $i$ with respect to the line of sight; and sedimentation parameter $\varepsilon$. The latter parameter is the dust-to-gas mass density ratio in the upper disk layers, normalized to the standard interstellar dust/gas mass ratio of 0.01. The results are shown in Figure 8. There we see that most of the Taurus data are concentrated in the region of the plot around $n_{13-31} = -0.8$ and $W_{10} = 1\,\mu m$, and are indicated by the models to have $\varepsilon = 0.001 - 0.01$, $M = 0.4 - 0.8 M_{\odot}$, $dM/dt = 10^{-9} - 10^{-8} M_{\odot}\ \text{year}^{-1}$, $i = 11 - 64°$, and $\alpha = 0.001 - 0.02$. Almost all the rest of the data points lie at larger $n_{13-31}$ *and* $W_{10}$, and are encompassed by the models simply by extending the range of $\varepsilon$ up to the interstellar value of unity. Apart from the known transitional disks, which are structurally different from the rest, only a half-dozen objects, with $W_{10} = 3 - 5\,\mu m$, lie outside the model range. These objects apparently possess a larger optically-thin dust component than expected for their optically-thick dust disk. Perhaps they, like the transitional disks, have a structure different from the rest: for example, they may have optically-thin radial gaps[14].

---

[14] A few of the outliers – RY Tau, GK Tau, and V836 Tau – have been suggested to have radial gaps on the

Thus, much of the correlation between $n_{13-31}$ and $W_{10}$ reflects the range of sedimentation of submicron grains, not their growth to larger sizes. As the dust settles to the midplane, the starlight-absorbing surface is illuminated more obliquely, and the silicate-feature emission, by small dust grains in the optically-thin upper layers of the disk, decreases relative to the emission by the cooler, optically-thick disk underneath. Surely grains do grow as the dust settles to the midplane and the disk structure changes, but this process is not evident in any of the indices we have constructed. A better way to search for grain growth is by detailed modeling of silicate-feature profiles with mineral opacities, such as that undertaken by S06, H06, van Boekel et al. (2005) or Bouwman et al. (2001, 2008); we will present such results for the present sample in a future paper (Sargent et al. 2008b, in preparation).

### 4.5    Search for trends among the crystalline emission features and system properties

The solid matter we observe originated in small, amorphous interstellar grains, and underwent their transition to a partially crystalline state while in their present disks; we can hope that trends among the observed crystallinity, and other properties of the disks and central stars, would reveal the mechanism of transition. Therefore we have searched for correlation between the crystalline indices and mass, luminosity, and accretion rate of the host star(s), and with the disk mass and disk/star mass ratio. A crystallinity trend with the former three properties might be traceable to radiative

---

basis of their infrared CO-line emission (Najita, Carr and Mathieu 2003).

vaporization of grains at the center, or annealing within the disk. A correlation among crystallinity and disk mass or disk/star mass ratio may point to spiral density waves and shocks as the means of dust processing, because the propensity for development of fluid instabilities in the disk is linked to these quantities. The Taurus cluster age, 2 Myr, is much longer than the time scale $(10^4 - 10^5$ year) for complete conversion to crystallinity by any of these mechanisms (Gail 2004, Harker and Desch 2002). Thus we can test most of the proposed grain-crystallization processes.

The results of the search are shown in Table 4, Figure 9 and Figure 10, and are easy to summarize: there is a striking lack of strong correlation between crystallinity and any of the global properties of the protoplanetary systems. The correlation coefficients $r$ are the lowest, and the probability of reproduction by a random distribution, $p_{\mathrm{rand}}(r, N)$, the highest, among all of the correlations we calculate in Table 4. Correlation between crystallinity and stellar mass or luminosity can be rejected with particularly high confidence. The lack of correlation with disk mass or disk/star mass ratio needs to be regarded more cautiously, as submillimeter-continuum-derived disk masses are probably substantial underestimates of the true disk masses (e.g. Hartmann et al. 2006). The lack of correlation we observe may still be significant, if the submillimeter-derived disk-mass estimates are *systematic* underestimates. Taken together, these lacks of correlation weaken the cases for central grain evaporation/condensation and radial mixing, and – with the caveats above – for annealing *in situ* due to spiral-shock heating.

Two sets of *weak* trends emerge, though, at the level of $r \sim 0.2$ and $p_{rand} = 5-10\%$, involving two system parameters which are uncorrelated themselves.

### 4.5.1  *Weak trends of crystallinity with stellar accretion rate*

The stellar accretion rate is negatively correlated with the pyroxene index $P_{10}$, and positively correlated with the olivine and silica indices $O_{10}$ and $S_{10}$. This is not as helpful as it sounds. The trends are weak enough to be difficult to discern in Figure 10, so we cannot place tremendous confidence in their reality. Taking them at face value, and taking present stellar accretion rate as a tracer of the past accretion–processing rate in the disk, they contradict the models of annealing by viscous heating in the inner disk, which would predict positive correlation for both pyroxenes and olivines (Gail 2001, 2004; Nuth & Johnson 2006). The simplest situation calling for opposite trends in pyroxenes and olivines would be the approach by different degrees to chemical equilibrium at high temperature, which would favor pyroxenes – orthoenstatite, in particular – over olivines (Gail 2004). In turn, such a situation would be related to accretion rate, if crystallization were a result of evaporation and recondensation close to the star, and if accretion luminosity were a significant fraction of the total.

### 4.5.2  *Weak trends involving stellar multiplicity*

Stellar multiplicity is correlated positively with with $O_{10}$ and $S_{10}$ (but not $P_{10}$ or $O_{33}$), and negatively with the 10 $\mu$m silicate feature width $\Delta\lambda_{10}$ (but not the equivalent width $W_{10}$) and the continuum spectral index $n_{6-13}$ (but not $n_{13-25}$ or $n_{13-31}$). In Figure 9 the correlation of multiplicity with $O_{10}$, and the lack of correlation with $O_{33}$, can be seen clearly. This could potentially be a trace of the stronger correlation noted above between crystallinity and sedimentation, and an indication that at least dust in the *inner* disks tends to be more settled in multiple systems; note, however, that there is no correlation between multiplicity and dust settling in the outer disk (Furlan et al. 2006; see also Table 4).

## 4.6    Origin of the dust-crystallinity variations

In summary, we can say that none of the standard models of dust-grain crystallization is consistent, by itself, with the present large body of silicate and silica data in Taurus. It would help to have a heating or mixing process capable of dominating the effects discussed above, that can exhibit wide variation among systems which appear similar at visible-IR wavelengths. Alternatively, it would do to have one or more of the standard dust crystallization/mixing mechanisms in operation, but along with an additional process within the disk that erases the trends among observable quantities that are produced thereby. We hereby offer one suggestion of each type.

### 4.6.1    *High-energy annealing of dust grains.*

The propensity for high X-ray luminosity varies substantially among Class II YSOs of given mass or spectral type (see, e.g., Preibisch et al. 2005). Even more unpredictably variable is the propensity for X-ray flaring, and X-ray flares can be extremely luminous. Given large enough X-ray luminosity, it is possible by absorption of multiple X-rays to anneal or vaporize small grains throughout the inner-disk surfaces, *in situ*. This mechanism would naturally produce crystalline dust mass fractions that are no better correlated with stellar, disk, or accretion properties than are the X-ray properties. It is possible to test this suggestion by a search for trends among silicate-grain crystalline mass fraction, and X-ray luminosity and flaring frequency. Some of these comparisons could be made between the present base of data and the XMM-*Newton* X-ray survey of Taurus (Güdel et al. 2007). One particularly intriguing experiment of this sort would be to acquire mid-infrared silicate-feature observations of the YSOs in the Orion clouds that were detected in the COUP X-ray survey (Getman et al. 2005), as this sample is large and contains many examples of X-ray-flaring YSOs (Favata et al. 2005).

**4.6.2** *Giant planet formation and migration: the "born-again" disk.*

This idea was discussed briefly by S06. Giant-planet or brown-dwarf formation in the disk within the first 1-2 Myr of system life, and the subsequent clearing of the disk interior to the new companion's orbit on ~3000-orbit time scales, are currently the best explanations for the remarkably sharp inner edges, the virtually dust-free central clearings or radial gaps, and the pristine state of the remaining optically-thin dust, in transitional-disk systems like DM Tau (Calvet et al. 2005, Varnière et al. 2006) and GM Aur (Calvet et al. 2005). However, the stability of the resulting configuration depends upon the relative masses of companion and outer disk. As soon as the inner edge of the disk accumulates a mass comparable to that of the companion, type II orbital migration (e.g. Nelson & Papaloizou 2004; Terquem 2004) sets in. Thereafter the planet would drift inward to be consumed or ejected by the central star, followed by the inner edge of the disk, all on a viscous time scale,

$$\tau_\nu = \frac{\tau_{orb}}{2\pi\alpha}\left(\frac{v_{orb}}{c_s}\right)^2 \approx 10^5 \text{ years} \ll \text{Taurus age} \quad , \quad (7)$$

for typical parameter values (viscosity parameter $\alpha = 0.01$, ratio of sound and orbital speeds $c_s/v_{orb} = h/r = 0.05$, orbital period $\tau_{orb} = 40$ years).

This suggests the following scenario:

1. A protoplanetary disk like that around IS Tau, the dust in which through its life has settled vertically and experienced compositional changes – in particular, has possibly developed a high crystalline mass fraction in its inner regions, by any of the means discussed above – gives rise to a giant planet, several AU away from the central star.

2. Torques from the new planet assist in the rapid ($\lesssim 10^4$ year; Varnière et al 2006) clearing of the inner disk, which eliminates virtually all of the crystalline dust. The massive outer disk remains, its dust essentially pristine, and its further progress toward the star is blocked by the planet's mean-motion orbital resonances. Its structure and mid-infrared spectrum now resemble those of DM Tau, which has a massive outer disk.

3. A few hundred thousand years later, the duration depending upon the mass of the planet, Type II orbital migration commences, and about $10^5$ years further, the planet is gone and the central clearing has been backfilled with pristine material. Now, although it may be well over 1 Myr old, and extends to small radii, the disk's dust is like new, throughout; it is "born again." Its spectrum resembles that of UY Aur.

4. Repeat, if possible, until the outer disk is no longer capable of migrating the planet. The end structure, and spectrum, might resemble those of CoKu Tau/4, for which the outer disk is apparently not very massive, though in this case the disk appears to be truncated by the binary star it contains (Ireland & Kraus 2008) instead of a giant planet.

In a population of disks in various stages of this process, a very wide range of inner-disk crystalline mass fraction would be evident, ranging from essentially zero in the "born again" disks to nearly unity in disks that have gone a long time between giant planets. We would thus see no strong trends of crystalline-dust mass fraction with stellar, accretion, or disk properties.

## 5 Conclusions

We have characterized the crystalline-silicate content of the protoplanetary disks of Taurus, and explored the relations between the contents and the other observable features of the disks and their central stars. The main conclusions are as follows.

1. The easily-calculated crystalline indices (Figure 3; Equations 1-2; Table 1) that we have constructed serve adequately as proxies of the mass fractions of crystalline pyroxenes, olivines, and silicas in the optically-thin surface layers of protoplanetary disks. The indices may also be sensitive to large-grain mass fractions, as has been found for what we call the $O_{10}$ index in Herbig Ae/Be systems (e.g. van Boekel et al. 2003), but we find in this sample of Class II YSOs that all of the indices are more tightly correlated with mineral mass fractions.

2. The sample exhibits a range of crystalline-silicate dust-mass fraction in its inner disks – from essentially none to nearly 100% – that is surprisingly wide, considering that the disks, their ages, and their central stars are so similar (Figure 2, Figure 5, Table 1). Especially worthy of note is a small fraction (2%) of the population that exhibits silicate features indistinguishable from the interstellar profile, but is otherwise quite similar to the more heavily-processed majority. The frequency of appearance of the long- and short-wavelength crystalline-silicate emission features indicates that the crystalline silicates are confined to the central several AU of the disks, as expected.

3. Correlations among the crystalline indices, the equivalent widths and full widths at half-maximum of the silicate complexes, and the continuum spectral indices of the underlying, optically-thick disks, are all consistent with a general increase in crystalline-silicate dust-mass fraction as more of the dust settles to the midplane (Table 4, Figure 7). Presumably this joint evolution in dust composition and disk structure is accompanied by grain growth, but our technique is not capable of revealing this.

4. The crystalline indices are uncorrelated with stellar mass, stellar luminosity, disk mass, and disk/star mass ratio (Table 4, Figure 9, Figure 10). Only weak correlations of some of the crystalline indices are seen with stellar accretion rate and stellar multiplicity. These results appear to contradict the predictions of all of the standard models of crystalline-silicate production and radial mixing of dust, in their current forms. Either another grain-crystallizing mechanism dominates over these, or another process must be at work within the disks to erase the correlations they produce.

5. Accordingly, we propose one of each sort that seem qualitatively not to be contradicted by the sum of the evidence, and thus to be worth further investigation. X-ray heating and annealing of dust grains is introduced as an alternative dust processing mechanism. Giant-planetary formation and migration, which would erase the correlations that the standard models of dust-grain processing would produce, is introduced as an adjunct to these models.

The next step is to extract from these spectra the details of grain composition and size, by modeling the silicate profiles with laboratory measurements of mineral optical constants. We will be able thereby to search for trends involving grain growth. This work will be presented in an upcoming article by Sargent et al. (2008b, in preparation).

We are grateful to Chat Hull and Don Barry for assistance with data processing and reduction, to Jeff Bary for access to results in advance of publication, to Jeroen Bouwman for a copy of the ISO spectrum of Comet Hale-Bopp, and to an anonymous referee for a very thorough review. This work is based on observations made with the Spitzer Space Telescope, which is operated by the Jet Propulsion Laboratory, California Institute of Technology, under NASA contract 1407. Our work is supported in part by NASA through grants to the *Spitzer*-IRS instrument team at Rochester and Cornell (JPL contract number 1257184), through grant NAG5-13210, STScI grant AR-09524.01-A, and Origins grant NAG5-9670 to U. Michigan, and through a *Spitzer* fellowship to C.H.C. E.F. acknowledges support from a NASA Postdoctoral Program Fellowship, administered by Oak Ridge Associated Universities through a contract with NASA. P.D. acknowledges grants from PAPIIT, UNAM and CONACyT. We have made use of data products from the Two Micron All Sky Survey, which is a joint project of the University of Massachusetts and the Infrared Processing and Analysis Center at Caltech, funded by NASA and the National Science Foundation. Like all astronomers, we have also depended upon the SIMBAD and VizieR databases, operated at CDS (Strasbourg, France); NASA's Astrophysics Data System Abstract Service; and the NASA/ IPAC Infrared Science Archive operated by JPL, under contract with NASA.

## Figure captions

Figure 1 (**color on-line only**): *Spitzer*-IRS spectrum of IS Tau. At the top are also indicated the positions of the strongest mineral features expected in submicron dust grains, calculated from optical constants for α quartz (Wenrich & Christensen 1996), orthoenstatite (Jaeger et al. 1998) and forsterite (Fabian et al. 2001). At bottom, for comparison, is an ISO SWS spectrum of Comet Hale-Bopp (Crovisier et al. 1997).

Figure 2: example silicate-feature spectra from our Taurus mid-infrared spectral survey. The calibrated spectrum is in red; in blue are the points used for the polynomial baseline fit, and the fit baseline itself is shown in a dotted blue line. In green is the equivalent width per channel – the spectrum, minus the baseline and divided by the baseline, scaled arbitrarily for the display. The dotted black line is the composite equivalent-width-per-channel spectrum that represents "pristine" (interstellar-like, submicron, amorphous silicate) grains, with the two silicate features scaled independently to match those from the target spectra at 9.9 and 21.7 μm. As discussed in section 3.2, the "pristine" features themselves are the averages of those features in LkCa 15 and UY Aur. The full width at half maximum of the 10 μm feature is indicated by magenta bars. Spectra have been chosen to illustrate the wide variety of crystalline content among the sample objects: from grains that are dominated by olivines, pyroxenes and silica (V955 Tau and IRAS 04187+1927), to grains with smaller concentrations of minerals (IRAS F04192+2647 and DR Tau), to grains with silicate features indistinguishable from those of interstellar grains (UY Aur and LkCa 15). Also included are two examples of unusual spectra: those of ZZ Tau, showing that its emission is heavily dominated by silica, and of UX Tau A, in which the silicate features are joined by PAH features at 11.2, 12.0 and 12.7 μm. The complete set of 84 Taurus Class II spectra, presented in the same manner, is available in the **online-only** supplementary material.

Figure 3: outline of the crystalline-silicate index extraction, using the spectrum of IS Tau as an example. In each frame is a calibrated spectrum, baseline fit, equivalent width per channel, and composite pristine spectrum, presented in the same manner as in Figure 2. The green bars are the bands within which the equivalent width per channel is summed for the indices centered at 9.2, 9.9, 11.1, 12.5, 21.7 and 33.6 μm, reading from left to right.

Figure 4 (**color on-line only**): relation between the crystalline-silicate indices, and the mass fractions of the corresponding crystalline minerals (upper) and amorphous large-grain material (lower), determined from the same spectra by fitting laboratory spectra (Sargent et al. 2006). Blue diamonds indicate the $P_{10}$ index and orthoenstatite, and magenta squares the $O_{10}$ index and forsterite. The quality of linear fits to the data are inticated by the square of the linear correlation coefficient, $r^2$ (see below, Equation 4).

Figure 5 (**color on-line only**): histograms of the crystalline indices $P_{10}$, $O_{10}$, and $S_{10}$, and of the equivalent widths of the 10 and 20 μm silicate complexes and the 33 μm olivine feature, $W_{10}$, $W_{20}$, and $W_{33}$.

Figure 6: luminosity-effective temperature relation for our Taurus sample, compared to isochrones and stellar tracks for 1-5 Myr ages calculated by Siess et al. (2000).

Figure 7 (**color on-line only**): trends among the emission features and their crystalline indices. Plotted are indices derived from features of pyroxene, olivine and silica ($P_{10}$, $O_{33}$, and $S_{10}$, respectively), the full width at half-maximum of the 10 μm silicate complex ($\Delta\lambda_{10}$), the equivalent widths of the 10 and 20 μm silicate complexes ($W_{10}$ and $W_{20}$), and two of the continuum spectral indices ($n_{6-13}$ and $n_{13-31}$), all plotted as functions of the olivine index $O_{10}$. Upper limits are indicated by open symbols.

Figure 8: the dependence of $n_{13-31}$ and $W_{10}$ in our sample of Class II YSOs (blue diamonds), compared to that of the YSO disk models by d'Alessio et al. (2006). Data for the transitional disks CoKu Tau/4, DM Tau, and GM Aur are plotted with yellow-centered symbols. A dotted line at $n_{13-31} = -4/3$, as obtained for a geometrically-thin, opaque passive disk, indicates the extreme of sedimentation. In the models plotted in the upper panel, small interstellar-like grains are assumed to populate the disk around

stars of mass 0.4 (squares), 0.5 (triangles) and $0.8 M_\odot$ (circles), with disk accretion rates of $10^{-9}$ (open symbols) and $10^{-8} M_\odot$ yr$^{-1}$ (filled symbols); the disk axis is oriented by $i = 45°$ with respect to the line of sight, and the dust settling parameter $\varepsilon$ – the depletion factor with respect to gas for dust in the upper layers of the disk – is 0.001, 0.01, 0.1, 0.2, 0.5 and 1 (large to small symbols). To illustrate the effect of a range of disk orientation, the upper panel's model track for $M = 0.5 M_\odot$ and $dM/dt = 10^{-9} M_\odot$ yr$^{-1}$ is reproduced in the lower panel, along with the model results for these parameters, $\varepsilon = 0.01$, and $i$ = 12, 17, 24, 30, 37, 45, 53, and 64° (light red to light violet centers). Note that the models for disks with smallest $\varepsilon$ all predict $n_{13-31}$ and $W_{10}$ values corresponding to the densest concentration of data points. Note also that for increasing $\varepsilon$, the models form tracks radiating from this dense concentration along positive slopes, to encompass most of the rest of the data points.

Figure 9 (**color on-line only**): histograms of the olivine indices, $O_{10}$ (upper panel) and $O_{33}$ (lower panel), for single stars and for multiple-star systems. Note the weak correlation of $O_{10}$ with stellar multiplicity.

Figure 10 (**color on-line only**): search for trends between the crystalline indices and the global properties of the systems. Plotted are the crystalline indices for pyroxene ($P_{10}$), olivine ($O_{10}$, $O_{33}$) and silica ($S_{10}$), and the equivalent widths of the 10 and 20 μm silicate complexes ($W_{10}$ and $W_{20}$), from Table 1, against the stellar luminosity, stellar mass, disk/star mass ratio, and accretion rate from Table 3.

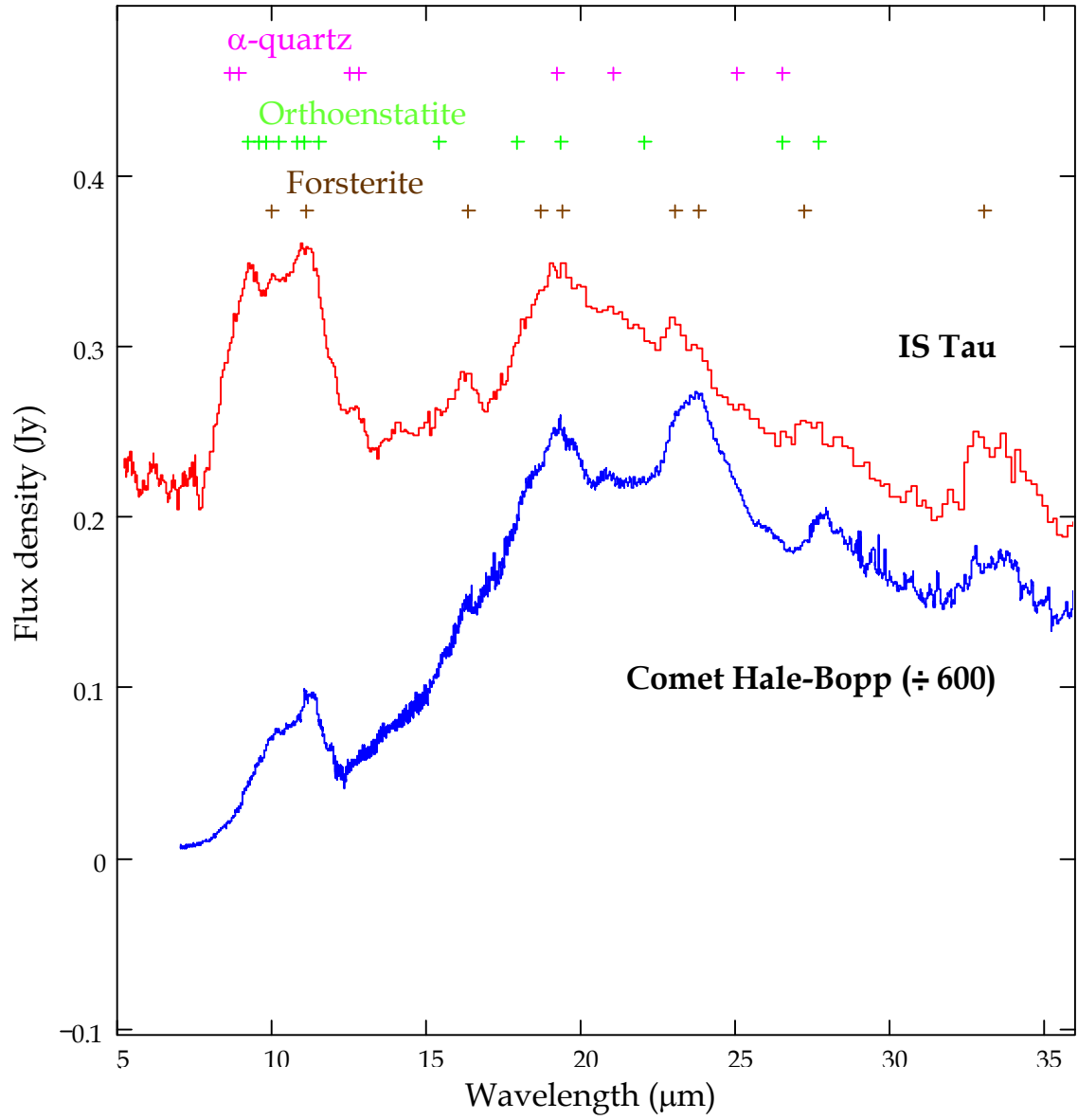

Figure 1

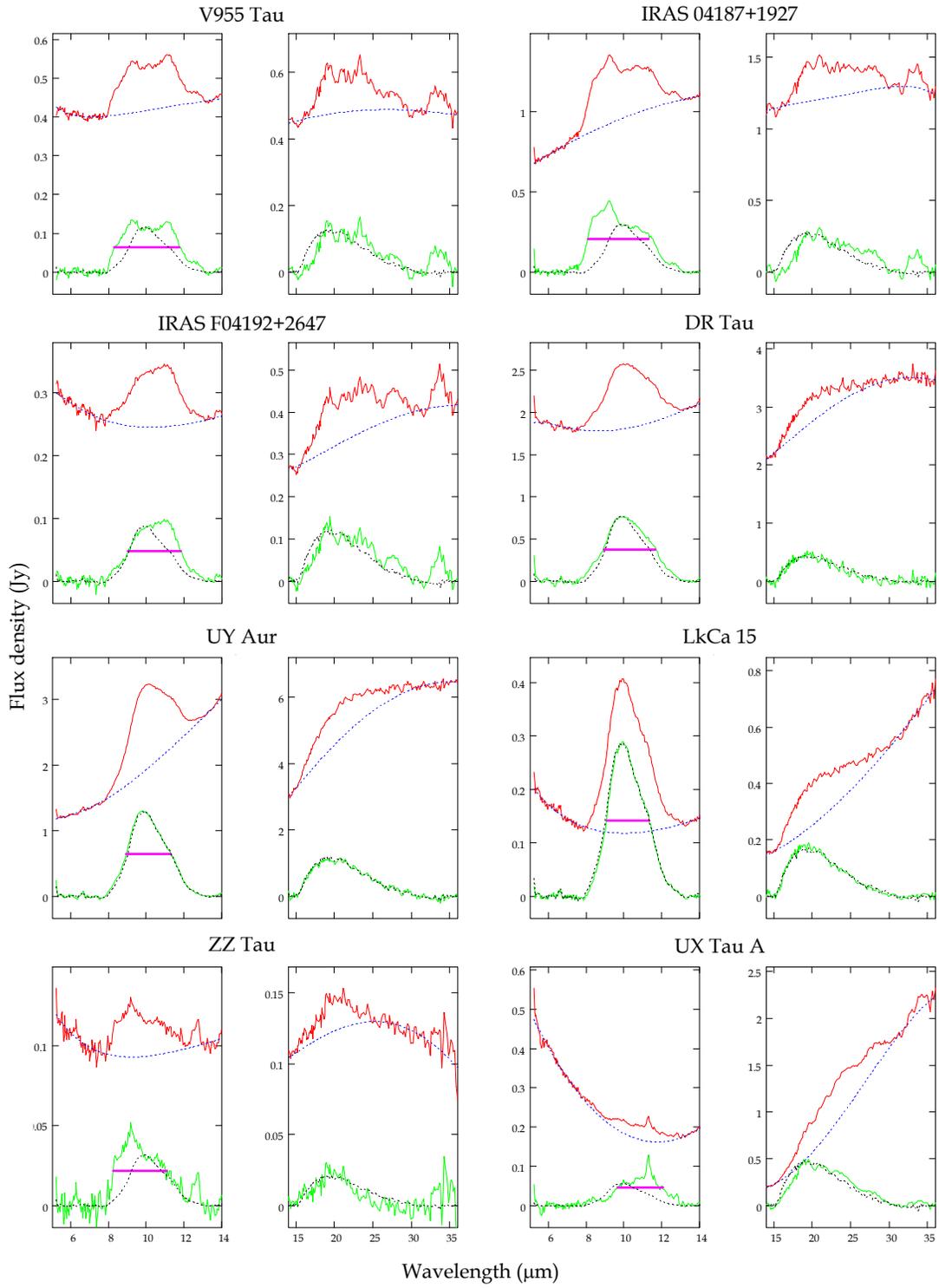

Figure 2

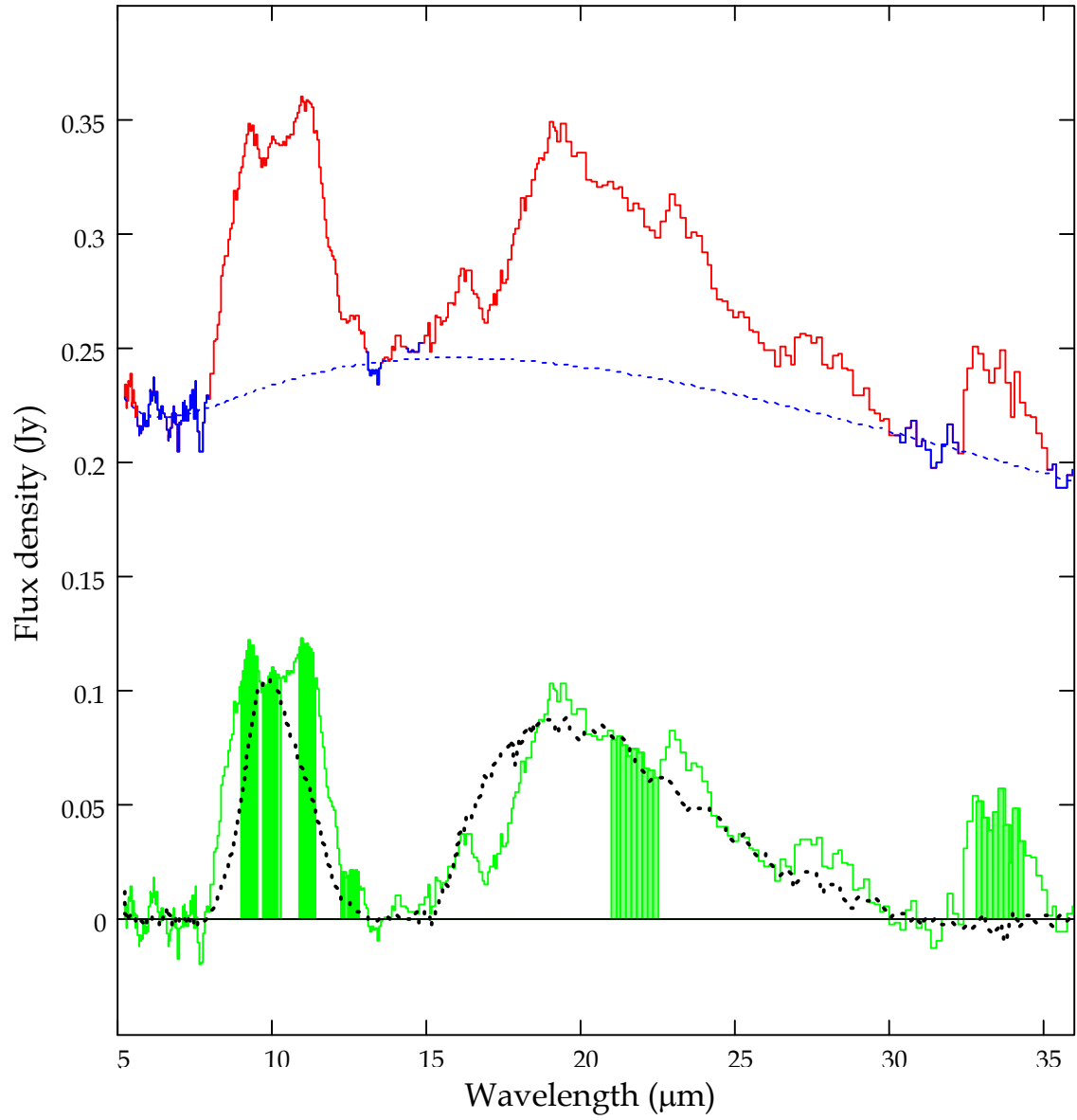

Figure 3

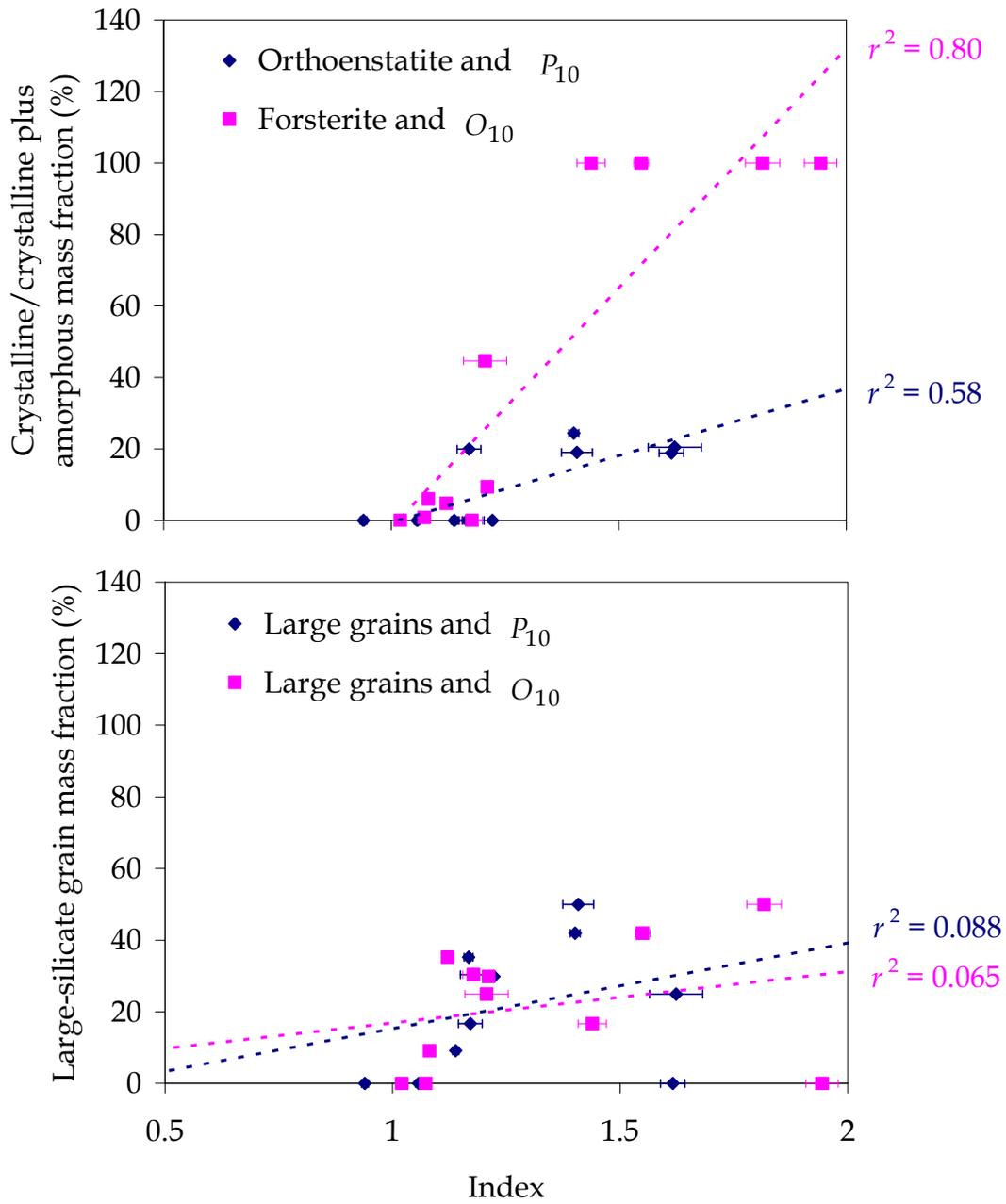

Figure 4

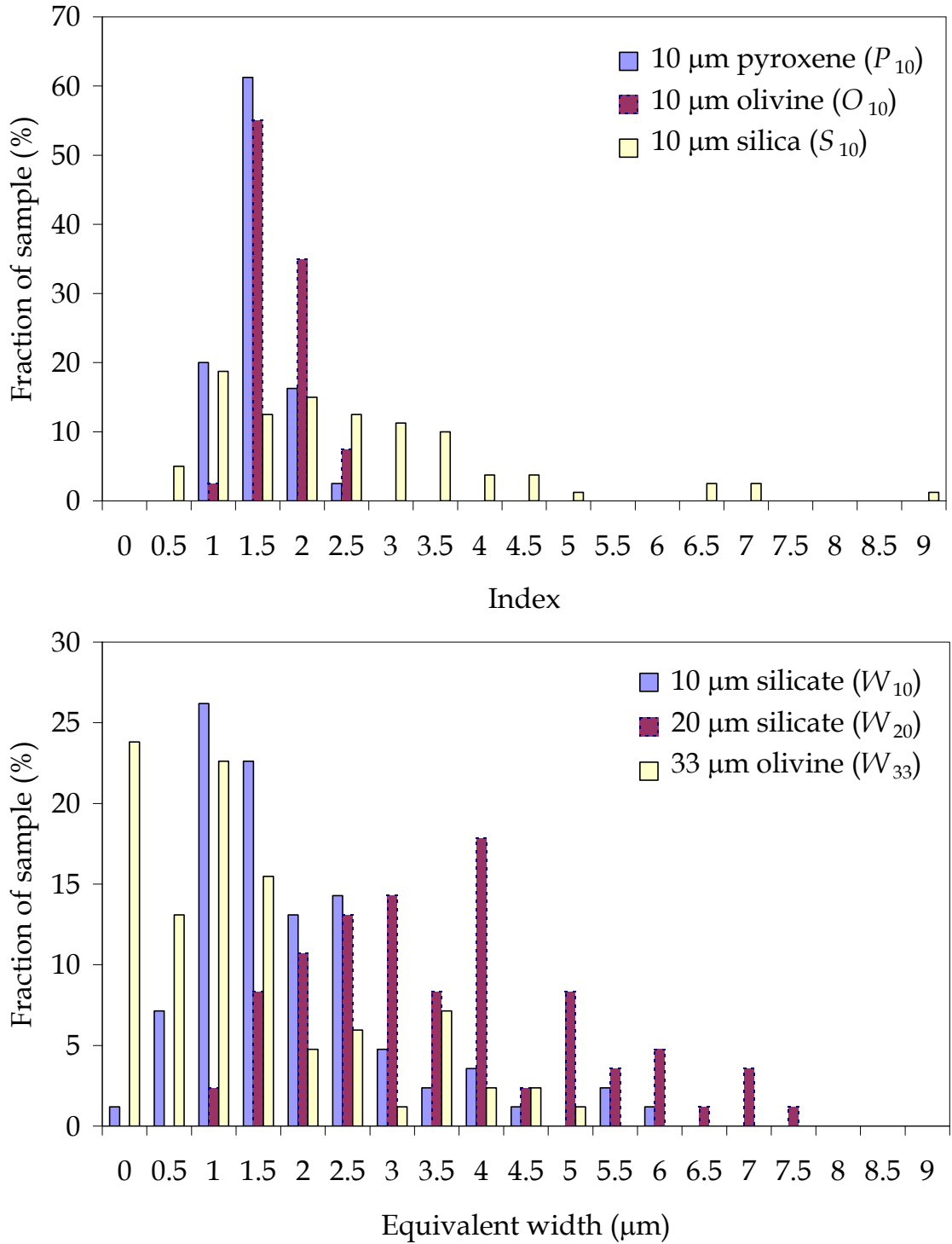

Figure 5

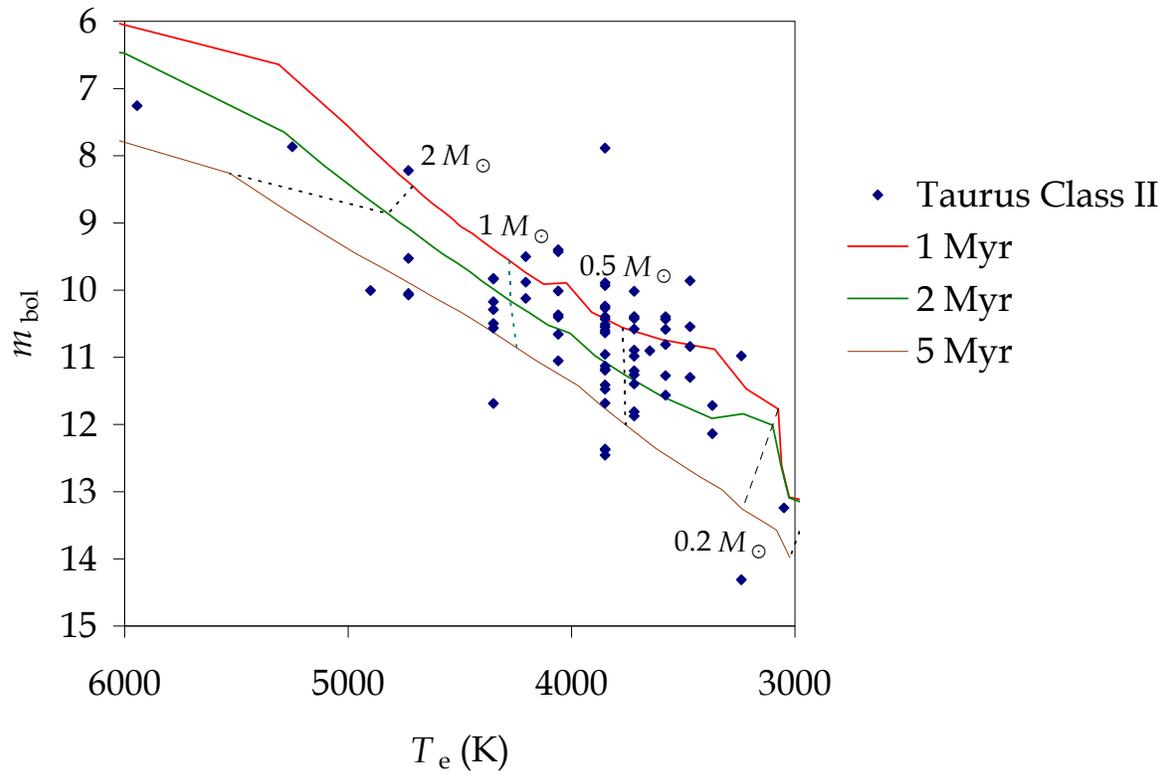

Figure 6

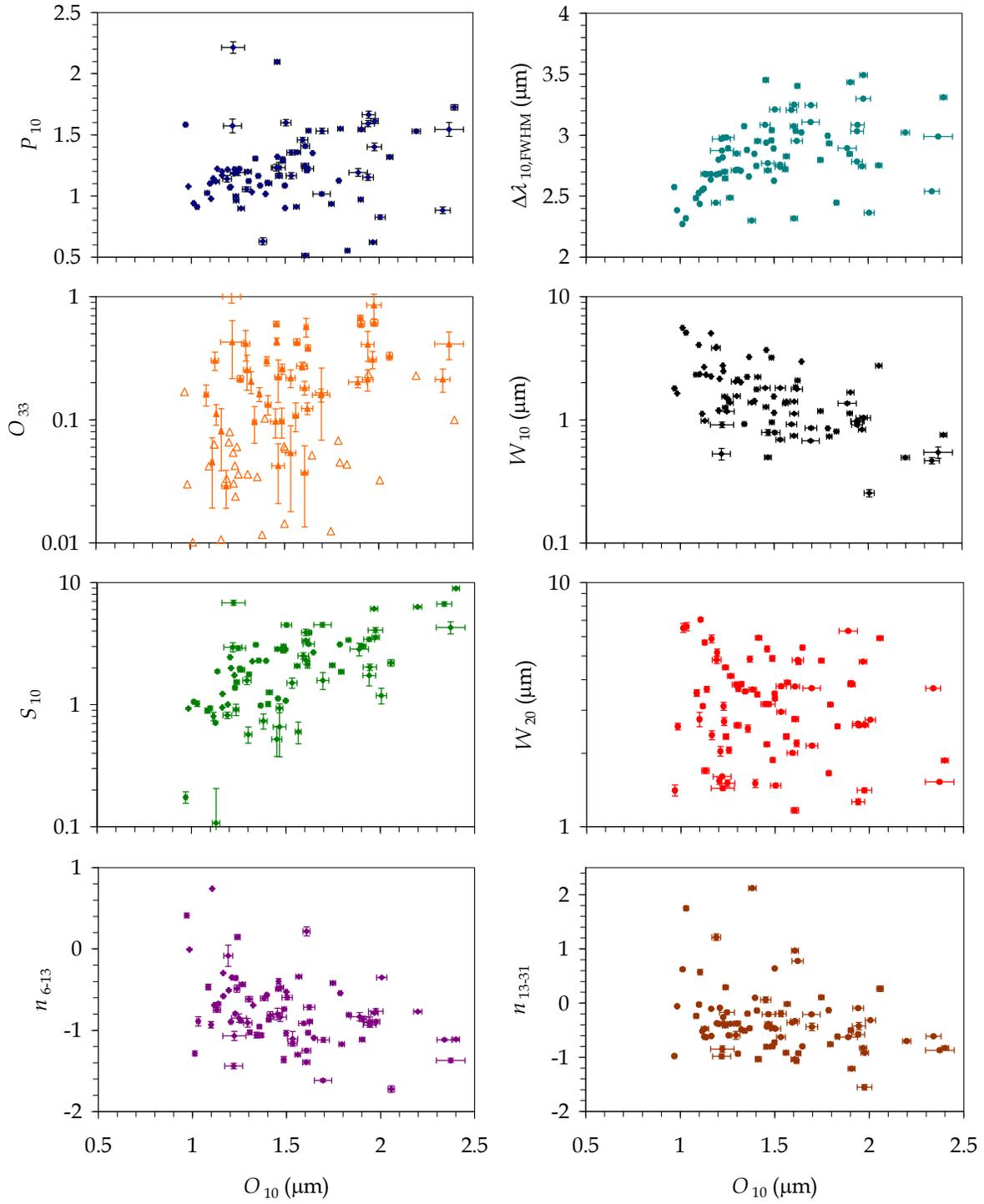

Figure 7

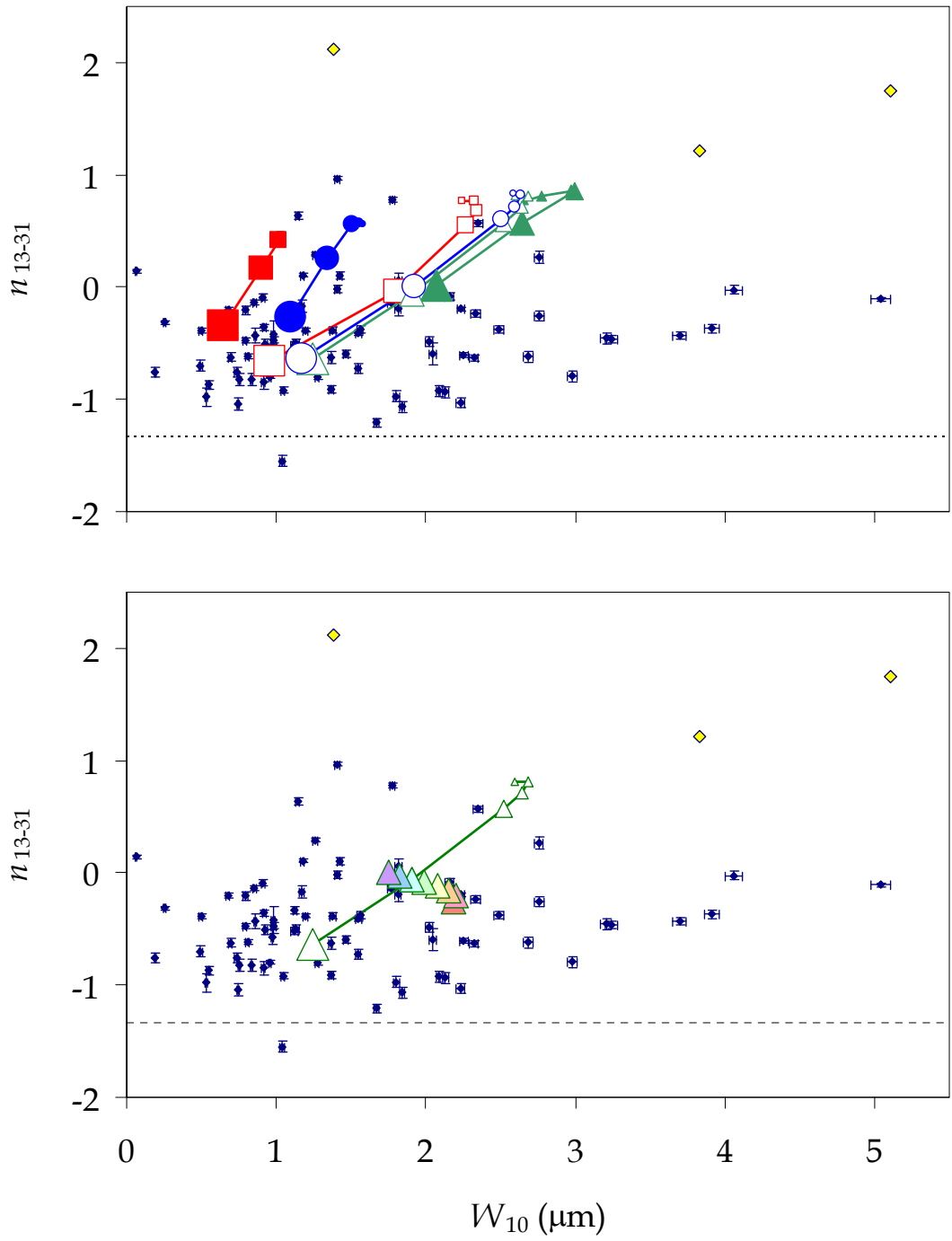

Figure 8

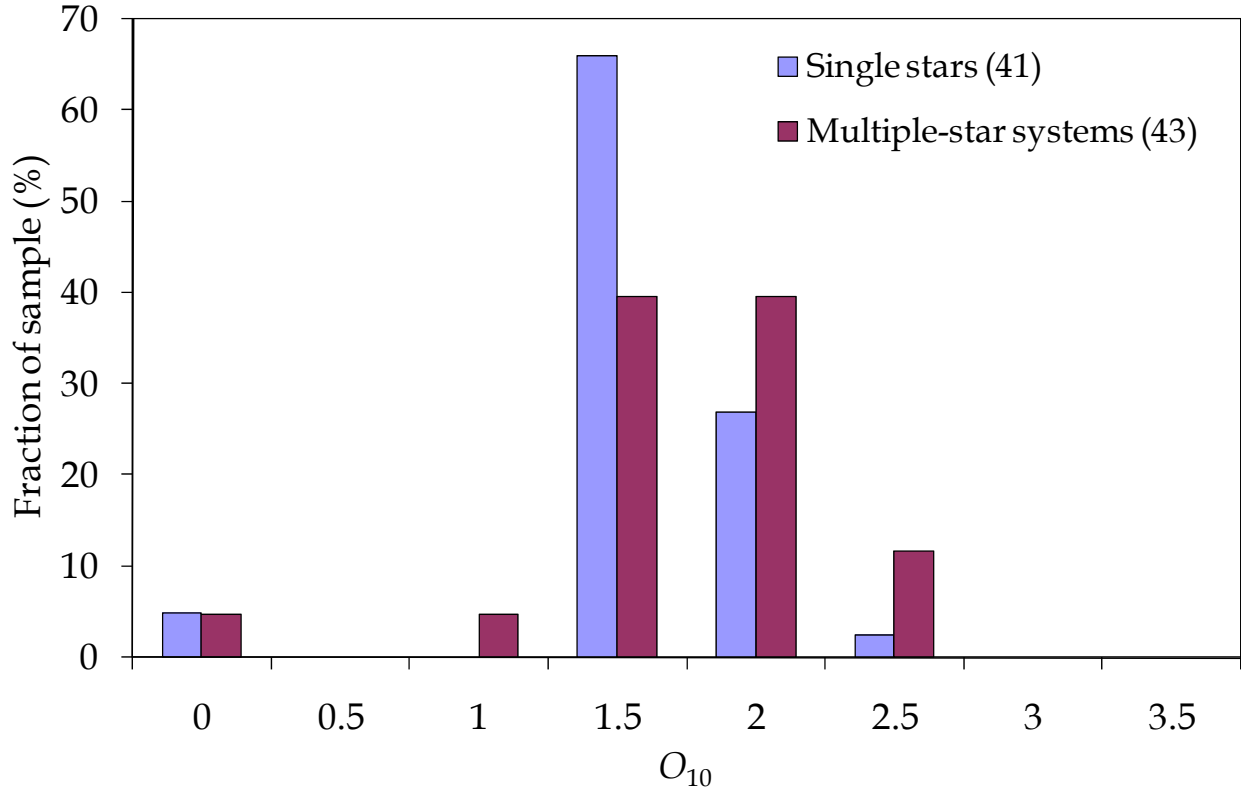

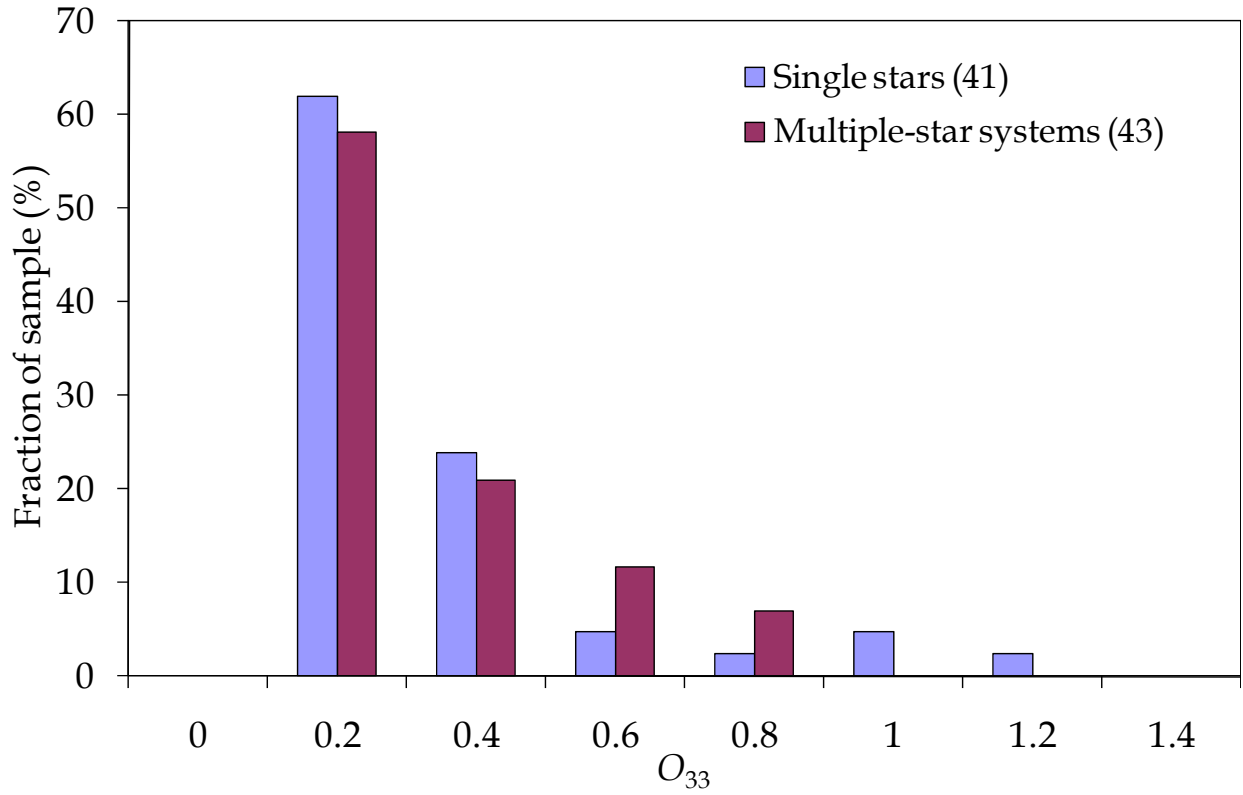

Figure 9

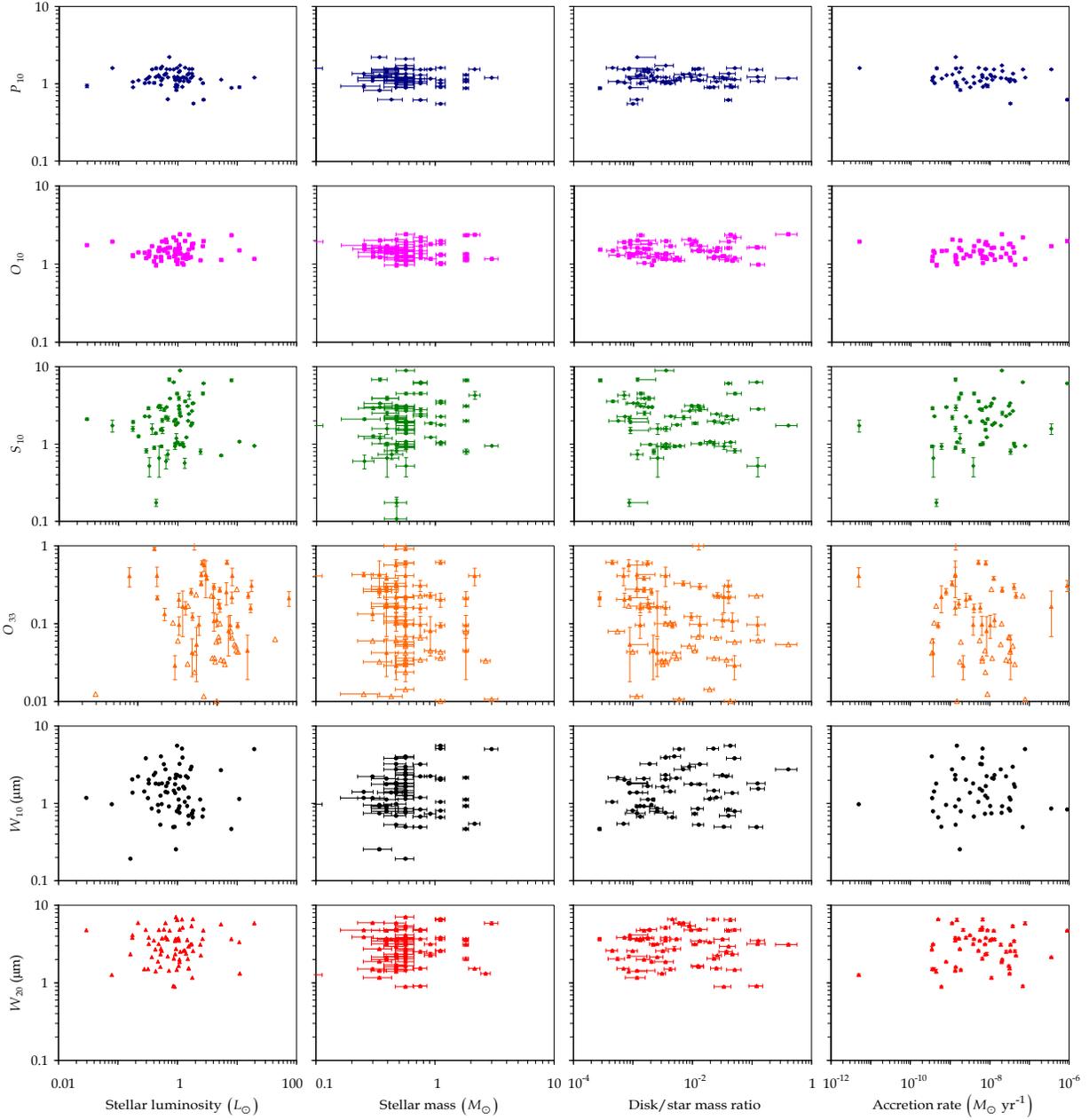

Figure 10

Table 1: silicate and silica features extracted from *Spitzer* IRS spectra of Taurus Class II objects §

| Object | 10 μm silicate feature equivalent width, $W_{10}$ (μm) † | 10 μm silicate feature FWHM, $\Delta\lambda_{10}$ (μm) | 20 μm silicate feature equivalent width, $W_{20}$ (μm) † | 33.6 μm olivine feature equivalent width, $W_{33}$ (μm) † | 9.2 μm pyroxene index, $P_{10}$ § | 11.1 μm olivine index, $O_{10}$ § | 12.5 μm silica index, $S_{10}$ § | 33.6 μm olivine index, $O_{33}$ § |
|---|---|---|---|---|---|---|---|---|
| 04108+2910 * | 0.192± 0.004 | 2.07 | 2.338± 0.023 | 0.411± 0.023 | | | | 0.918± 0.041 |
| 04187+1927 | 1.274± 0.012 | 3.45 | 2.175± 0.040 | 0.209± 0.040 | 2.096± 0.014 | 1.456± 0.014 | 2.852± 0.092 | 0.601± 0.031 |
| 04200+2759 | 1.549± 0.020 | 2.70 | 4.490± 0.073 | -0.229± 0.073 | 0.999± 0.010 | 1.237± 0.015 | 0.911± 0.097 | -0.321± 0.042 |
| 04216+2603 | 0.984± 0.016 | 2.68 | 1.697± 0.046 | 0.101± 0.046 | 1.118± 0.009 | 1.130± 0.019 | 0.107± 0.098 | 0.303± 0.051 |
| 04248+2612 | 1.410± 0.017 | 2.32 | 2.758± 0.062 | 0.054± 0.062 | 0.514± 0.016 | 1.607± 0.018 | 2.304± 0.170 | 0.037± 0.024 |
| 04303+2240 | 0.835± 0.010 | 2.75 | 4.736± 0.049 | 0.360± 0.049 | 0.621± 0.013 | 1.967± 0.019 | 6.094± 0.114 | 0.309± 0.050 |
| 04370+2559 | 2.488± 0.031 | 2.98 | 2.700± 0.100 | -0.014± 0.100 | 1.219± 0.007 | 1.230± 0.007 | 1.376± 0.043 | -0.048± 0.030 |
| 04385+2550 | 1.258± 0.016 | 2.65 | 2.341± 0.056 | -0.043± 0.056 | 0.963± 0.009 | 1.239± 0.014 | 1.529± 0.057 | -0.145± 0.024 |
| AA Tau | 1.465± 0.018 | 2.89 | 2.061± 0.060 | 0.010± 0.060 | 1.222± 0.010 | 1.256± 0.007 | 1.976± 0.063 | -0.104± 0.036 |
| BP Tau | 2.322± 0.029 | 2.68 | 3.653± 0.102 | 0.073± 0.102 | 1.223± 0.005 | 1.137± 0.004 | 1.870± 0.050 | 0.112± 0.021 |
| CI Tau | 2.235± 0.027 | 2.88 | 2.525± 0.089 | 0.026± 0.089 | 1.165± 0.005 | 1.356± 0.010 | 2.298± 0.036 | 0.028± 0.034 |
| CoKu Tau/3 | 1.842± 0.020 | 3.04 | 2.196± 0.069 | 0.202± 0.069 | 1.208± 0.010 | 1.615± 0.011 | 2.128± 0.143 | 0.568± 0.099 |
| CoKu Tau/4 | 1.384± 0.020 | 2.30 | 3.642± 0.072 | 0.024± 0.072 | 0.629± 0.029 | 1.381± 0.020 | 0.734± 0.104 | -0.001± 0.012 |
| CW Tau | 0.926± 0.013 | 3.07 | 3.580± 0.053 | 0.114± 0.053 | 1.306± 0.012 | 1.341± 0.010 | 3.094± 0.056 | 0.096± 0.032 |
| CX Tau | 0.792± 0.015 | 2.77 | 3.173± 0.042 | 0.052± 0.042 | 1.232± 0.043 | 1.467± 0.033 | 0.658± 0.285 | 0.042± 0.021 |
| CY Tau | 0.530± 0.009 | 2.97 | 1.606± 0.025 | 0.314± 0.025 | 1.573± 0.057 | 1.219± 0.048 | 2.953± 0.249 | 1.002± 0.116 |
| CZ Tau | 1.803± 0.022 | 2.57 | 1.409± 0.074 | 0.133± 0.074 | 1.583± 0.009 | 0.969± 0.008 | 0.174± 0.019 | -1.301± 0.169 |
| DD Tau | 0.957± 0.011 | 3.04 | 1.876± 0.038 | 0.147± 0.038 | 1.288± 0.013 | 1.488± 0.012 | 3.001± 0.083 | 0.259± 0.024 |
| DE Tau | 1.767± 0.021 | 2.75 | 3.477± 0.076 | 0.177± 0.076 | 1.110± 0.006 | 1.406± 0.010 | 1.011± 0.044 | 0.300± 0.026 |
| DF Tau | 0.744± 0.011 | 3.07 | 1.167± 0.031 | -0.035± 0.031 | 1.249± 0.017 | 1.605± 0.019 | 3.344± 0.080 | -1.122± 0.277 |
| DG Tau * | 0.062± 0.004 | | 2.261± 0.023 | 0.081± 0.023 | | | | 0.231± 0.019 |
| DH Tau | 2.755± 0.029 | 2.75 | 5.915± 0.104 | 0.465± 0.104 | 1.319± 0.015 | 2.057± 0.017 | 2.199± 0.135 | 0.329± 0.023 |

Table 1: silicate and silica features extracted from *Spitzer* IRS spectra of Taurus Class II objects §

| Object | 10 μm silicate feature equivalent width, $W_{10}$ (μm) † | 10 μm silicate feature FWHM, $\Delta\lambda_{10}$ (μm) | 20 μm silicate feature equivalent width, $W_{20}$ (μm) † | 33.6 μm olivine feature equivalent width, $W_{33}$ (μm) † | 9.2 μm pyroxene index, $P_{10}$ § | 11.1 μm olivine index, $O_{10}$ § | 12.5 μm silica index, $S_{10}$ § | 33.6 μm olivine index, $O_{33}$ § |
|---|---|---|---|---|---|---|---|---|
| DK Tau | 2.977± 0.034 | 3.02 | 5.413± 0.116 | -0.003± 0.116 | 1.351± 0.007 | 1.646± 0.006 | 2.675± 0.042 | -0.133± 0.052 |
| DL Tau | 0.495± 0.013 | 3.02 | 0.909± 0.020 | 0.087± 0.020 | 1.529± 0.016 | 2.198± 0.022 | 6.305± 0.111 | -0.670± 0.228 |
| DM Tau | 3.833± 0.049 | 2.45 | 4.820± 0.170 | 0.057± 0.170 | 1.140± 0.027 | 1.190± 0.023 | 0.818± 0.049 | 0.029± 0.010 |
| DN Tau | 0.790± 0.011 | 3.21 | 1.475± 0.029 | 0.058± 0.029 | 1.599± 0.026 | 1.504± 0.027 | 4.488± 0.166 | 0.001± 0.058 |
| DO Tau | 0.855± 0.009 | 3.00 | 1.656± 0.032 | 0.051± 0.032 | 1.126± 0.006 | 1.785± 0.008 | 3.110± 0.048 | 0.007± 0.068 |
| DP Tau | 1.378± 0.018 | 2.49 | 4.141± 0.069 | 0.164± 0.069 | 0.899± 0.009 | 1.265± 0.017 | 1.926± 0.079 | 0.215± 0.013 |
| DQ Tau | 0.498± 0.011 | 2.71 | 0.893± 0.022 | 0.058± 0.022 | 1.164± 0.016 | 1.467± 0.018 | 0.934± 0.079 | 0.222± 0.084 |
| DR Tau | 1.194± 0.014 | 2.80 | 1.540± 0.050 | 0.007± 0.050 | 1.070± 0.005 | 1.204± 0.007 | 2.443± 0.036 | -0.034± 0.066 |
| DS Tau | 2.125± 0.026 | 2.72 | 3.671± 0.091 | 0.027± 0.091 | 1.122± 0.008 | 1.305± 0.010 | 1.771± 0.048 | -0.111± 0.036 |
| F04147+2822 | 2.232± 0.027 | 2.95 | 5.933± 0.102 | 0.054± 0.102 | 1.103± 0.010 | 1.413± 0.016 | 1.260± 0.048 | 0.133± 0.024 |
| F04192+2647 | 1.132± 0.018 | 2.85 | 3.873± 0.052 | 0.342± 0.052 | 0.972± 0.011 | 1.902± 0.014 | 2.988± 0.132 | 0.672± 0.034 |
| F04262+2654 | 0.912± 0.013 | 2.78 | 2.640± 0.037 | 0.094± 0.037 | 1.153± 0.026 | 1.941± 0.028 | 3.426± 0.128 | 0.213± 0.042 |
| F04297+2246A | 3.695± 0.041 | 2.94 | 5.349± 0.143 | 0.304± 0.143 | 1.321± 0.006 | 1.458± 0.007 | 1.124± 0.026 | 0.432± 0.028 |
| F04297+2246B | 1.363± 0.016 | 2.89 | 6.326± 0.067 | 0.224± 0.067 | 1.192± 0.032 | 1.889± 0.049 | 2.843± 0.334 | 0.202± 0.020 |
| F04570+2520 | 1.038± 0.017 | 3.30 | 1.410± 0.032 | 0.186± 0.032 | 1.402± 0.031 | 1.974± 0.038 | 4.067± 0.203 | 0.855± 0.193 |
| FM Tau | 2.331± 0.032 | 2.48 | 3.528± 0.110 | -0.013± 0.110 | 1.025± 0.007 | 1.084± 0.012 | 0.894± 0.042 | 0.161± 0.031 |
| FN Tau | 1.411± 0.017 | 2.83 | 3.897± 0.059 | 0.235± 0.059 | 1.358± 0.011 | 1.566± 0.016 | 0.599± 0.121 | 0.428± 0.028 |
| FO Tau | 0.920± 0.012 | 3.21 | 2.004± 0.036 | 0.088± 0.036 | 1.458± 0.020 | 1.592± 0.028 | 2.504± 0.154 | 0.273± 0.020 |
| FP Tau | 1.169± 0.014 | 2.98 | 1.507± 0.049 | 0.056± 0.049 | 1.192± 0.025 | 1.247± 0.040 | 2.903± 0.129 | 0.058± 0.060 |
| FQ Tau | 0.860± 0.010 | 3.25 | 2.147± 0.035 | -0.002± 0.035 | 1.532± 0.037 | 1.696± 0.031 | 1.578± 0.249 | 0.165± 0.097 |
| FS Tau | 1.425± 0.018 | 2.85 | 1.503± 0.057 | 0.032± 0.057 | 1.017± 0.005 | 1.395± 0.006 | 2.285± 0.045 | -0.248± 0.102 |
| FT Tau | 1.559± 0.019 | 2.85 | 2.601± 0.068 | 0.064± 0.068 | 1.199± 0.013 | 1.301± 0.019 | 0.570± 0.086 | 0.254± 0.081 |

Table 1: silicate and silica features extracted from *Spitzer* IRS spectra of Taurus Class II objects §

| Object | 10 μm silicate feature equivalent width, $W_{10}$ (μm) † | 10 μm silicate feature FWHM, $\Delta\lambda_{10}$ (μm) | 20 μm silicate feature equivalent width, $W_{20}$ (μm) † | 33.6 μm olivine feature equivalent width, $W_{33}$ (μm) † | 9.2 μm pyroxene index, $P_{10}$ § | 11.1 μm olivine index, $O_{10}$ § | 12.5 μm silica index, $S_{10}$ § | 33.6 μm olivine index, $O_{33}$ § |
|---|---|---|---|---|---|---|---|---|
| FV Tau | 0.807± 0.011 | 2.45 | 2.576± 0.038 | -0.038± 0.038 | 0.554± 0.009 | 1.831± 0.012 | 3.378± 0.045 | -0.215± 0.043 |
| FX Tau | 3.240± 0.041 | 2.66 | 4.856± 0.139 | 0.120± 0.139 | 1.085± 0.008 | 1.366± 0.007 | 0.984± 0.028 | 0.161± 0.021 |
| FZ Tau | 0.755± 0.008 | 3.31 | 1.869± 0.030 | 0.120± 0.030 | 1.724± 0.022 | 2.401± 0.018 | 8.941± 0.177 | -0.137± 0.100 |
| GG Tau A | 2.756± 0.034 | 2.82 | 3.112± 0.115 | -0.001± 0.115 | 1.184± 0.005 | 1.227± 0.005 | 1.737± 0.019 | -0.245± 0.054 |
| GH Tau | 1.122± 0.012 | 3.25 | 3.749± 0.050 | 0.116± 0.050 | 1.408± 0.016 | 1.607± 0.023 | 3.906± 0.236 | 0.182± 0.023 |
| GI Tau | 2.253± 0.030 | 2.68 | 2.372± 0.098 | 0.042± 0.098 | 1.168± 0.005 | 1.164± 0.005 | 1.224± 0.018 | 0.081± 0.042 |
| GK Tau | 3.911± 0.049 | 2.68 | 5.167± 0.170 | 0.064± 0.170 | 1.216± 0.003 | 1.193± 0.005 | 1.003± 0.014 | -0.071± 0.033 |
| GM Aur | 5.110± 0.074 | 2.32 | 6.594± 0.253 | -0.041± 0.253 | 0.910± 0.007 | 1.031± 0.009 | 1.017± 0.053 | -0.030± 0.007 |
| GN Tau | 2.090± 0.022 | 3.40 | 4.724± 0.079 | 0.274± 0.079 | 1.534± 0.009 | 1.624± 0.013 | 3.886± 0.084 | 0.383± 0.022 |
| GO Tau | 1.815± 0.023 | 3.08 | 3.179± 0.074 | 0.053± 0.074 | 1.231± 0.028 | 1.452± 0.027 | 0.520± 0.145 | 0.097± 0.026 |
| Haro 6-13 | 1.143± 0.015 | 2.63 | 3.349± 0.053 | -0.126± 0.053 | 0.901± 0.006 | 1.500± 0.006 | 1.075± 0.022 | -0.231± 0.014 |
| Haro 6-28 | 0.980± 0.012 | 3.08 | 2.598± 0.041 | -0.149± 0.041 | 1.664± 0.030 | 1.944± 0.035 | 2.027± 0.118 | -1.203± 0.235 |
| Haro 6-37 | 0.734± 0.011 | 2.93 | 3.163± 0.038 | 0.053± 0.038 | 1.550± 0.009 | 1.793± 0.014 | 1.857± 0.065 | -0.139± 0.045 |
| HK Tau | 1.776± 0.020 | 2.95 | 4.809± 0.073 | 0.104± 0.073 | 1.224± 0.015 | 1.621± 0.030 | 3.125± 0.086 | 0.124± 0.013 |
| HN Tau | 2.022± 0.027 | 2.71 | 3.840± 0.089 | 0.360± 0.089 | 1.033± 0.005 | 1.323± 0.004 | 2.264± 0.022 | 0.205± 0.042 |
| HO Tau | 2.045± 0.027 | 2.71 | 3.817± 0.091 | 0.311± 0.091 | 1.055± 0.020 | 1.294± 0.022 | 1.575± 0.116 | 0.415± 0.115 |
| HP Tau | 2.157± 0.028 | 2.69 | 2.037± 0.092 | 0.041± 0.092 | 1.075± 0.004 | 1.210± 0.006 | 1.991± 0.027 | -0.148± 0.080 |
| HQ Tau | 2.686± 0.036 | 2.56 | 5.679± 0.126 | -0.012± 0.126 | 1.126± 0.004 | 1.127± 0.004 | 0.712± 0.020 | -0.178± 0.063 |
| IP Tau | 4.058± 0.055 | 2.53 | 2.752± 0.185 | 0.030± 0.185 | 1.102± 0.007 | 1.099± 0.007 | 0.933± 0.033 | 0.031± 0.042 |
| IQ Tau | 1.364± 0.017 | 2.72 | 2.343± 0.058 | 0.100± 0.058 | 0.912± 0.010 | 1.559± 0.015 | 2.073± 0.053 | 0.109± 0.029 |
| IS Tau | 1.672± 0.015 | 3.43 | 3.819± 0.060 | 0.443± 0.060 | 1.544± 0.014 | 1.906± 0.019 | 3.006± 0.054 | 0.602± 0.038 |
| IT Tau | 0.545± 0.010 | 2.99 | 1.526± 0.020 | 0.006± 0.020 | 1.544± 0.057 | 2.373± 0.076 | 4.279± 0.491 | 0.412± 0.105 |

Table 1: silicate and silica features extracted from *Spitzer* IRS spectra of Taurus Class II objects §

| Object | 10 μm silicate feature equivalent width, $W_{10}$ (μm) † | 10 μm silicate feature FWHM, $\Delta\lambda_{10}$ (μm) | 20 μm silicate feature equivalent width, $W_{20}$ (μm) † | 33.6 μm olivine feature equivalent width, $W_{33}$ (μm) † | 9.2 μm pyroxene index, $P_{10}$ § | 11.1 μm olivine index, $O_{10}$ § | 12.5 μm silica index, $S_{10}$ § | 33.6 μm olivine index, $O_{33}$ § |
|---|---|---|---|---|---|---|---|---|
| LkCa 15 | 5.582± 0.081 | 2.27 | 6.515± 0.277 | 0.032± 0.277 | 0.940± 0.006 | 1.012± 0.007 | 1.054± 0.029 | -0.009± 0.010 |
| MHO-3 | 2.350± 0.033 | 2.44 | 7.050± 0.120 | -0.036± 0.120 | 0.977± 0.005 | 1.105± 0.004 | <0.017 | -0.043± 0.009 |
| RW Aur A | 1.122± 0.022 | 2.55 | 3.117± 0.058 | 0.081± 0.058 | 1.144± 0.005 | 1.116± 0.006 | 0.801± 0.062 | 0.045± 0.026 |
| RY Tau | 5.041± 0.064 | 2.64 | 5.882± 0.217 | -0.022± 0.217 | 1.201± 0.002 | 1.163± 0.004 | 0.948± 0.008 | -0.043± 0.011 |
| T Tau * | -0.158± 0.007 | | 1.320± 0.018 | -0.032± 0.018 | | | | -0.038± 0.033 |
| UX Tau A * | 0.660± 0.008 | 2.42 | 6.635± 0.067 | 0.128± 0.067 | | | | 0.095± 0.008 |
| UY Aur | 1.642± 0.029 | 2.39 | 2.573± 0.079 | -0.039± 0.079 | 1.078± 0.004 | 0.984± 0.003 | 0.929± 0.016 | -0.097± 0.030 |
| UZ Tau/e | 1.548± 0.018 | 2.89 | 3.506± 0.066 | -0.003± 0.066 | 1.085± 0.011 | 1.498± 0.008 | 2.828± 0.066 | -0.321± 0.061 |
| V410 Anon 13 | 0.974± 0.012 | 3.03 | 1.266± 0.035 | 0.170± 0.035 | 1.592± 0.026 | 1.941± 0.034 | 1.732± 0.309 | 0.410± 0.112 |
| V710 Tau | 0.692± 0.010 | 2.76 | 2.954± 0.040 | 0.091± 0.040 | 1.356± 0.020 | 1.532± 0.024 | <0.078 | 0.218± 0.036 |
| V773 Tau | 0.467± 0.009 | 2.54 | 3.685± 0.040 | 0.347± 0.040 | 0.883± 0.027 | 2.339± 0.039 | 6.677± 0.267 | 0.213± 0.046 |
| V807 Tau | 0.676± 0.013 | 3.11 | 3.691± 0.041 | 0.107± 0.041 | 1.017± 0.015 | 1.694± 0.046 | 4.508± 0.192 | 0.159± 0.019 |
| V836 Tau | 3.207± 0.037 | 2.96 | 4.889± 0.126 | 0.139± 0.126 | 1.302± 0.007 | 1.486± 0.012 | 2.787± 0.145 | 0.098± 0.027 |
| V955 Tau | 1.045± 0.013 | 3.49 | 2.612± 0.040 | 0.316± 0.040 | 1.613± 0.019 | 1.976± 0.020 | 3.562± 0.119 | 0.615± 0.039 |
| VY Tau | 1.815± 0.022 | 2.74 | 3.758± 0.074 | 0.121± 0.074 | 1.166± 0.025 | 1.533± 0.027 | 1.505± 0.145 | 0.054± 0.036 |
| XZ Tau | 0.254± 0.007 | 2.36 | 2.738± 0.026 | -0.015± 0.026 | 0.827± 0.017 | 2.006± 0.027 | 1.186± 0.174 | -0.129± 0.032 |
| ZZ Tau | 0.913± 0.011 | 2.87 | 1.439± 0.031 | 0.245± 0.031 | 2.214± 0.048 | 1.222± 0.062 | 6.827± 0.324 | 0.428± 0.212 |
| ZZ Tau IRS | 1.177± 0.013 | 2.80 | 4.786± 0.059 | -0.041± 0.059 | 0.935± 0.011 | 1.746± 0.014 | 2.096± 0.069 | -0.076± 0.012 |

* In these objects, difficulties associated with stellar multiplicity or interfering spectral features prevent reliable calculation of certain indices, which are therefore left blank.

† Uncertainties in the equivalent widths are determined from the measured noise in $W_\nu$, in each spectral channel that comprise the feature, after smoothing by the spectral resolution, by adding these noises in quadrature. The noise in each channel is determined either from the standard deviation of the mean, for observations comprised of more than one nod-pair cycle, or set to half the difference between the independent measurements, for observations with only one nod-pair of spectra.

§ Uncertainties in the crystalline indices are calculated by adding in quadrature the values of standard deviation of the mean of $W_\nu$ within each index's wavelength, after removing a low-order polynomial fit to the points within each index's band.

## Table 2: mid-infrared continuum spectral indices of Taurus Class II objects

| Name | $n$(6-13) | | | $n$(13-25) | | | $n$(13-31) | | |
|---|---|---|---|---|---|---|---|---|---|
| 04108+2910 | -1.17 | ± | 0.02 | -0.53 | ± | 0.04 | -0.76 | ± | 0.04 |
| 04187+1927 | -0.49 | ± | 0.02 | -0.59 | ± | 0.03 | -0.81 | ± | 0.01 |
| 04200+2759 | -0.49 | ± | 0.04 | -0.13 | ± | 0.04 | -0.41 | ± | 0.04 |
| 04216+2603 | -0.75 | ± | 0.03 | -0.41 | ± | 0.04 | -0.47 | ± | 0.03 |
| 04248+2612 | 0.22 | ± | 0.06 | 1.29 | ± | 0.03 | 0.97 | ± | 0.02 |
| 04303+2240 | -0.79 | ± | 0.01 | -0.49 | ± | 0.01 | -0.83 | ± | 0.01 |
| 04370+2559 | -0.36 | ± | 0.02 | -0.08 | ± | 0.03 | -0.38 | ± | 0.02 |
| 04385+2550 | 0.15 | ± | 0.03 | 0.51 | ± | 0.03 | 0.29 | ± | 0.01 |
| AA Tau | -0.88 | ± | 0.02 | -0.47 | ± | 0.03 | -0.59 | ± | 0.02 |
| BP Tau | -0.67 | ± | 0.02 | -0.29 | ± | 0.03 | -0.63 | ± | 0.02 |
| CI Tau | -0.95 | ± | 0.02 | -0.04 | ± | 0.03 | -0.19 | ± | 0.02 |
| CoKu Tau/3 | -1.03 | ± | 0.02 | -0.79 | ± | 0.04 | -1.07 | ± | 0.03 |
| CoKu Tau/4 | -0.61 | ± | 0.03 | 2.95 | ± | 0.02 | 2.12 | ± | 0.02 |
| CW Tau | -1.06 | ± | 0.04 | -0.18 | ± | 0.05 | -0.51 | ± | 0.04 |
| CX Tau | -0.84 | ± | 0.05 | 0.11 | ± | 0.04 | -0.21 | ± | 0.03 |
| CY Tau | -1.44 | ± | 0.03 | -0.93 | ± | 0.05 | -0.98 | ± | 0.04 |
| CZ Tau | 0.41 | ± | 0.03 | -0.68 | ± | 0.02 | -0.98 | ± | 0.02 |
| DD Tau | -0.74 | ± | 0.01 | -0.62 | ± | 0.03 | -0.80 | ± | 0.01 |
| DE Tau | -0.87 | ± | 0.02 | 0.09 | ± | 0.03 | -0.14 | ± | 0.02 |
| DF Tau | -1.39 | ± | 0.02 | -0.86 | ± | 0.04 | -1.04 | ± | 0.04 |
| DG Tau | -0.22 | ± | 0.01 | 0.34 | ± | 0.01 | 0.14 | ± | 0.01 |
| DH Tau | -1.72 | ± | 0.04 | 0.72 | ± | 0.05 | 0.27 | ± | 0.05 |
| DK Tau | -1.09 | ± | 0.01 | -0.38 | ± | 0.02 | -0.80 | ± | 0.03 |
| DL Tau | -0.77 | ± | 0.01 | -0.53 | ± | 0.01 | -0.70 | ± | 0.03 |
| DM Tau | -0.08 | ± | 0.13 | 1.83 | ± | 0.07 | 1.22 | ± | 0.06 |
| DN Tau | -0.60 | ± | 0.03 | -0.34 | ± | 0.03 | -0.47 | ± | 0.02 |
| DO Tau | -0.54 | ± | 0.01 | 0.06 | ± | 0.01 | -0.13 | ± | 0.01 |
| DP Tau | -0.44 | ± | 0.02 | 0.00 | ± | 0.02 | -0.38 | ± | 0.01 |
| DQ Tau | -0.47 | ± | 0.02 | -0.26 | ± | 0.03 | -0.39 | ± | 0.01 |
| DR Tau | -0.90 | ± | 0.01 | -0.23 | ± | 0.01 | -0.38 | ± | 0.01 |
| DS Tau | -1.02 | ± | 0.03 | -0.69 | ± | 0.03 | -0.93 | ± | 0.03 |
| F04147+2822 | -0.82 | ± | 0.05 | -0.61 | ± | 0.03 | -1.03 | ± | 0.05 |
| F04192+2647 | -1.11 | ± | 0.02 | -0.19 | ± | 0.04 | -0.50 | ± | 0.03 |
| F04262+2654 | -0.91 | ± | 0.03 | 0.20 | ± | 0.04 | -0.09 | ± | 0.03 |
| F04297+2246A | -0.40 | ± | 0.03 | 0.00 | ± | 0.03 | -0.43 | ± | 0.02 |
| F04297+2246B | -0.83 | ± | 0.05 | -0.19 | ± | 0.04 | -0.63 | ± | 0.03 |

| Name | $n$(6-13) | | | $n$(13-25) | | | $n$(13-31) | | |
|------|------|---|------|------|---|------|------|---|------|
| F04570+2520 | -0.77 | ± | 0.03 | -1.31 | ± | 0.03 | -1.55 | ± | 0.05 |
| FM Tau | -0.47 | ± | 0.03 | 0.05 | ± | 0.03 | -0.24 | ± | 0.03 |
| FN Tau | -0.34 | ± | 0.01 | 0.30 | ± | 0.02 | -0.02 | ± | 0.01 |
| FO Tau | -0.91 | ± | 0.02 | -0.11 | ± | 0.04 | -0.36 | ± | 0.02 |
| FP Tau | -0.85 | ± | 0.05 | 0.00 | ± | 0.05 | -0.17 | ± | 0.04 |
| FQ Tau | -1.12 | ± | 0.03 | -0.30 | ± | 0.05 | -0.44 | ± | 0.07 |
| FS Tau | -0.57 | ± | 0.01 | 0.34 | ± | 0.02 | 0.10 | ± | 0.02 |
| FT Tau | -0.62 | ± | 0.03 | -0.18 | ± | 0.04 | -0.38 | ± | 0.05 |
| FV Tau | -0.81 | ± | 0.01 | -0.37 | ± | 0.01 | -0.62 | ± | 0.01 |
| FX Tau | -1.06 | ± | 0.03 | 0.01 | ± | 0.05 | -0.46 | ± | 0.03 |
| FZ Tau | -1.11 | ± | 0.01 | -0.63 | ± | 0.02 | -0.83 | ± | 0.03 |
| GG Tau A | -0.80 | ± | 0.02 | 0.03 | ± | 0.02 | -0.26 | ± | 0.02 |
| GH Tau | -1.25 | ± | 0.03 | -0.09 | ± | 0.04 | -0.33 | ± | 0.03 |
| GI Tau | -0.58 | ± | 0.01 | -0.32 | ± | 0.03 | -0.61 | ± | 0.02 |
| GK Tau | -0.51 | ± | 0.01 | 0.07 | ± | 0.01 | -0.37 | ± | 0.02 |
| GM Aur | -0.89 | ± | 0.06 | 2.27 | ± | 0.06 | 1.75 | ± | 0.04 |
| GN Tau | -0.89 | ± | 0.02 | -0.56 | ± | 0.03 | -0.93 | ± | 0.02 |
| GO Tau | -0.80 | ± | 0.07 | 0.22 | ± | 0.07 | 0.06 | ± | 0.05 |
| Haro 6-13 | -0.53 | ± | 0.01 | 1.02 | ± | 0.03 | 0.64 | ± | 0.01 |
| Haro 6-28 | -0.90 | ± | 0.06 | -0.13 | ± | 0.06 | -0.42 | ± | 0.07 |
| Haro 6-37 | -1.17 | ± | 0.01 | -0.51 | ± | 0.02 | -0.76 | ± | 0.02 |
| HK Tau | -0.72 | ± | 0.03 | 1.16 | ± | 0.04 | 0.78 | ± | 0.02 |
| HN Tau | -0.69 | ± | 0.01 | -0.19 | ± | 0.01 | -0.49 | ± | 0.01 |
| HO Tau | -0.91 | ± | 0.05 | -0.32 | ± | 0.07 | -0.59 | ± | 0.07 |
| HP Tau | -0.35 | ± | 0.01 | 0.14 | ± | 0.02 | -0.09 | ± | 0.01 |
| HQ Tau | -0.70 | ± | 0.01 | -0.22 | ± | 0.02 | -0.62 | ± | 0.02 |
| IP Tau | -0.93 | ± | 0.04 | 0.30 | ± | 0.05 | -0.03 | ± | 0.03 |
| IQ Tau | -1.30 | ± | 0.02 | -0.78 | ± | 0.05 | -0.91 | ± | 0.03 |
| IS Tau | -0.86 | ± | 0.03 | -0.88 | ± | 0.03 | -1.21 | ± | 0.03 |
| IT Tau | -1.37 | ± | 0.03 | -0.78 | ± | 0.04 | -0.87 | ± | 0.03 |
| LkCa 15 | -1.28 | ± | 0.03 | 0.90 | ± | 0.04 | 0.62 | ± | 0.03 |
| MHO-3 | 0.74 | ± | 0.01 | 1.47 | ± | 0.02 | 0.57 | ± | 0.05 |
| RW Aur A | -0.69 | ± | 0.01 | -0.27 | ± | 0.01 | -0.52 | ± | 0.01 |
| RY Tau | -0.30 | ± | 0.01 | 0.34 | ± | 0.01 | -0.11 | ± | 0.01 |
| T Tau | -0.50 | ± | 0.01 | 0.66 | ± | 0.02 | 0.51 | ± | 0.01 |
| UX Tau A | -1.98 | ± | 0.03 | 2.30 | ± | 0.04 | 1.69 | ± | 0.02 |
| UY Aur | -0.01 | ± | 0.01 | 0.23 | ± | 0.02 | -0.06 | ± | 0.01 |
| UZ Tau/e | -1.04 | ± | 0.03 | -0.47 | ± | 0.04 | -0.73 | ± | 0.04 |

| Name | $n$(6-13) | | | $n$(13-25) | | | $n$(13-31) | | |
|------|------|------|------|------|------|------|------|------|------|
| V410 Anon 13 | -0.93 | ± | 0.03 | -0.45 | ± | 0.04 | -0.58 | ± | 0.04 |
| V710 Tau | -1.16 | ± | 0.02 | -0.48 | ± | 0.03 | -0.63 | ± | 0.02 |
| V773 Tau | -1.12 | ± | 0.02 | -0.25 | ± | 0.05 | -0.61 | ± | 0.02 |
| V807 Tau | -1.62 | ± | 0.02 | 0.01 | ± | 0.03 | -0.21 | ± | 0.02 |
| V836 Tau | -1.36 | ± | 0.04 | -0.17 | ± | 0.04 | -0.46 | ± | 0.04 |
| V955 Tau | -0.89 | ± | 0.02 | -0.65 | ± | 0.04 | -0.92 | ± | 0.02 |
| VY Tau | -1.11 | ± | 0.09 | -0.07 | ± | 0.08 | -0.20 | ± | 0.06 |
| XZ Tau | -0.35 | ± | 0.01 | -0.09 | ± | 0.01 | -0.32 | ± | 0.01 |
| ZZ Tau | -1.07 | ± | 0.06 | -0.62 | ± | 0.07 | -0.85 | ± | 0.06 |
| ZZ Tau IRS | -0.42 | ± | 0.02 | 0.40 | ± | 0.04 | 0.11 | ± | 0.02 |

Table 3: assumed and derived parameters of the Taurus Class II YSOs

| Object | Spectral type | Number of stars, $N$ | Luminosity, $L_{star}$ ($L_\odot$) | Stellar mass,* $M_{star}$ ($M_\odot$) | Disk/star mass ratio, $M_{disk}/M_{star}$, ×1000 | | | Accretion rate, $dM_{star}/dt$ ($M_\odot$ yr$^{-1}$) | Ref. † |
|---|---|---|---|---|---|---|---|---|---|
| 04108+2910 | M0 | 1 | 0.16 | 0.57± 0.10 | | | | | 1,13 |
| 04187+1927 | M0 | 1 | | 0.57± 0.10 | | | | | 1,25 |
| 04200+2759 | | 1 | | | | | | | 1,20 |
| 04216+2603 | M1 | 1 | | 0.47± 0.10 | | | | | 1,13 |
| 04248+2612 | | 1 | | | | | | | |
| 04303+2240 | K7 | 1 | 2.71 | 0.75± 0.10 | 39.9 | ± | 5.3 | 8.9E-07 | 11 |
| 04370+2559 | M0 | 2 | 0.42 | 0.57± 0.10 | | | | | 21 |
| 04385+2550 | M0 | 2 | 0.52 | 0.57± 0.10 | | | | 7.8E-09 | 1,11 |
| AA Tau | M0 | 1 | 0.87 | 0.57± 0.10 | 23.0 | ± | 5.4 | 6.5E-09 | 1,2 |
| BP Tau | K7 | 1 | 1.08 | 0.75± 0.10 | 23.9 | ± | 6.2 | 1.3E-08 | 1,2 |
| | | | | 1.24± 0.28 | | | | | 30 |
| CI Tau | K7 | 1 | 1.11 | 0.75± 0.10 | 37.2 | ± | 9.4 | 2.6E-08 | 1,2 |
| CoKu Tau/3 | M1 | 2 | 0.68 | 0.47± 0.10 | 0.8 | ± | 0.9 | | |
| CoKu Tau/4 | M1.5 | 2 | 0.68 | 0.43± 0.10 | 1.2 | ± | 0.3 | | 10,13,31 |
| CW Tau | K3 | 1 | 1.46 | 1.82± 0.10 | 1.3 | ± | 0.2 | 1.0E-08 | 1,2 |
| CX Tau | M2 | 1 | 0.48 | 0.39± 0.10 | 2.6 | ± | 0.6 | 3.7E-10 | 1,2 |
| CY Tau | M1 | 1 | 0.52 | 0.47± 0.10 | 12.7 | ± | 2.7 | 1.4E-09 | 1,2 |
| | | | | 0.55± 0.33 | | | | | 30 |
| CZ Tau | M1 | 2 | 0.43 | 0.47± 0.10 | 0.8 | ± | 0.9 | 4.5E-10 | 1,18 |
| DD Tau | M3 | 2 | 0.47 | 0.34± 0.09 | 2.1 | ± | 0.5 | 7.9E-10 | 1,2 |
| DE Tau | M2 | 1 | 0.91 | 0.39± 0.10 | 13.3 | ± | 3.9 | 4.1E-08 | 1,2 |
| DF Tau | M3 | 2 | 1.77 | 0.34± 0.09 | 1.2 | ± | 0.4 | 1.1E-08 | 1,2 |
| DG Tau | K6 | 1 | 2.46 | 0.91± 0.10 | 26.3 | ± | 4.4 | 4.6E-08 | 1,2 |
| DH Tau | M1 | 2 | 0.63 | 0.47± 0.10 | 7.0 | ± | 2.1 | 1.1E-09 | 1,2 |

Table 3: assumed and derived parameters of the Taurus Class II YSOs

| Object | Spectral type | Number of stars, $N$ | Luminosity, $L_{star}$ ($L_\odot$) | Stellar mass,* $M_{star}$ ($M_\odot$) | | Disk/star mass ratio, $M_{disk}/M_{star}$, ×1000 | | | Accretion rate, $dM_{star}/dt$ ($M_\odot$ yr$^{-1}$) | Ref. † |
|---|---|---|---|---|---|---|---|---|---|---|
| DK Tau | M0 | 2 | 1.72 | 0.57± | 0.10 | 8.8 | ± | 2.4 | 3.8E-08 | 1,3 |
| DL Tau | K7 | 1 | 0.85 | 0.75± | 0.10 | 119.7 | ± | 31.0 | 6.8E-08 | 1,2 |
| | | | | 0.72± | 0.11 | | | | | 30 |
| DM Tau | M1 | 1 | 0.29 | 0.47± | 0.10 | 51.0 | ± | 13.8 | 2.1E-09 | 1,2,29 |
| | | | | 0.55± | 0.03 | | | | | 30 |
| DN Tau | M0 | 1 | 0.97 | 0.57± | 0.10 | 51.3 | ± | 14.0 | 1.9E-09 | 1,2,29 |
| DO Tau | M0 | 1 | 1.05 | 0.57± | 0.10 | 12.4 | ± | 2.8 | 3.0E-08 | 1,2,29 |
| DP Tau | M0 | 1 | 0.18 | 0.57± | 0.10 | 0.9 | ± | 0.9 | 3.2E-09 | 1,2,29 |
| DQ Tau | M0 | 2 | 0.90 | 0.57± | 0.10 | 33.6 | ± | 10.7 | 6.0E-10 | 1,3,29 |
| DR Tau | K7 | 1 | 1.54 | 0.75± | 0.10 | 25.3 | ± | 5.2 | 3.2E-08 | 1,2,29 |
| DS Tau | K5 | 1 | 0.93 | 1.11± | 0.10 | 5.4 | ± | 1.0 | 1.1E-08 | 1,2,29 |
| F04147+2822 | M4 | 1 | 0.22 | 0.30± | 0.08 | | | | | 1,13 |
| F04192+2647 | | 2 | | | | | | | | 1 |
| F04262+2654 | | 1 | | | | | | | | 1 |
| F04297+2246A | | 2 | | | | | | | | 1 |
| F04297+2246B | | 1 | | | | | | | | |
| F04570+2520 | | 1 | | | | | | | | 1 |
| FM Tau | M0 | 2 | 0.40 | 0.57± | 0.10 | 3.5 | ± | 0.6 | 1.3E-09 | 1,2,29 |
| FN Tau | M5 | 1 | 0.63 | 0.25± | 0.05 | | | | | 1,13 |
| FO Tau | M2 | 2 | 1.08 | 0.39± | 0.10 | 1.5 | ± | 0.4 | 2.1E-08 | 2,29 |
| FP Tau | M4 | 1 | 0.32 | 0.30± | 0.08 | | | | 3.5E-10 | 2 |
| FQ Tau | M2 | 2 | 0.37 | 0.39± | 0.10 | 2.6 | ± | 0.6 | 3.5E-07 | 1,18,29 |
| FS Tau | M1 | 2 | 0.28 | 0.47± | 0.10 | 4.2 | ± | 0.9 | 4.0E-10 | 1,2,29 |
| FT Tau | C | 1 | 1.30 | | | | | | | 1 |

Table 3: assumed and derived parameters of the Taurus Class II YSOs

| Object | Spectral type | Number of stars, $N$ | Luminosity, $L_{star}$ ($L_\odot$) | Stellar mass,* $M_{star}$ ($M_\odot$) | Disk/star mass ratio, $M_{disk}/M_{star}$, ×1000 | | | Accretion rate, $dM_{star}/dt$ ($M_\odot$yr$^{-1}$) | Ref. † |
|---|---|---|---|---|---|---|---|---|---|
| FV Tau | K5 | 2 | 1.83 | 1.11± 0.10 | 1.0 | ± | 0.2 | 3.3E-08 | 2,29 |
| FX Tau | M1 | 2 | 1.08 | 0.47± 0.10 | 1.9 | ± | 0.6 | 5.8E-09 | 1,5,29 |
| FZ Tau | M0 | 2 | 1.08 | 0.57± 0.10 | 3.5 | ± | 1.2 | 2.0E-08 | 1,2,29 |
| GG Tau A | M0 | 4 | 1.66 | 0.57± 0.10 | 407.1 | ± | 158.9 | 1.9E-08 | 1,2,29 |
| GH Tau | M2 | 2 | 1.04 | 0.39± 0.10 | 1.8 | ± | 0.5 | 1.7E-09 | 1,2,29 |
| GI Tau | K6 | 2 | 1.39 | 0.91± 0.10 | | | | 8.3E-09 | 1,2 |
| GK Tau | M0 | 2 | 1.22 | 0.57± 0.10 | 3.5 | ± | 0.6 | 6.5E-09 | 1,2,29 |
| GM Aur | K5 | 1 | 1.19 | 1.11± 0.10 | 22.5 | ± | 4.9 | 6.6E-09 | 2,29 |
| | | | | 0.84± 0.05 | | | | | 30 |
| GN Tau | M2 | 2 | 0.74 | 0.39± 0.10 | 1.5 | ± | 0.4 | 1.3E-08 | 7,8,29 |
| GO Tau | M0 | 1 | 0.33 | 0.57± 0.10 | 123.9 | ± | 41.6 | 3.8E-09 | 1,2,29 |
| Haro 6-13 | M0 | 1 | 10.90 | 0.57± 0.10 | 19.5 | ± | 3.9 | | 29 |
| Haro 6-28 | | 1 | 0.54 | | | | | | |
| Haro 6-37 | K6 | 3 | 1.74 | 0.91± 0.10 | 10.9 | ± | 1.2 | 7.6E-09 | 1,2,5,29 |
| HK Tau | M1 | 2 | 0.49 | 0.47± 0.10 | 9.6 | ± | 2.3 | | 29 |
| HN Tau | K5 | 2 | 0.33 | 1.11± 0.10 | 0.7 | ± | 0.2 | 2.5E-09 | 1,2,29 |
| HO Tau | M0 | 2 | 0.17 | 0.57± 0.10 | 3.5 | ± | 0.6 | 1.3E-09 | 1,2,29 |
| HP Tau | K3 | 1 | 1.49 | 1.82± 0.10 | 0.6 | ± | 0.2 | | 29 |
| HQ Tau | | 1 | 5.37 | | | | | | 1 |
| IP Tau | M0 | 1 | 0.53 | 0.57± 0.10 | 5.0 | ± | 1.4 | 3.5E-10 | 2,29 |
| IQ Tau | M1 | 1 | 0.91 | 0.47± 0.10 | 46.7 | ± | 11.8 | | 2,29 |
| IS Tau | M0 | 2 | 0.65 | 0.57± 0.10 | 1.8 | ± | 0.3 | 7.9E-09 | 1,2,29 |
| IT Tau | K2 | 2 | 1.55 | 2.15± 0.23 | 0.7 | ± | 0.2 | | 1,29 |
| LkCa 15 | K5 | 1 | 0.99 | 1.11± 0.10 | 43.2 | ± | 9.0 | 1.5E-09 | 1,2,29 |

Table 3: assumed and derived parameters of the Taurus Class II YSOs

| Object | Spectral type | Number of stars, $N$ | Luminosity, $L_{star}$ ($L_\odot$) | Stellar mass,* $M_{star}$ ($M_\odot$) | Disk/star mass ratio, $M_{disk}/M_{star}$, ×1000 | | | Accretion rate, $dM_{star}/dt$ ($M_\odot\,yr^{-1}$) | Ref. † |
|---|---|---|---|---|---|---|---|---|---|
| | | | | 0.97± 0.03 | | | | | 30 |
| MHO-3 | M0 | 1 | 0.94 | 0.57± 0.10 | | | | | 1,21 |
| RW Aur A | K3 | 3 | 2.41 | 1.82± 0.10 | 2.2 | ± | 0.1 | 3.3E-08 | 1,2,29 |
| RY Tau | G1 | 1 | 19.49 | 2.99± 0.39 | 6.0 | ± | 1.3 | 7.8E-08 | 17,13,29 |
| T Tau | K0 | 3 | 11.10 | 2.67± 0.25 | 3.1 | ± | 0.4 | 3.2E-08 | 1,2,29 |
| UX Tau A | K5 | 4 | 1.81 | 1.11± 0.10 | 4.6 | ± | 0.8 | 5.0E-10 | 1,2,7,29 |
| UY Aur | M0 | 2 | 1.25 | 0.57± 0.10 | 3.2 | ± | 0.8 | 4.3E-08 | 1,6,29 |
| UZ Tau/e | M1 | 4 | 1.05 | 0.47± 0.10 | 127.4 | ± | 34.4 | 5.6E-09 | 2,29 |
| V410 Anon 13 | M6 | 1 | 0.08 | 0.08± 0.03 | | | | 5.0E-12 | 1,12,2,14 |
| V710 Tau | M1 | 2 | 1.53 | 0.47± 0.10 | 40.3 | ± | 12.1 | | 1,2,29 |
| V773 Tau | K3 | 4 | 8.02 | 1.82± 0.10 | 0.3 | ± | 0.0 | | 1,2,29 |
| V807 Tau | K7 | 2 | 2.64 | 0.75± 0.10 | 1.3 | ± | 0.2 | 4.0E-09 | 1,2,29 |
| V836 Tau | K7 | 1 | 0.59 | 0.75± 0.10 | 13.3 | ± | 4.4 | 6.3E-09 | 1,4,29 |
| V955 Tau | K5 | 2 | 1.33 | 1.11± 0.10 | 0.5 | ± | 0.1 | 5.2E-09 | 2,29 |
| VY Tau | M0 | 2 | 0.55 | 0.57± 0.10 | 0.9 | ± | 0.9 | | 1,29 |
| XZ Tau | M3 | 2 | 0.94 | 0.34± 0.09 | | | | 1.8E-09 | 1,2 |
| ZZ Tau | M3 | 2 | 0.72 | 0.34± 0.05 | 1.2 | ± | 1.2 | 1.3E-09 | 1,5,29 |
| ZZ Tau IRS | M5 | 2 | 0.03 | 0.25± 0.09 | | | | 8.7E-09 | 1,11 |

* When two masses are given, the first was determined from isochrones as in §4.1. The second is the mass determined from disk kinematics as revealed in molecular-line observations. The two are generally in good agreement. The latter should be regarded as more accurate, but for consistency with the other objects we use the former in our analysis.

† References: (1) Kenyon & Hartmann 1995; (2) White & Ghez 2001; (3) Gullbring et al 1998; (4) Hartigan, Edwards, & Ghandour 1995; (5) Valenti et al 1993; (6) Hartigan & Kenyon 2003; (7) White & Basri 2003; (8) Luhman 2004; (9) Muzerolle et al. 2003b; (10) d'Alessio et al. 2005; (11) White & Hillenbrand 2004; (12) Furlan et al. 2005; (13) Kenyon et al. 1998; (14) Muzerolle et al. 2000; (15) Muzerolle et al. 2003a; (16) Muzerolle et al. 1998; (17) Calvet et al. 2004; (18) Hartmann et al. 1998; (19) White & Hillenbrand 2005; (20) Kenyon et al. 1994a; (21) Briceño et al. 1998; (22) DeWarf et al. 2003; (23) Smith et al. 2005; (24) Kenyon et al. 1990; (25) Gullbring et al. 2000; (26) Briceño et al. 2002; (27) Itoh et al. 2002; (28) Böhm & Catala 1993; (29) Andrews & Williams 2005; (30) Simon, Dutrey & Guilloteau 2001; (31) Ireland & Kraus 2008.

Table 4: linear correlation coefficients between the Class II YSO system parameters from tables 2 and 3 (below diagonal), and the probabilities (%) that these correlations could have been generated from a random distribution (above diagonal).

| | $M_{\rm star}$ | $M_{\rm disk}$ | $\dfrac{M_{\rm disk}}{M_{\rm star}}$ | $L_{\rm star}$ | $\dfrac{dM_{\rm star}}{dt}$ | $N$ | $P_{10}$ | $O_{10}$ | $S_{10}$ | $O_{33}$ | $W_{10}$ | $W_{20}$ | $W_{33}$ | $\Delta\lambda_{10}$ | $n_{6-13}$ | $n_{13-25}$ | $n_{13-31}$ |
|---|---|---|---|---|---|---|---|---|---|---|---|---|---|---|---|---|---|
| $M_{\rm star}$ | | 70 | 36 | 0 | 77 | 15 | 37 | 80 | 38 | 17 | 43 | 46 | 38 | 0 | 78 | 50 | 46 |
| $M_{\rm disk}$ | -0.05 | | 0 | 97 | 71 | 10 | 70 | 59 | 93 | 19 | 24 | 89 | 19 | 85 | 75 | 80 | 67 |
| $M_{\rm disk}/M_{\rm star}$ | -0.12 | 0.99 | | 70 | 86 | 5 | 81 | 61 | 78 | 25 | 42 | 76 | 17 | 71 | 75 | 94 | 89 |
| $L_{\rm star}$ | 0.68 | 0.00 | -0.05 | | 40 | 43 | 22 | 97 | 91 | 11 | 34 | 23 | 13 | 0 | 36 | 38 | 33 |
| $dM_{\rm star}/dt$ | 0.04 | 0.05 | 0.02 | 0.11 | | 40 | 9 | 7 | 4 | 58 | 39 | 53 | 16 | 94 | 98 | 22 | 36 |
| $N$ | 0.17 | 0.20 | 0.25 | 0.09 | -0.11 | | 67 | 9 | 11 | 52 | 16 | 84 | 42 | 8 | 6 | 97 | 75 |
| $P_{10}$ | -0.11 | -0.05 | -0.03 | -0.14 | -0.23 | 0.05 | | 13 | 1 | 0 | 15 | 0 | 3 | 0 | 12 | 0 | 0 |
| $O_{10}$ | 0.03 | -0.07 | -0.07 | 0.00 | 0.24 | 0.19 | 0.17 | | 0 | 0 | 0 | 15 | 0 | 0 | 0 | 1 | 1 |
| $S_{10}$ | 0.10 | -0.01 | -0.04 | 0.01 | 0.27 | 0.18 | 0.31 | 0.64 | | 2 | 0 | 2 | 0 | 0 | 0 | 0 | 0 |
| $O_{33}$ | -0.16 | -0.17 | -0.15 | -0.19 | 0.07 | -0.07 | 0.45 | 0.33 | 0.27 | | 1 | 3 | 0 | 13 | 2 | 0 | 0 |
| $W_{10}$ | 0.09 | 0.15 | 0.11 | 0.11 | 0.09 | 0.02 | -0.16 | -0.51 | -0.41 | -0.28 | | 0 | 34 | 44 | 27 | 0 | 0 |
| $W_{20}$ | 0.09 | -0.02 | -0.04 | 0.14 | 0.09 | 0.02 | -0.32 | -0.16 | -0.26 | -0.23 | 0.61 | | 51 | 62 | 81 | 0 | 2 |
| $W_{33}$ | -0.10 | -0.17 | -0.18 | -0.18 | 0.19 | 0.09 | 0.24 | 0.32 | 0.35 | 0.70 | -0.11 | 0.07 | | 17 | 0 | 1 | 0 |
| $\Delta\lambda_{10}$ | -0.44 | 0.02 | 0.05 | -0.40 | 0.01 | -0.19 | 0.70 | 0.45 | 0.47 | 0.16 | 0.08 | 0.05 | 0.15 | | 42 | 5 | 0 |
| $n_{6-13}$ | -0.03 | 0.04 | 0.04 | 0.11 | 0.00 | -0.21 | -0.17 | -0.41 | -0.44 | -0.26 | 0.12 | -0.03 | -0.32 | -0.09 | | 14 | 2 |
| $n_{13-25}$ | 0.08 | 0.03 | -0.01 | 0.10 | -0.16 | 0.00 | -0.44 | -0.30 | -0.36 | -0.45 | 0.32 | 0.47 | -0.28 | -0.21 | 0.16 | | 7 |
| $n_{13-31}$ | 0.09 | 0.05 | 0.02 | 0.12 | -0.12 | -0.04 | -0.47 | -0.28 | -0.35 | -0.48 | -0.46 | 0.26 | 0.36 | -0.30 | -0.25 | 0.20 | |

Key:

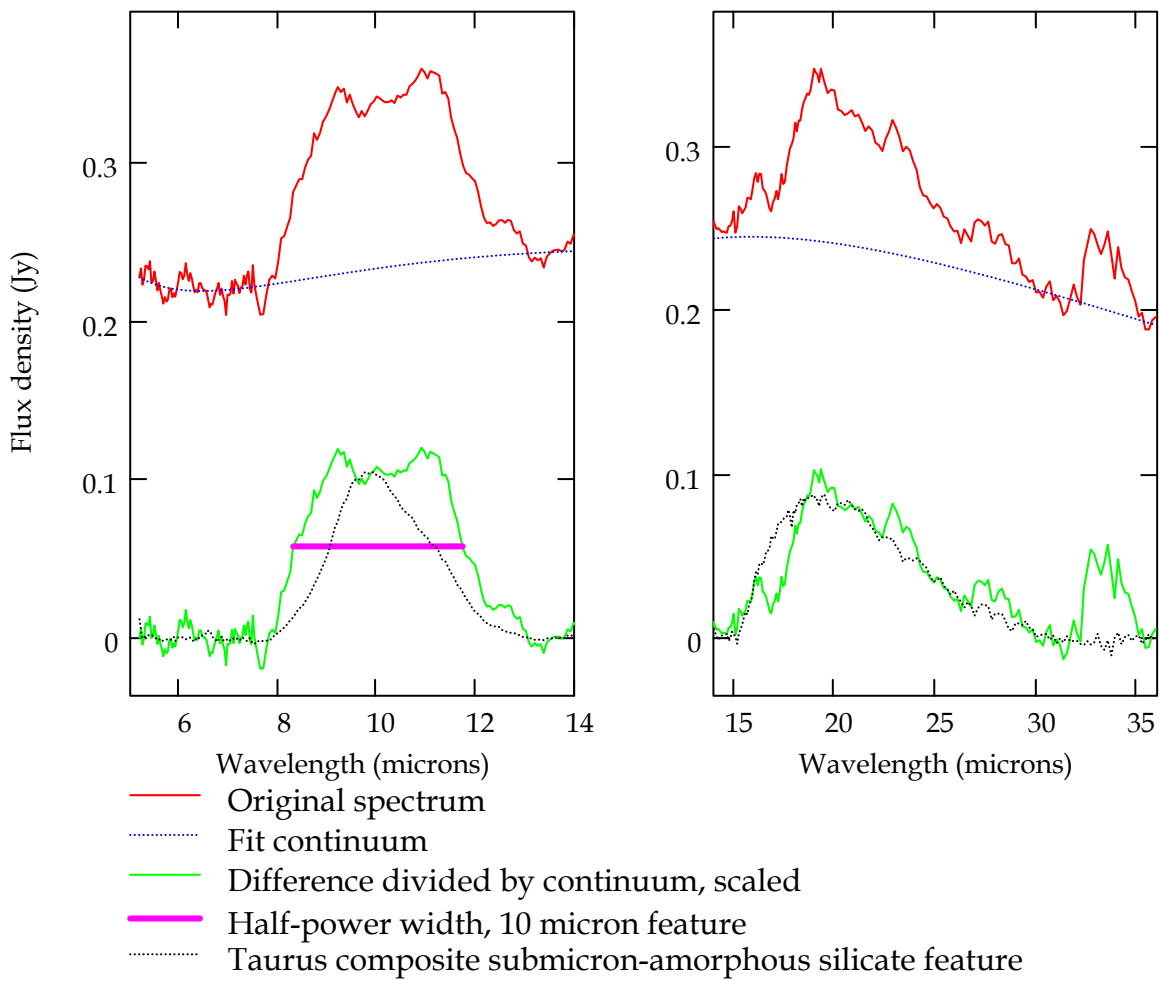

Original spectrum
Fit continuum
Difference divided by continuum, scaled
Half-power width, 10 micron feature
Taurus composite submicron-amorphous silicate feature

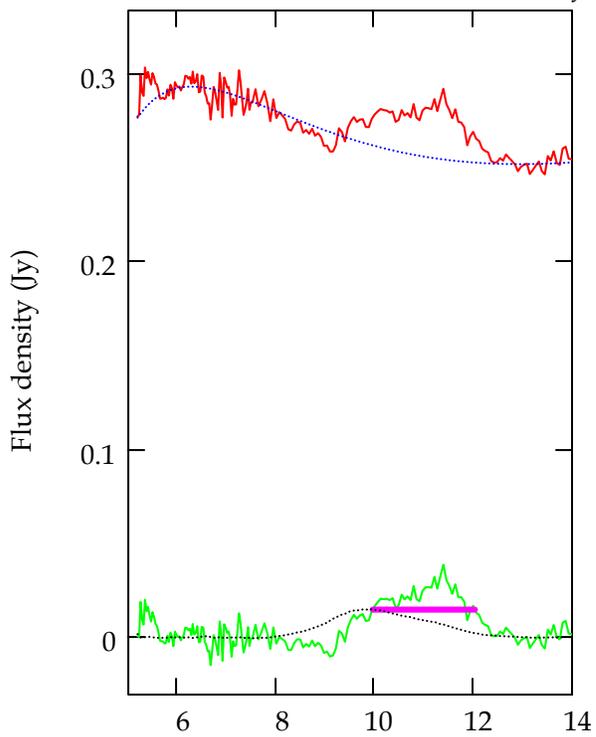
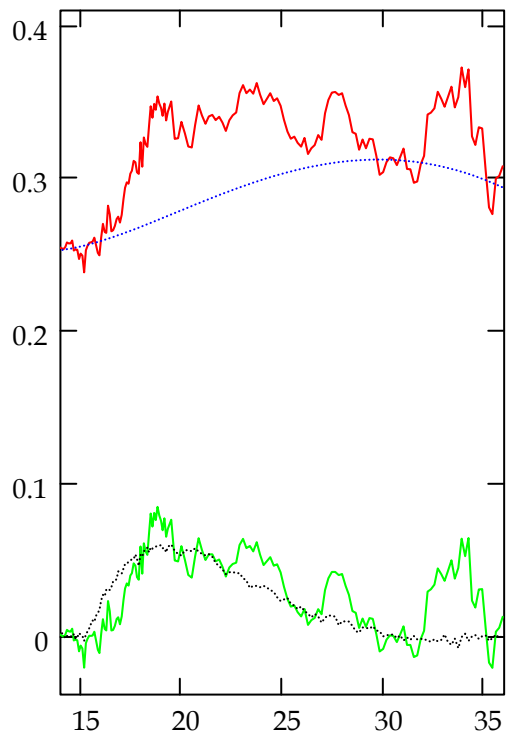
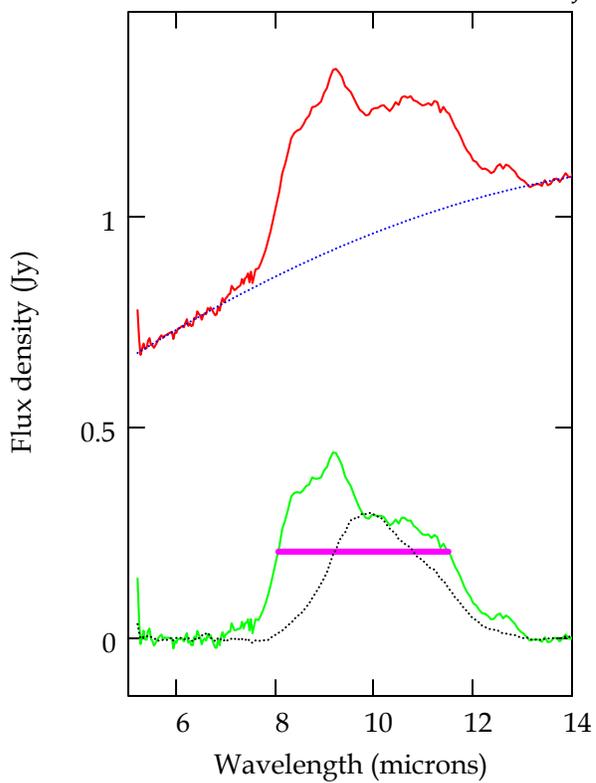
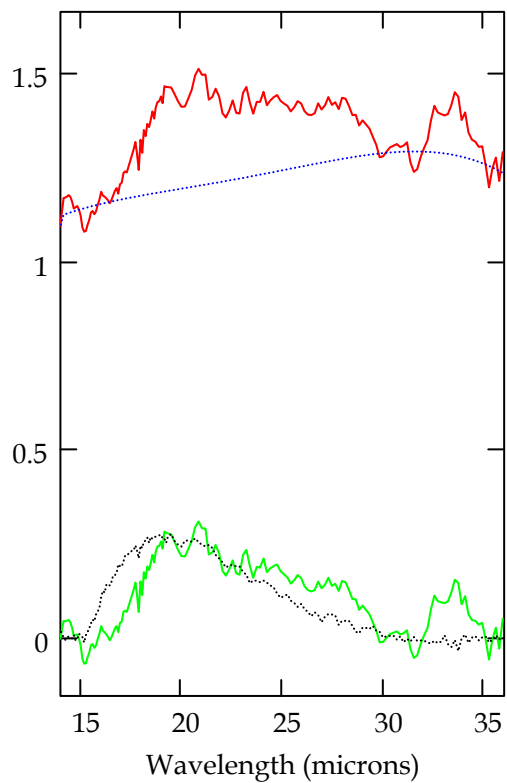

Names$_J$ = "04108+2910"

Names$_J$ = "04187+1927"

Flux density (Jy)

Wavelength (microns)

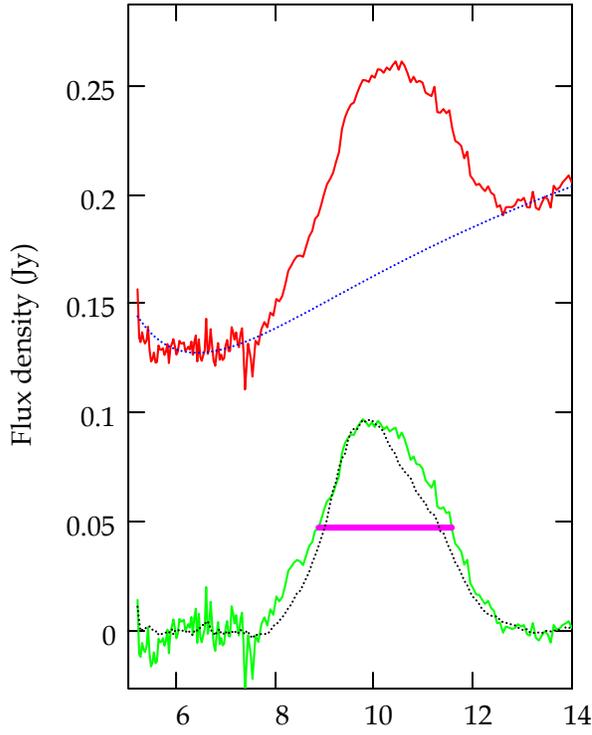
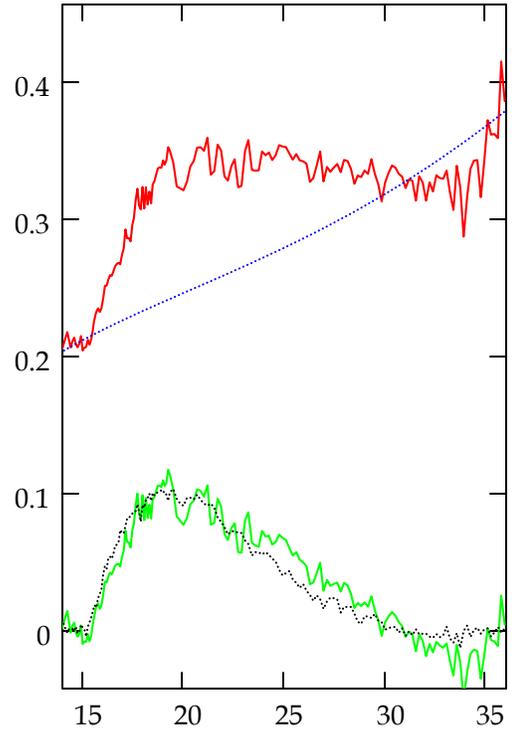

Names$_J$ = "04200+2759"

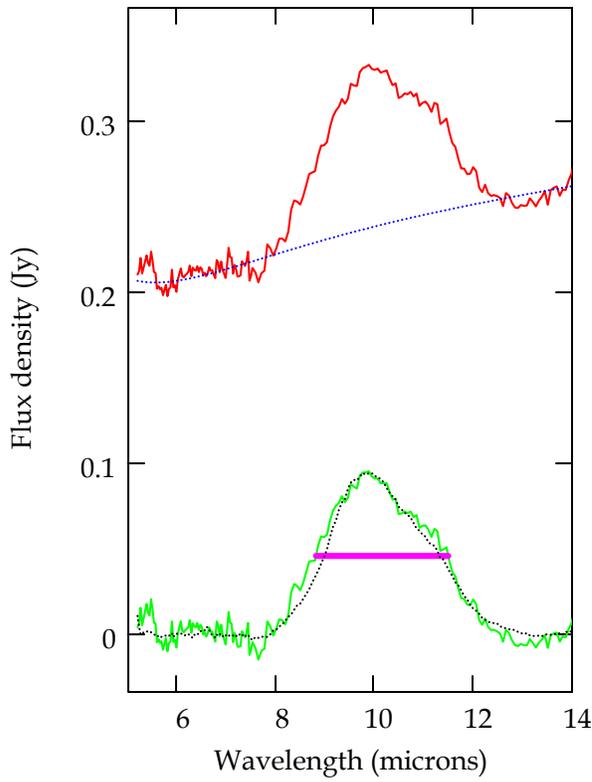
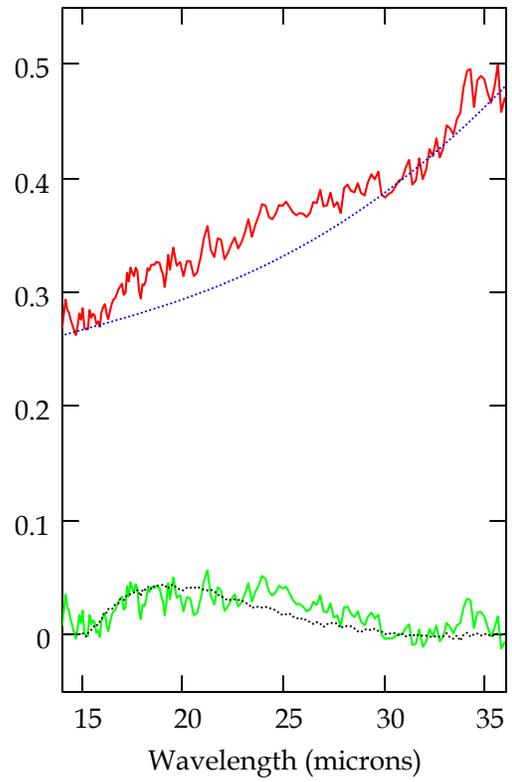

Names$_J$ = "04216+2603"

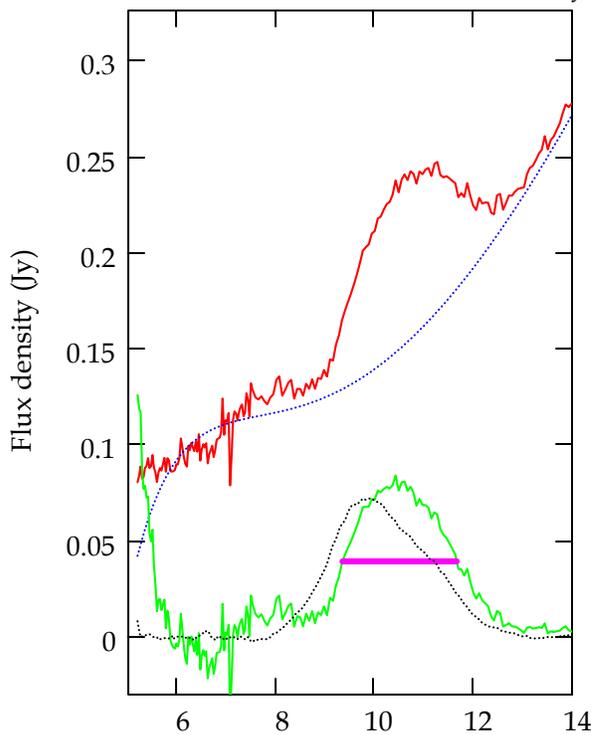
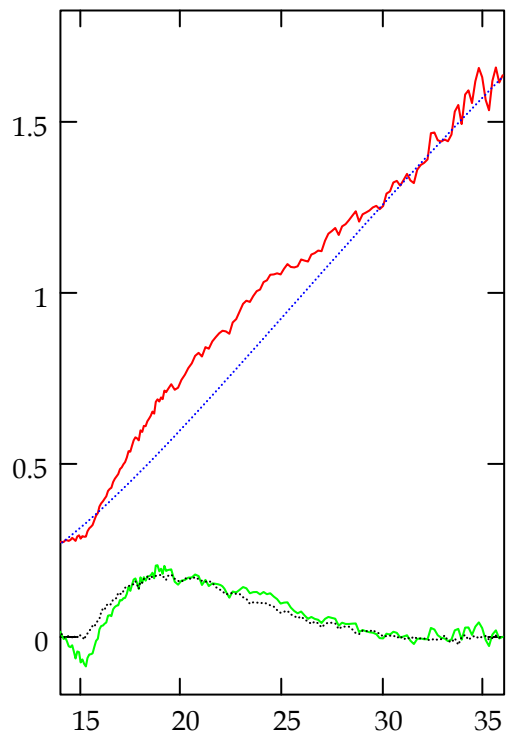

Names$_J$ = "04248+2612"

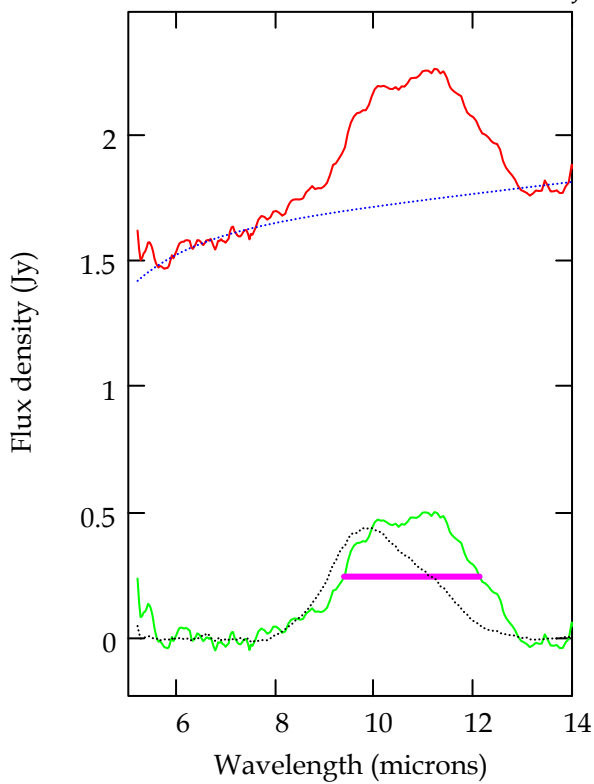
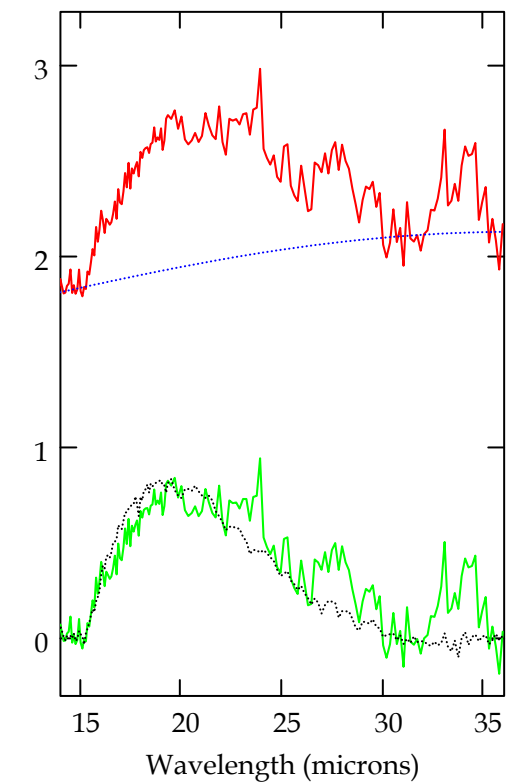

Names$_J$ = "04303+2240"

Names$_J$ = "04370+2559"

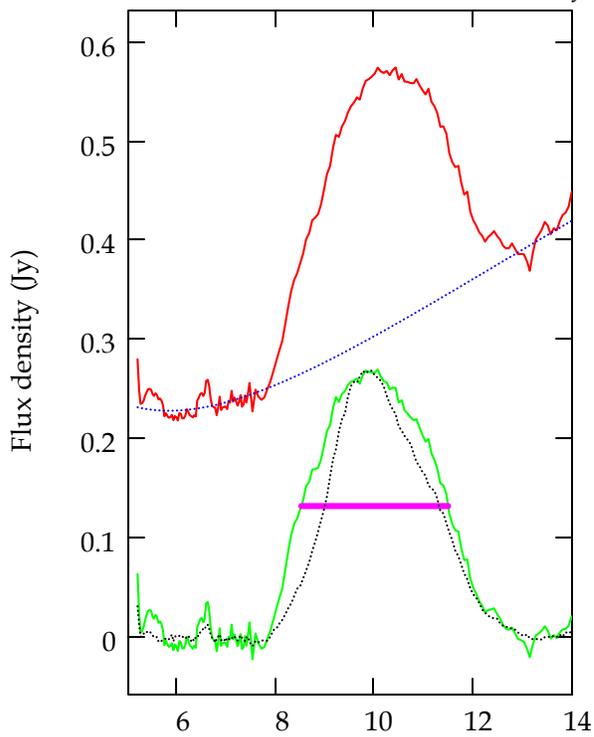
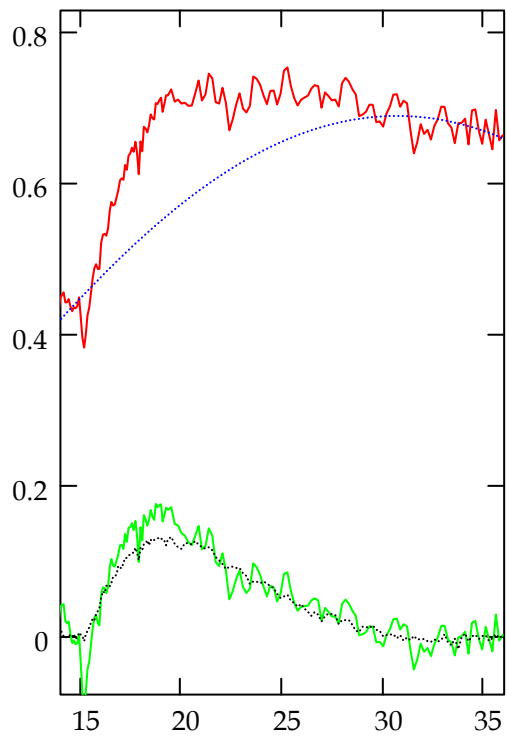

Names$_J$ = "04385+2550"

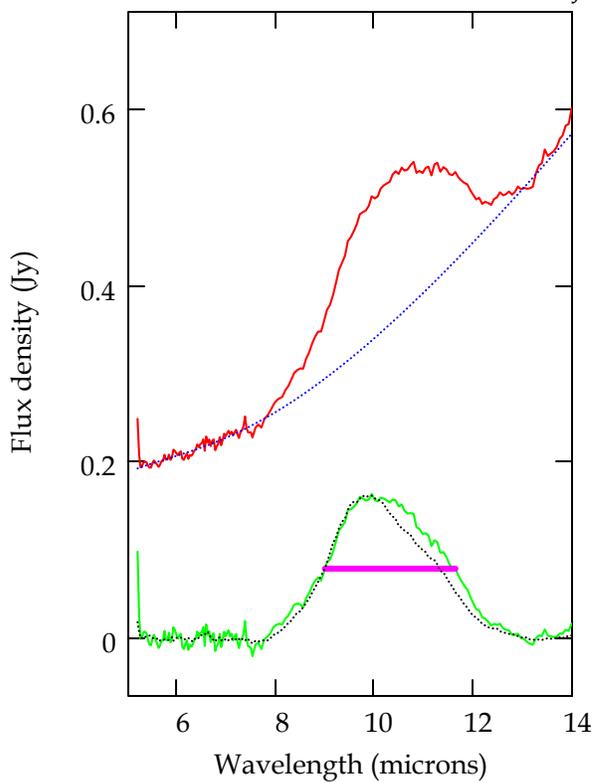
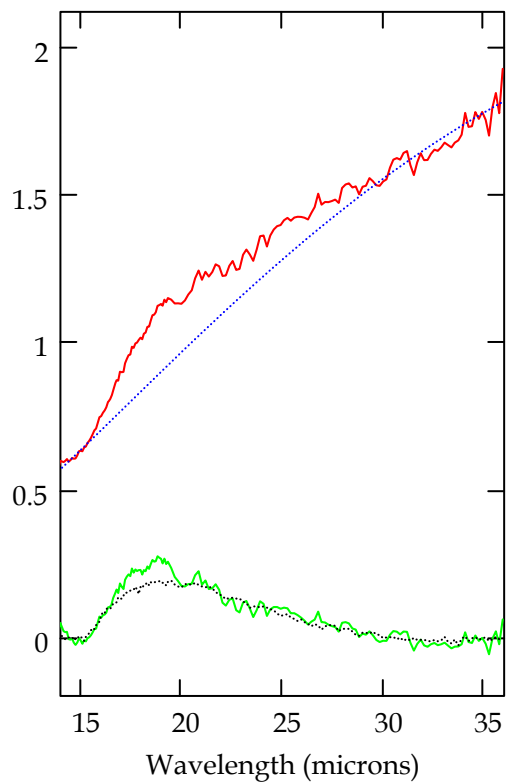

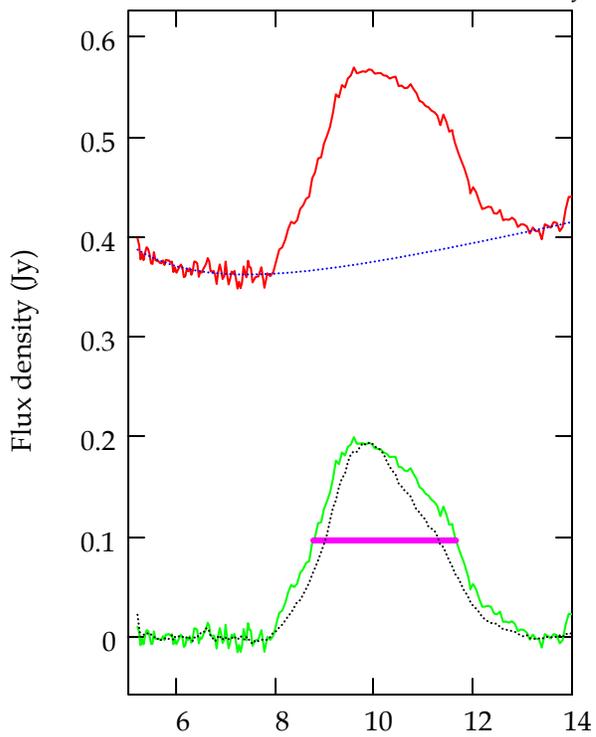
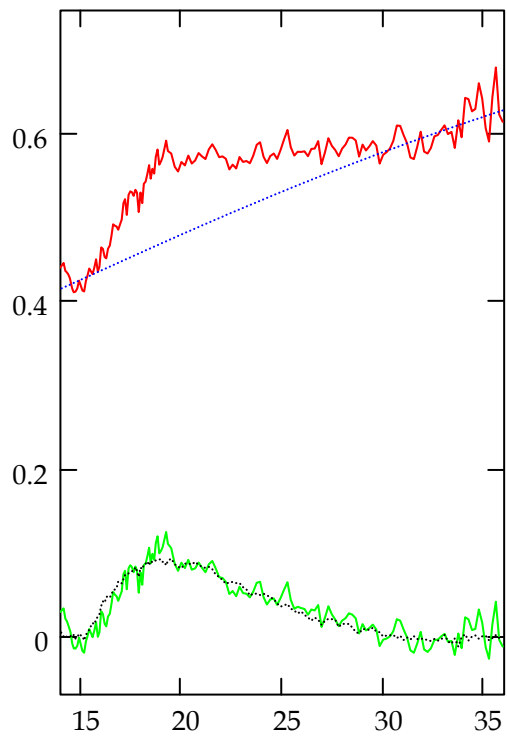

Names$_J$ = "AA Tau"

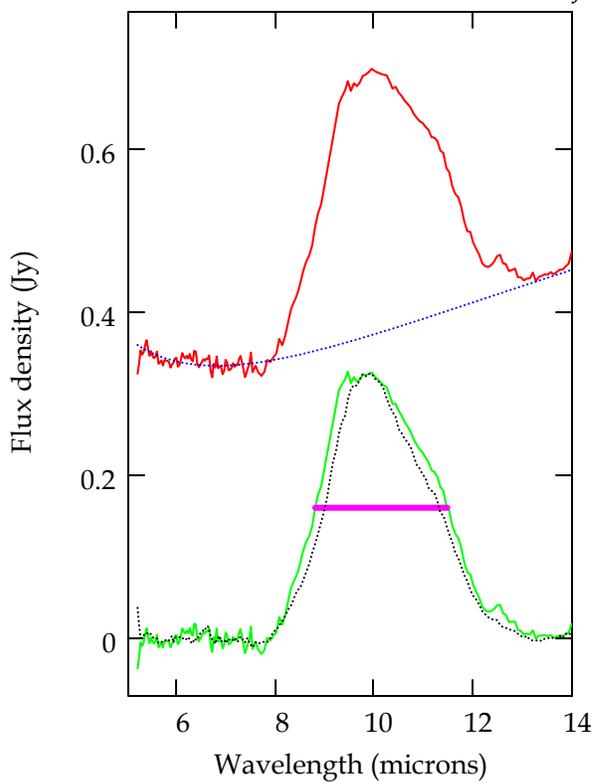
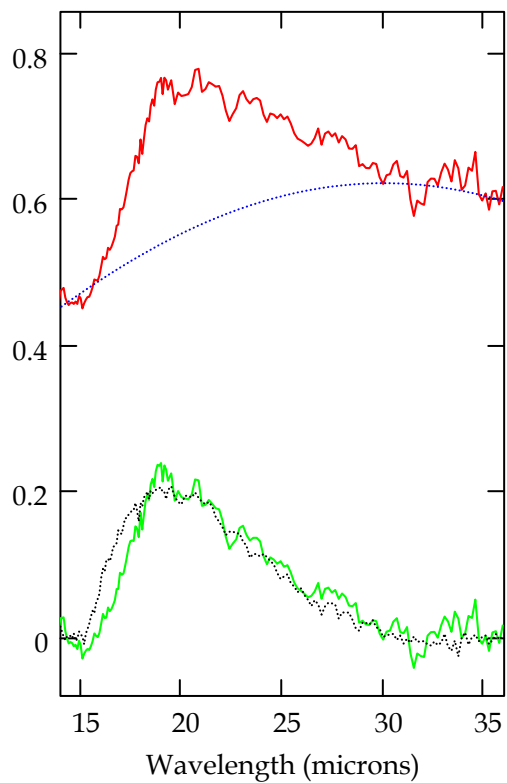

Names$_J$ = "BP Tau"

Names$_J$ = "CI Tau"

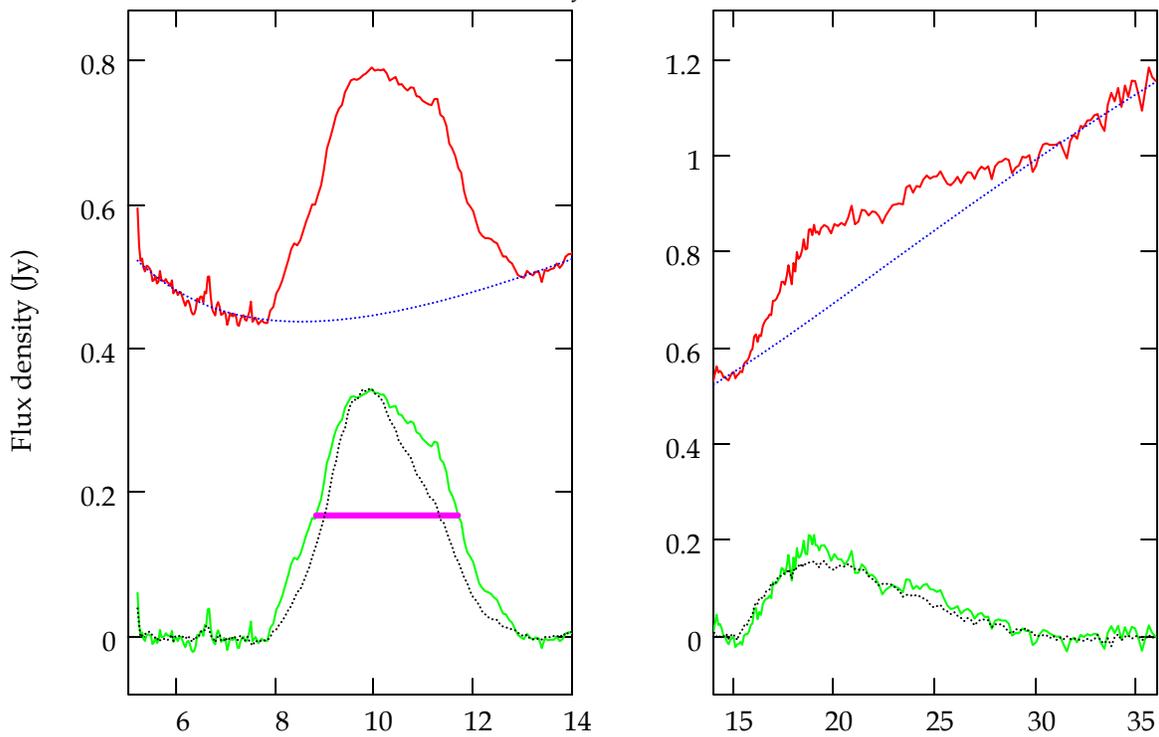

Names$_J$ = "CoKu Tau/3"

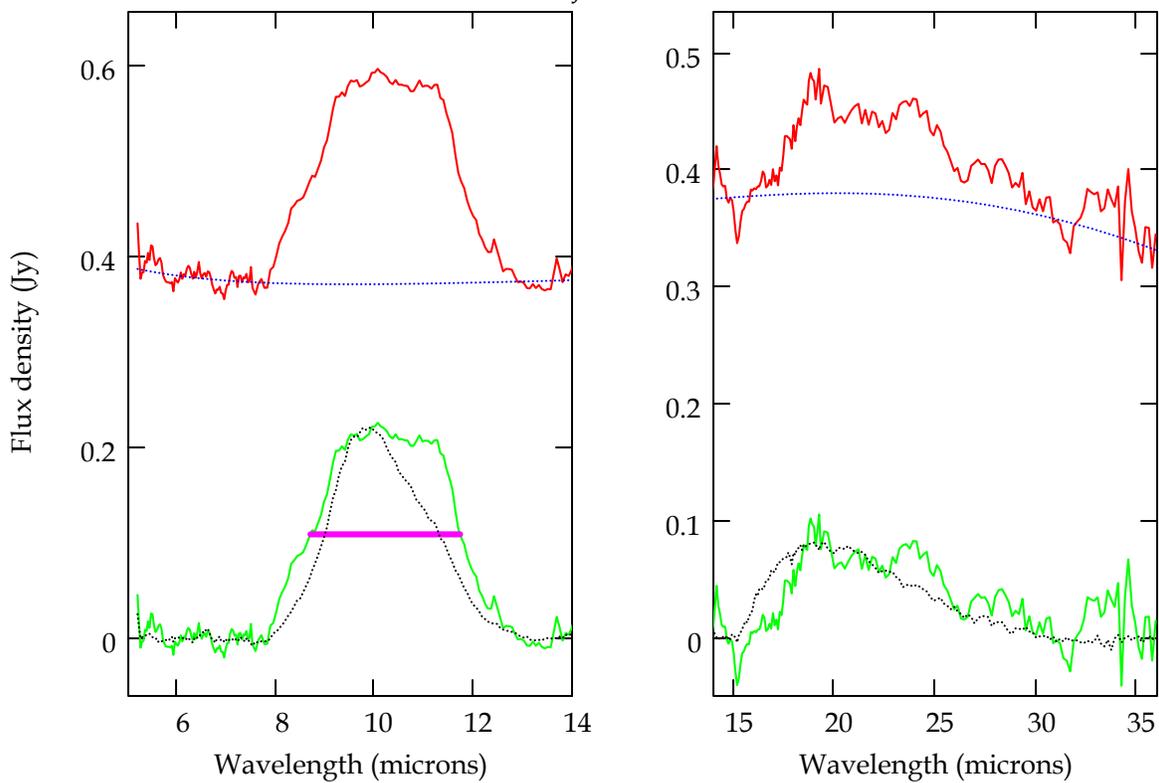

Names$_J$ = "CoKu Tau/4"

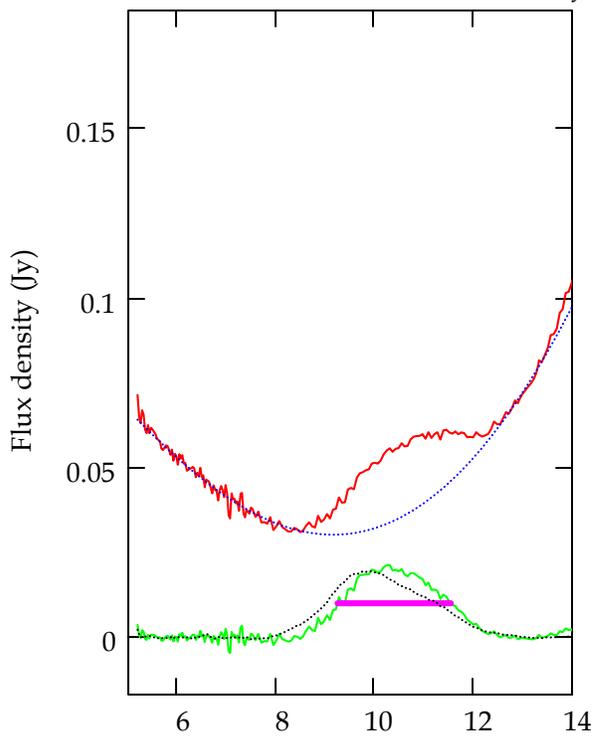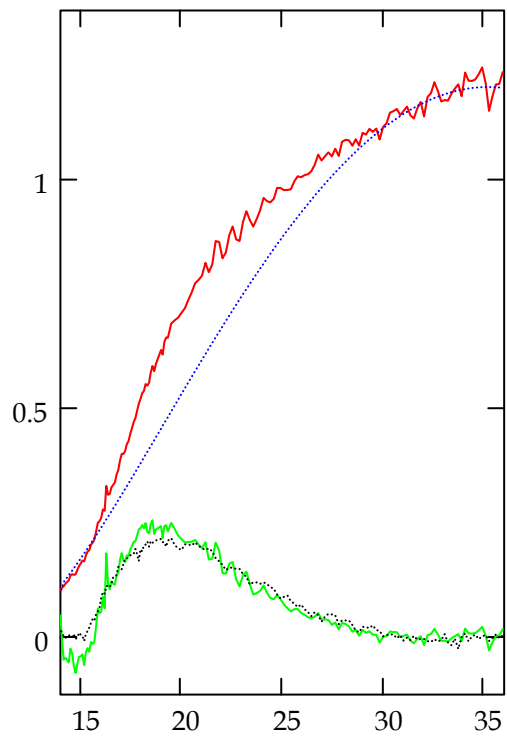

Names$_J$ = "CW Tau"

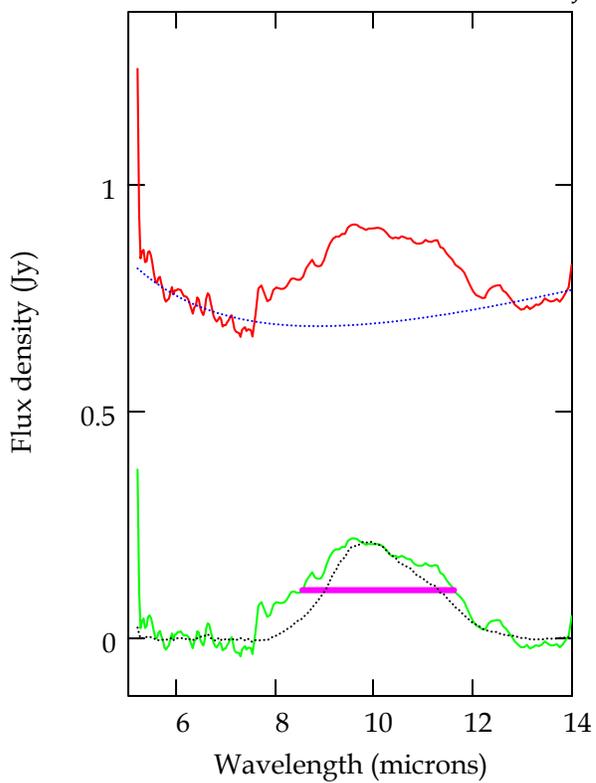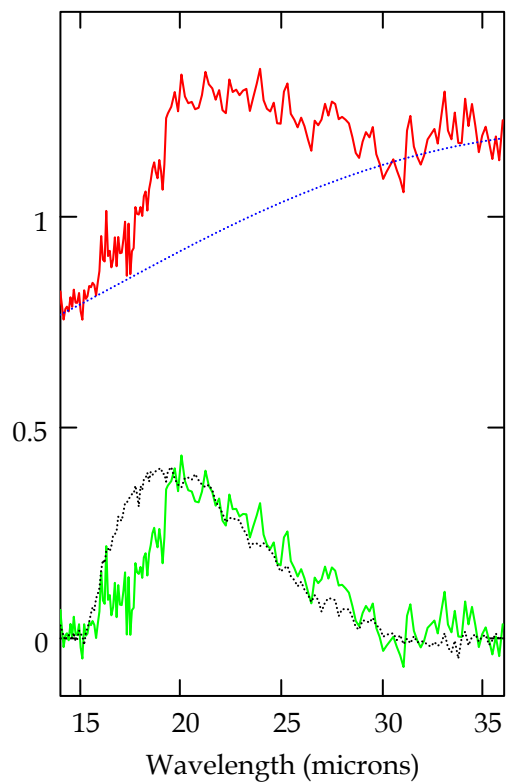

Flux density (Jy)

Wavelength (microns)

Names$_J$ = "CX Tau"

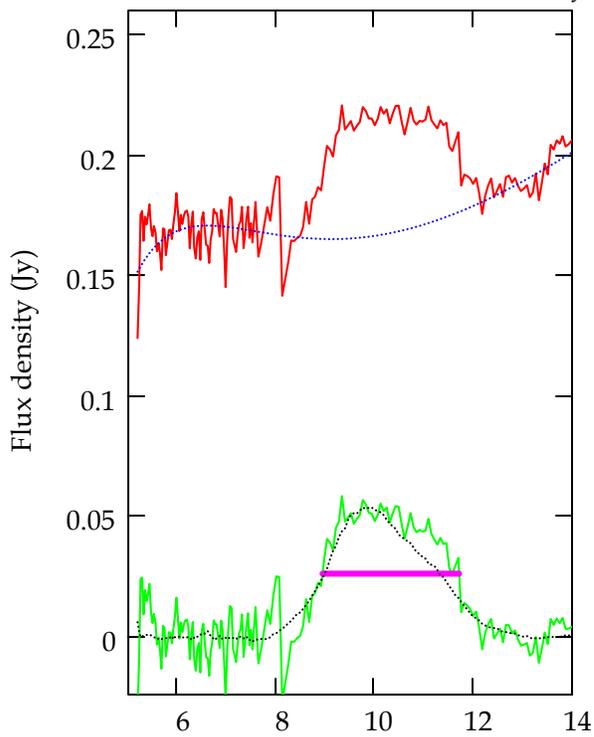
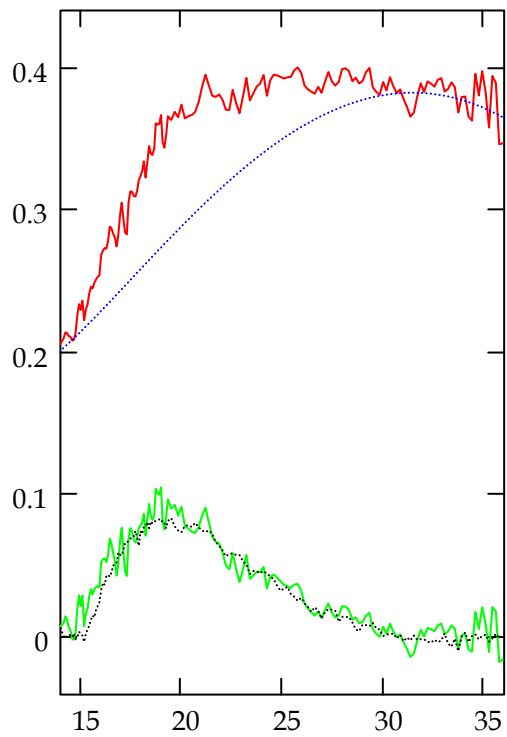

Names$_J$ = "CY Tau"

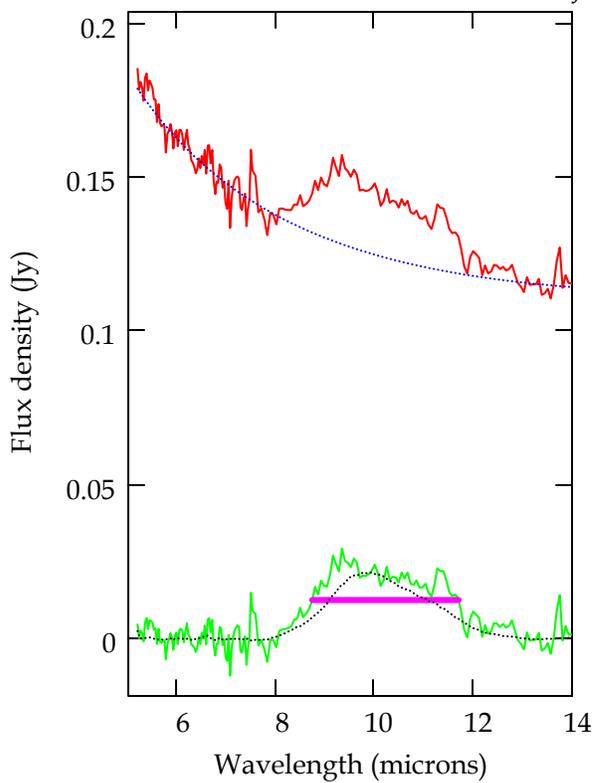
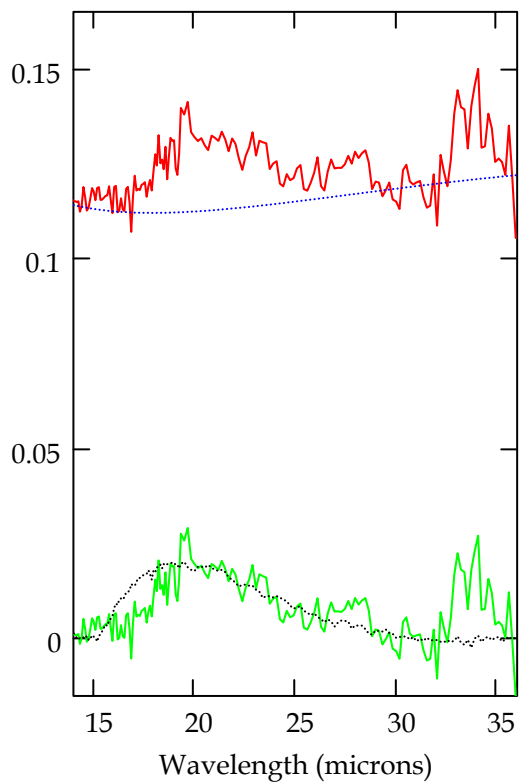

Wavelength (microns)

## Names$_J$ = "CZ Tau"

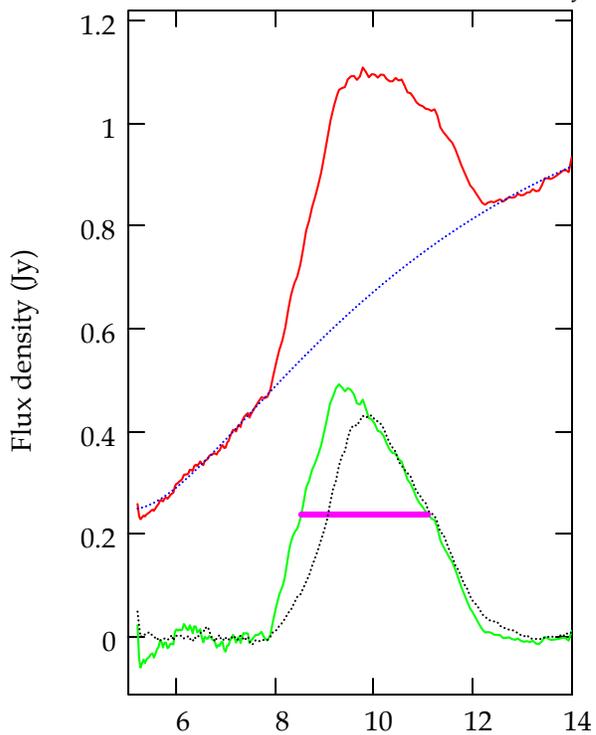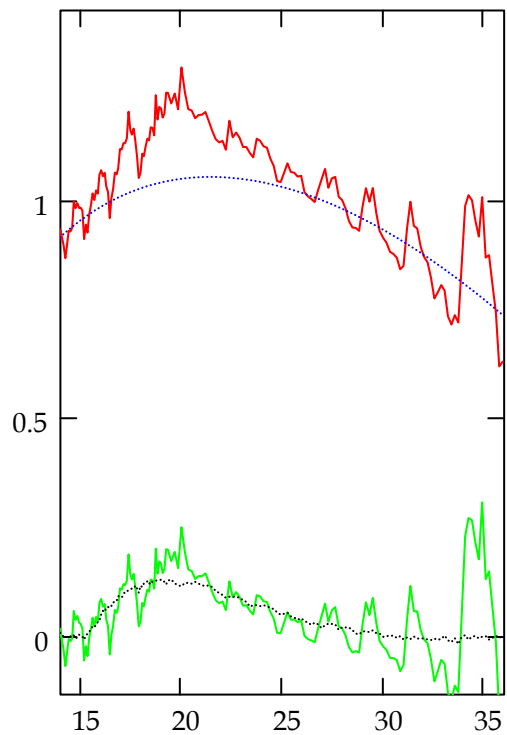

## Names$_J$ = "DD Tau"

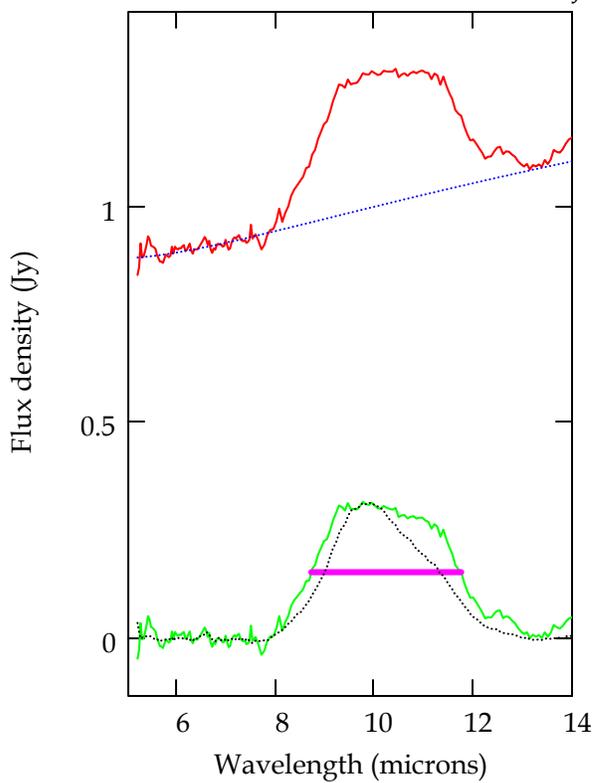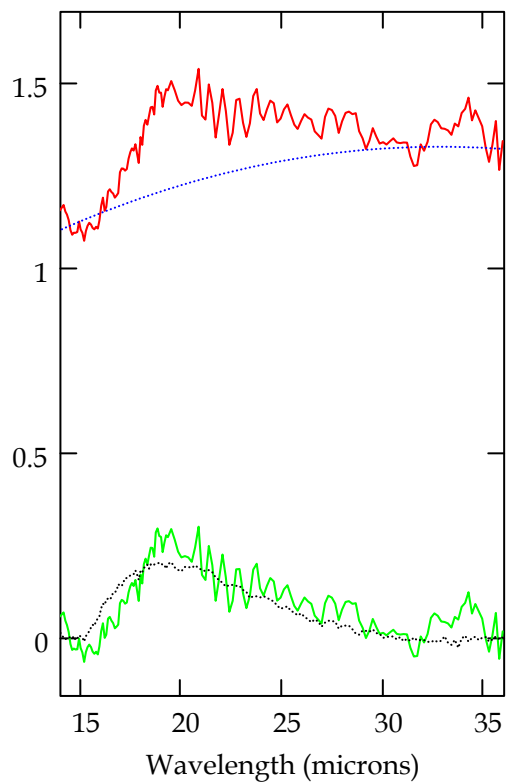

Wavelength (microns)

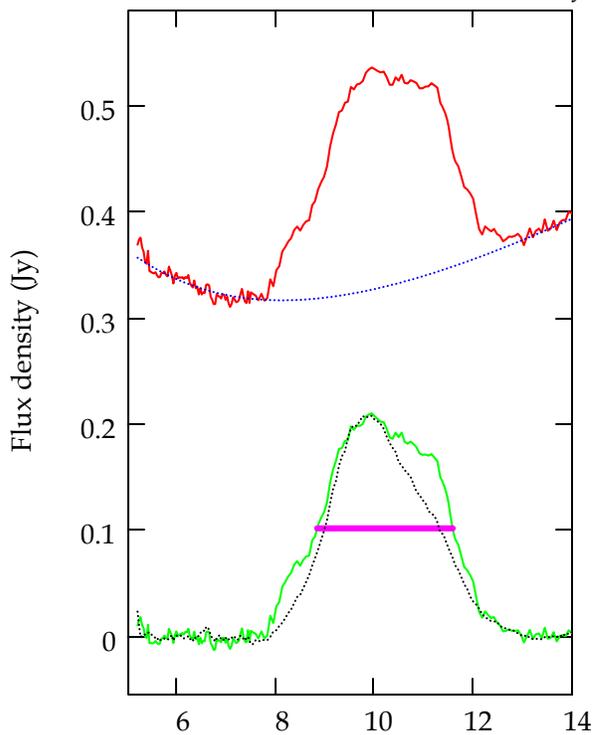
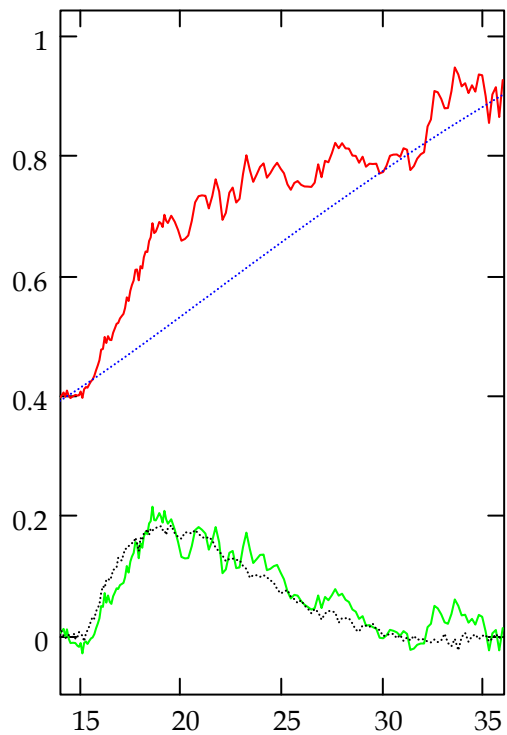

Names$_J$ = "DE Tau"

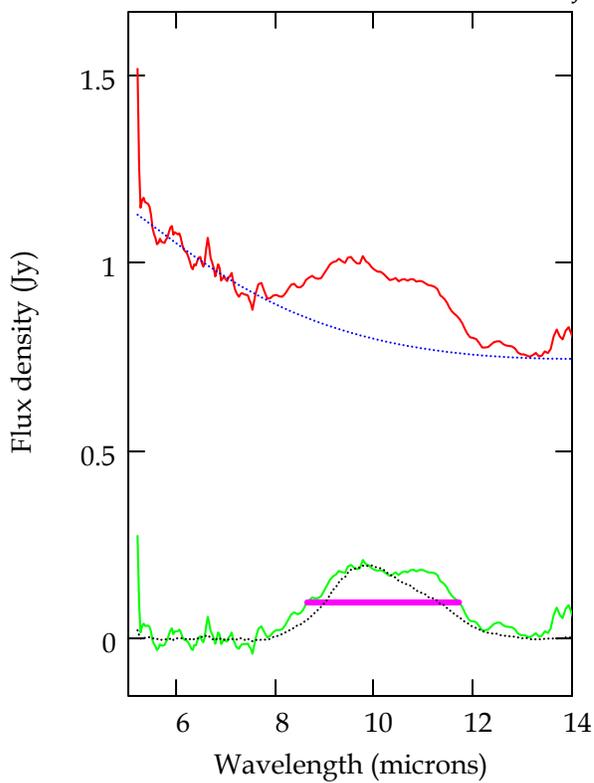
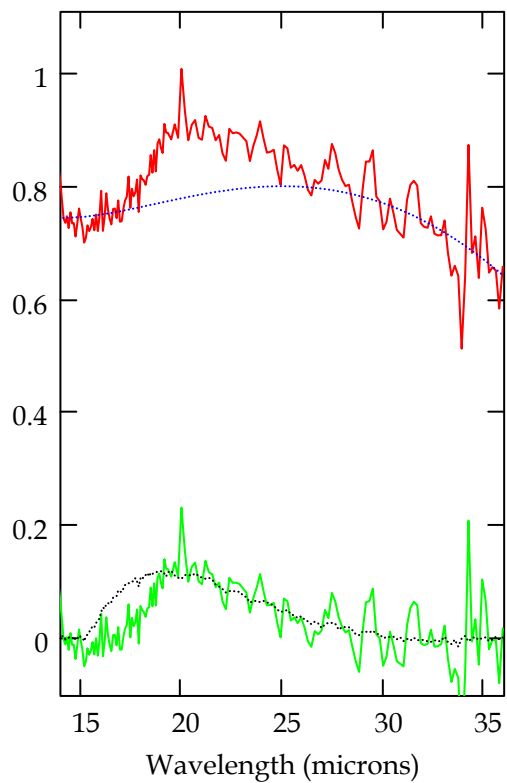

Names$_J$ = "DF Tau"

Flux density (Jy)

Wavelength (microns)

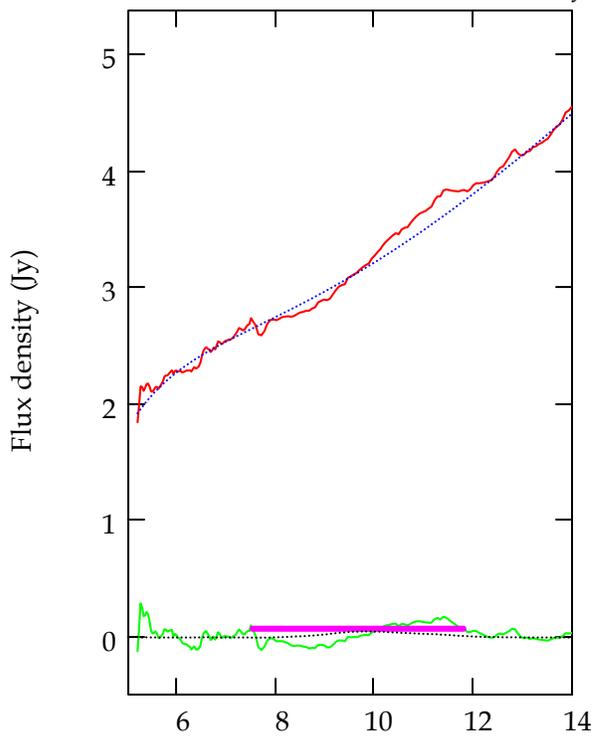
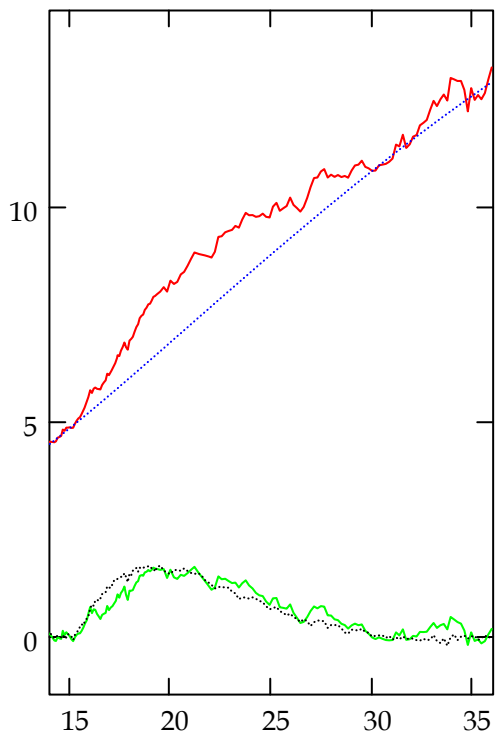
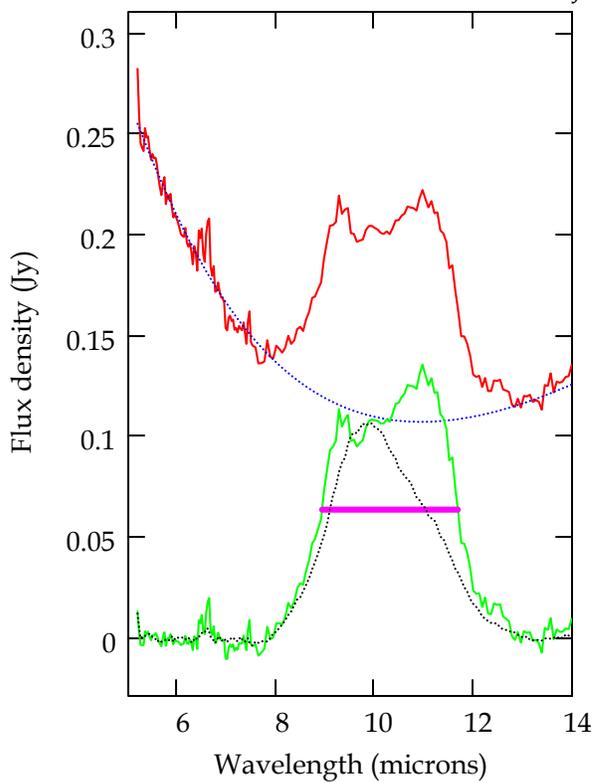
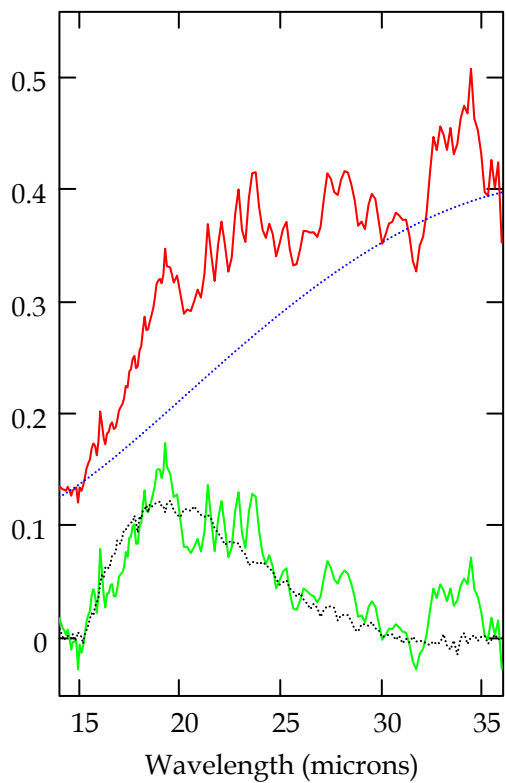

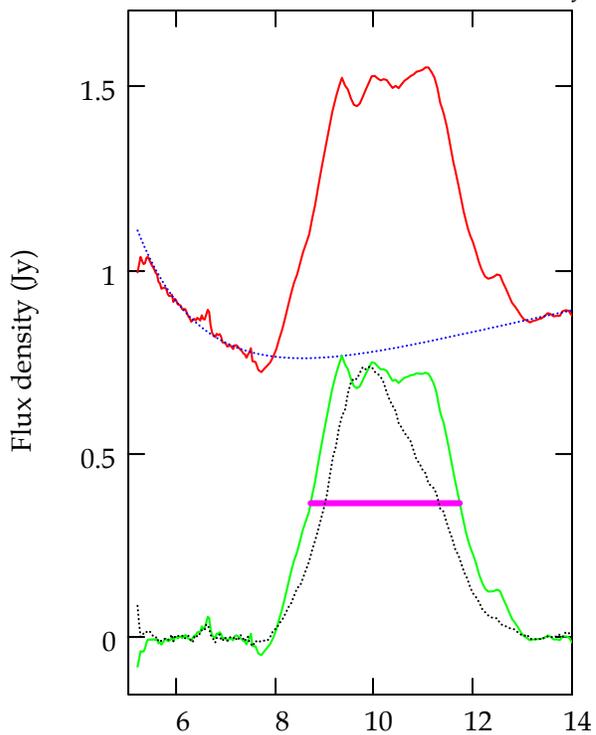
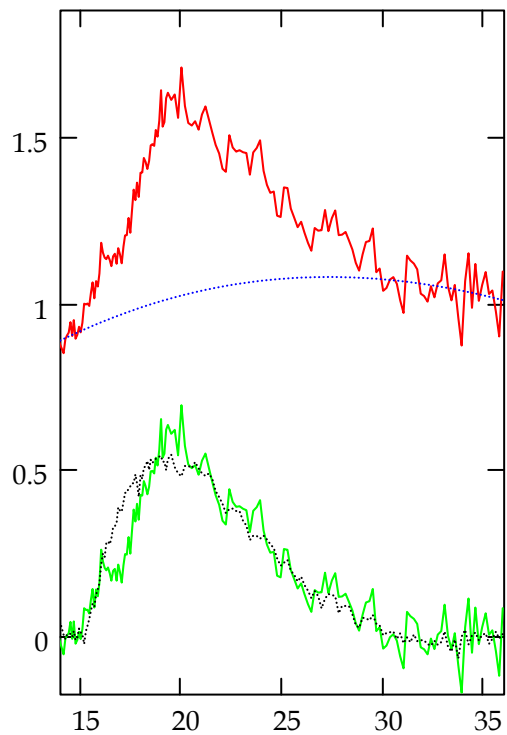
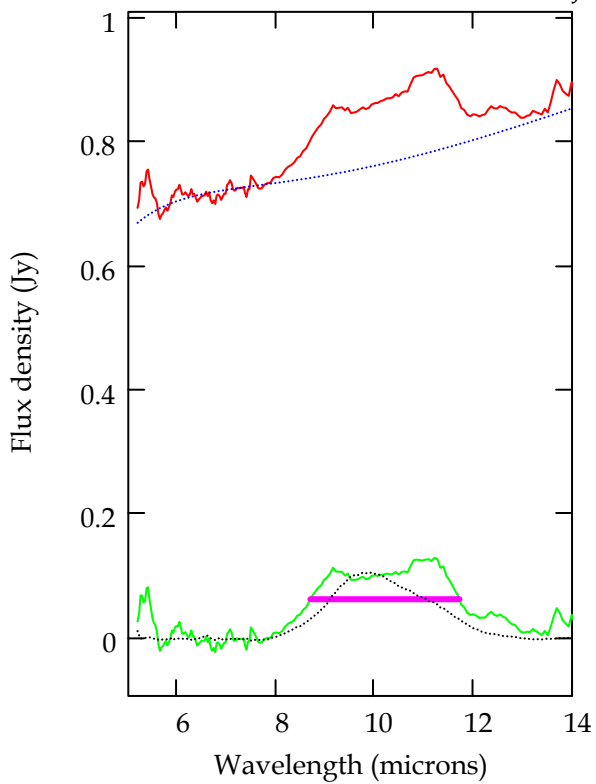
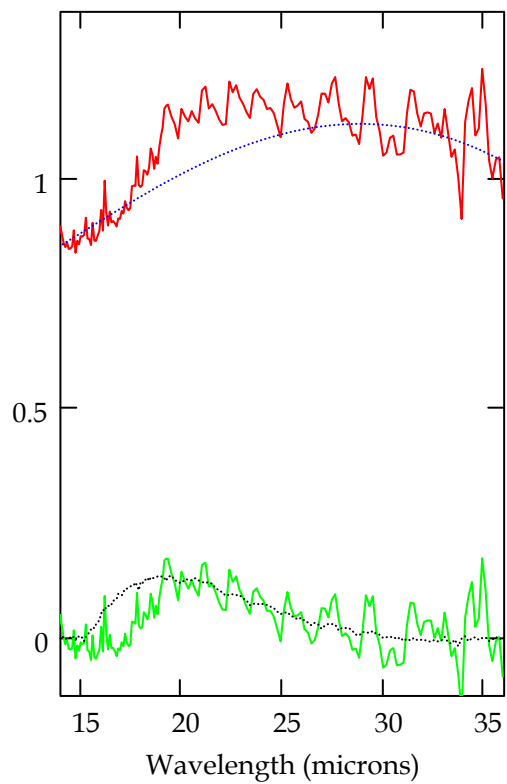

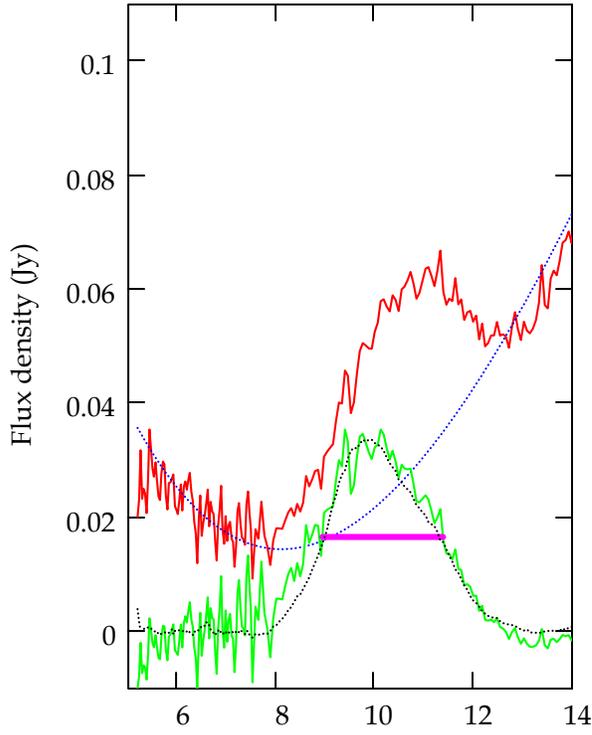
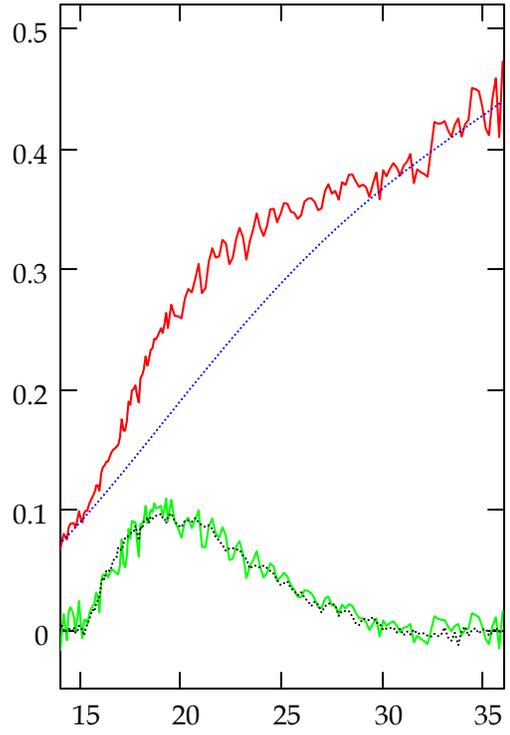
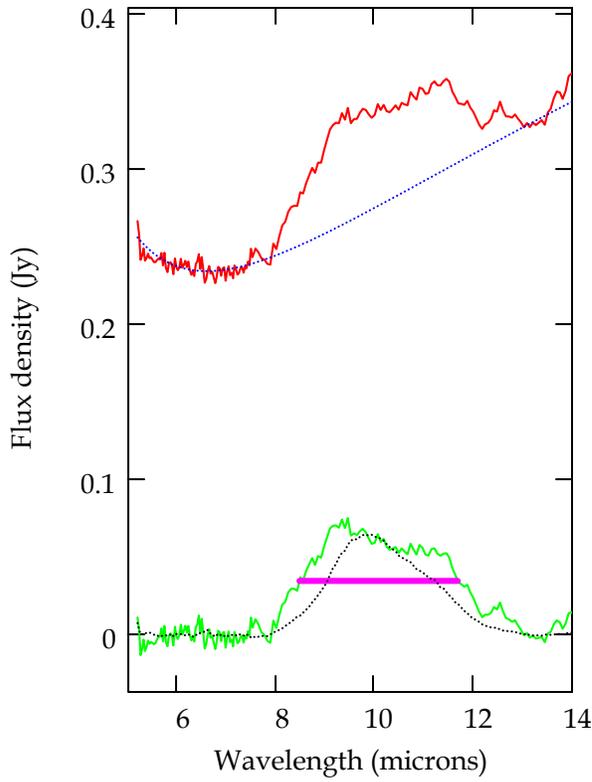
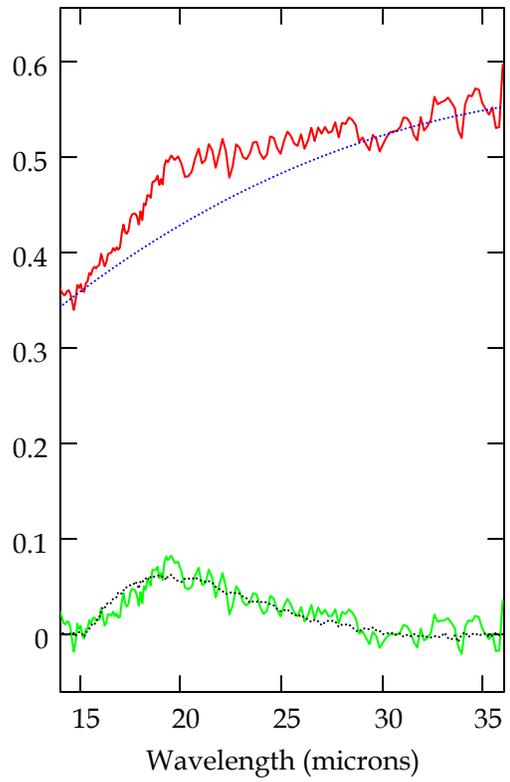

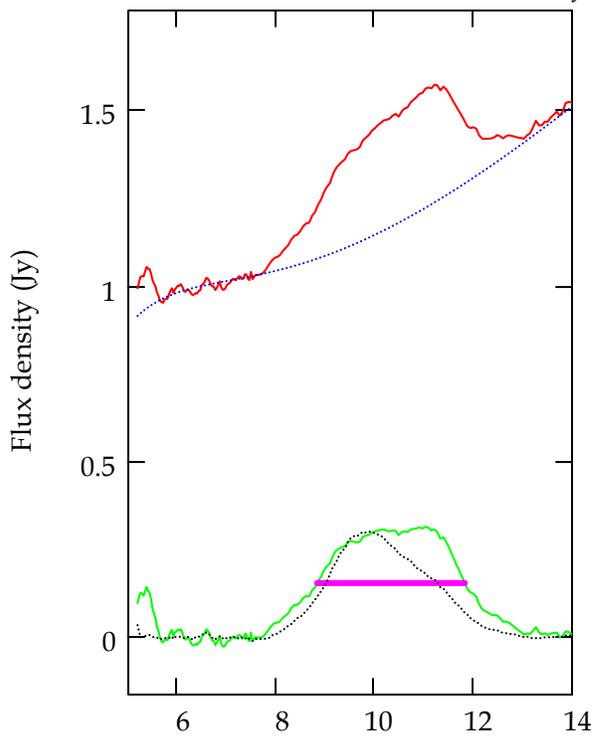
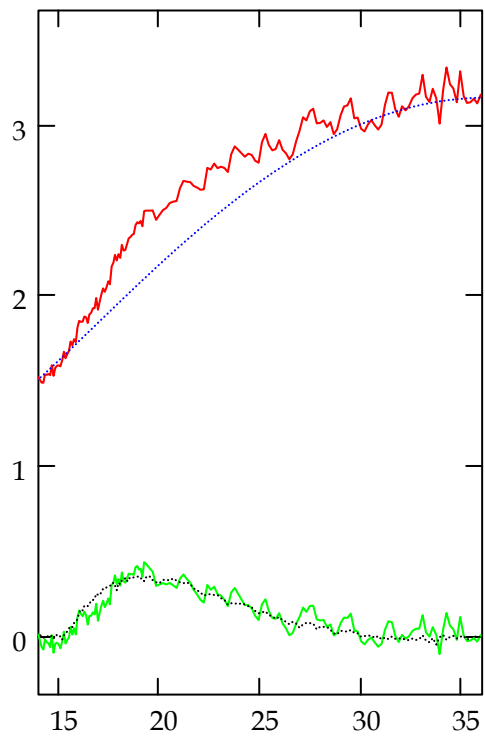
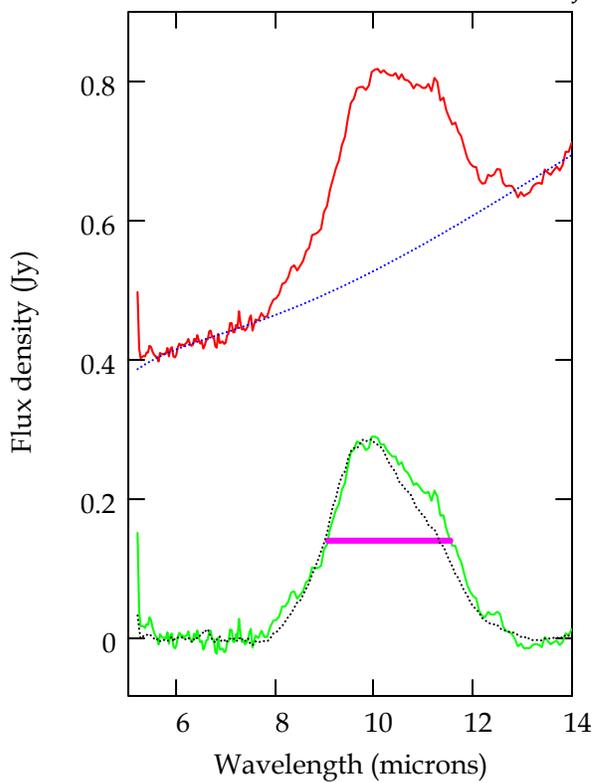
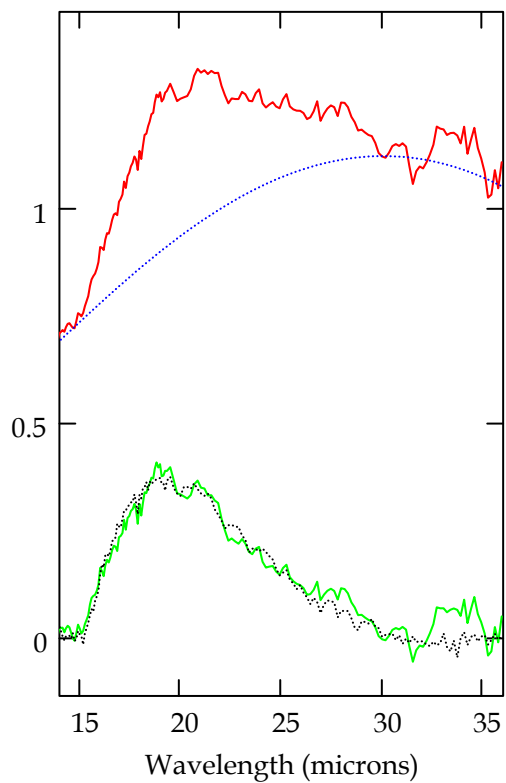

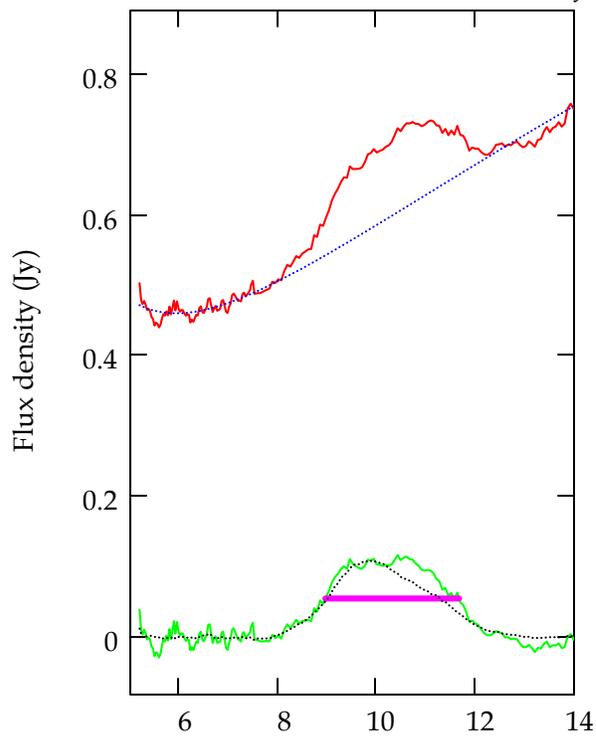
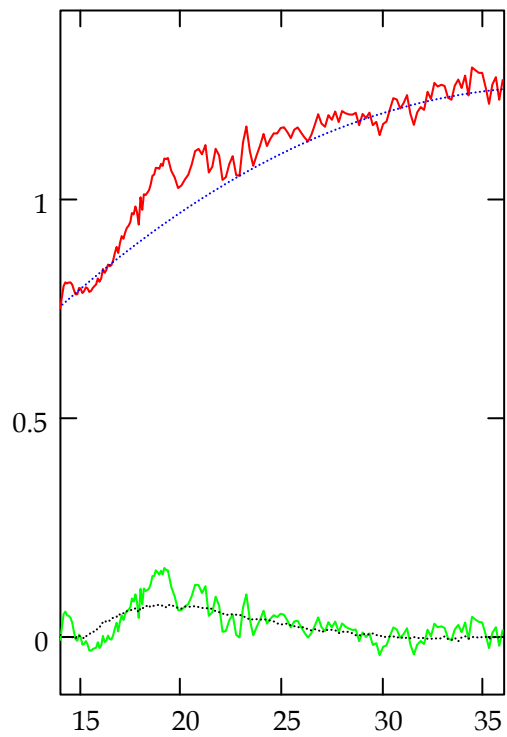
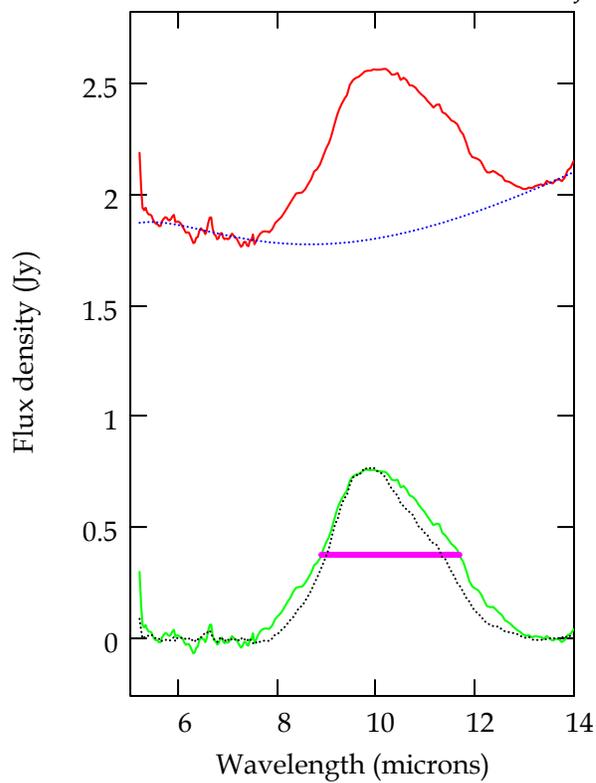
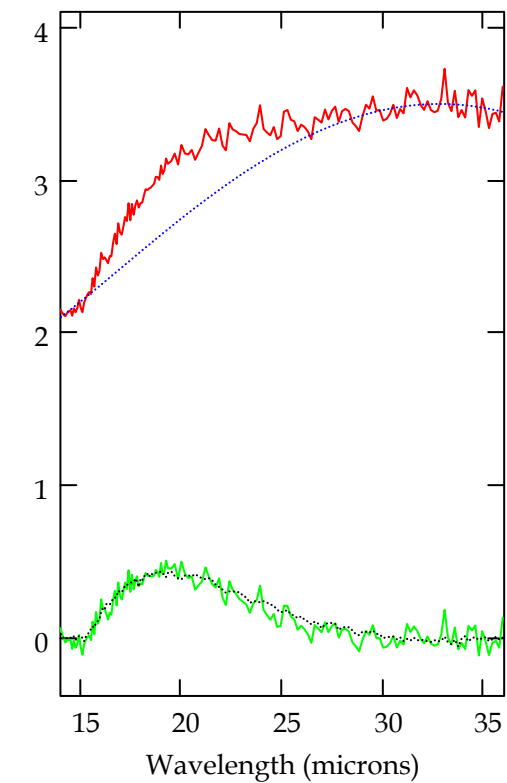

Names$_J$ = "DQ Tau"

Names$_J$ = "DR Tau"

Flux density (Jy)

Wavelength (microns)

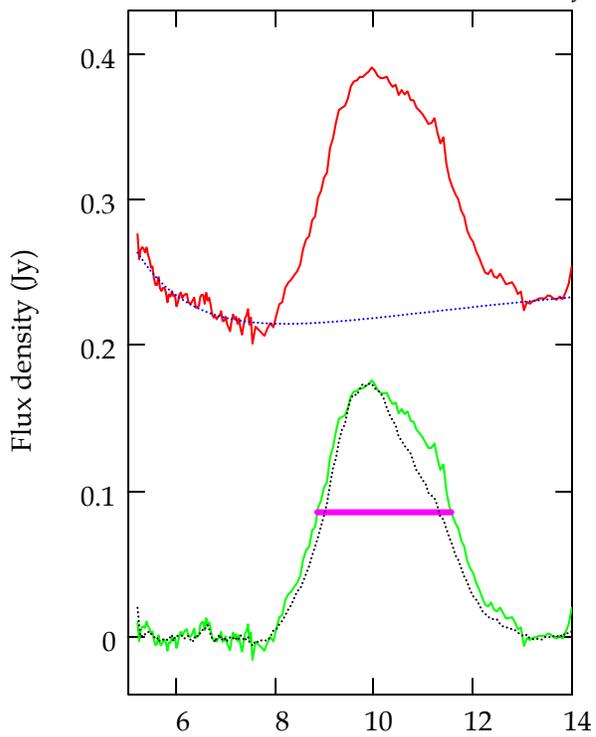
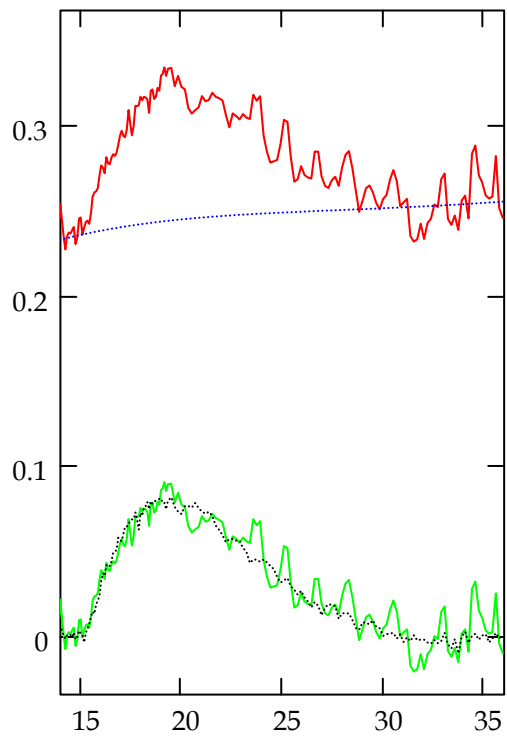

Names_J = "DS Tau"

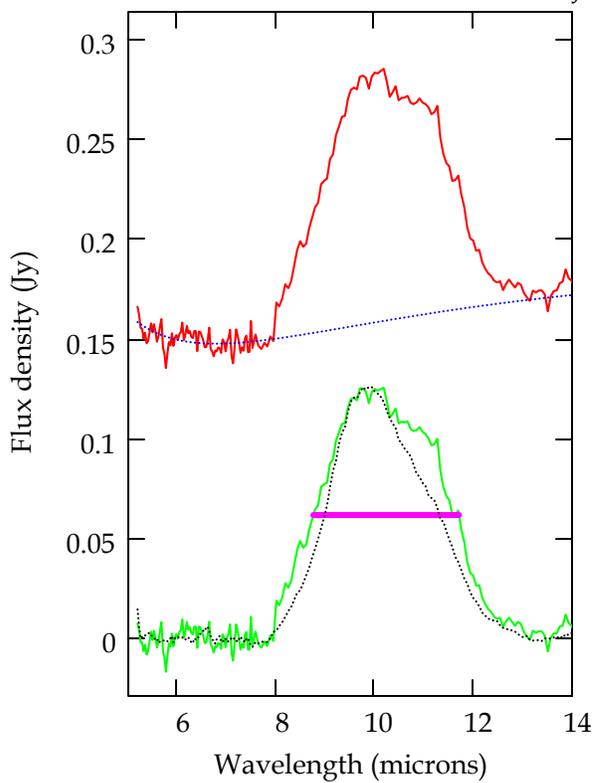
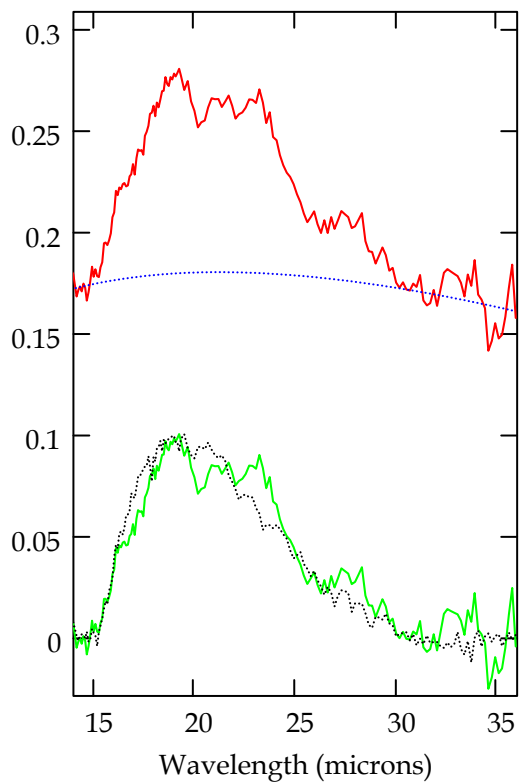

Names_J = "F04147+2822"

Names$_J$ = "F04192+2647"

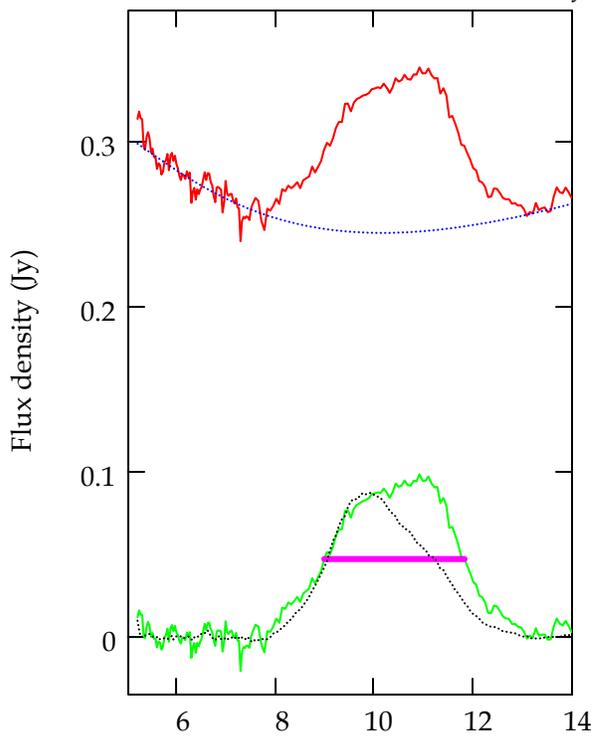
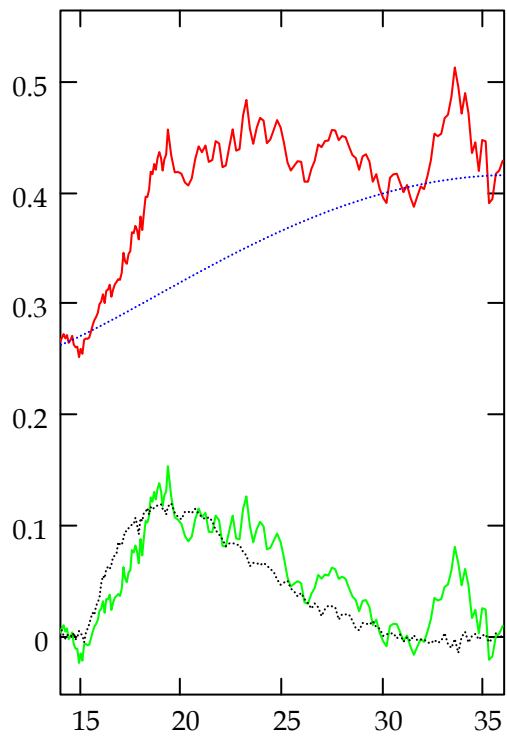

Names$_J$ = "F04262+2654"

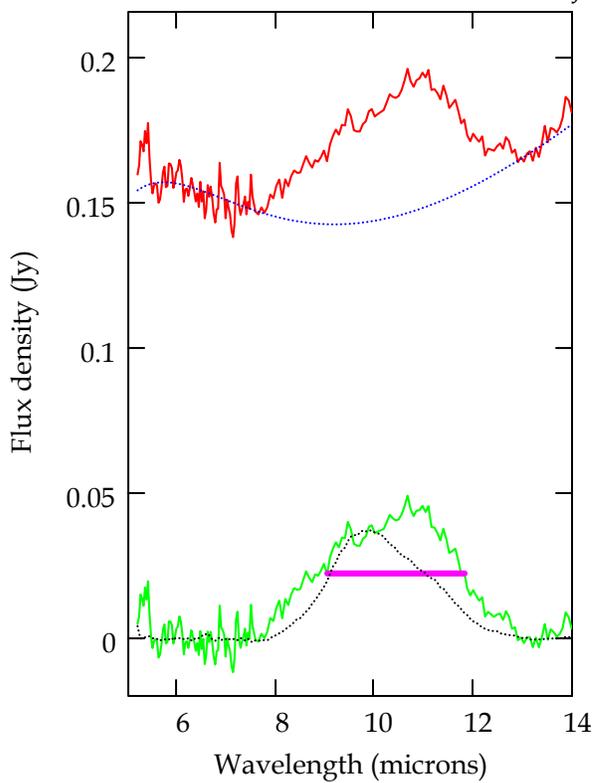
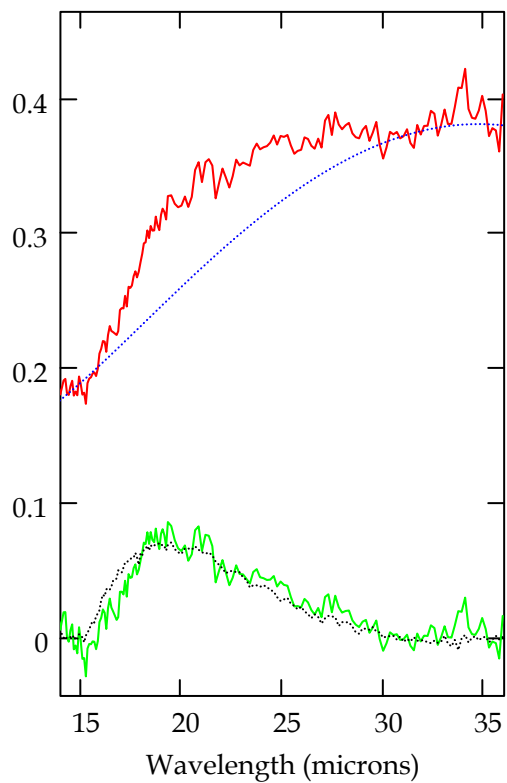

Names$_J$ = "F04297+2246A"

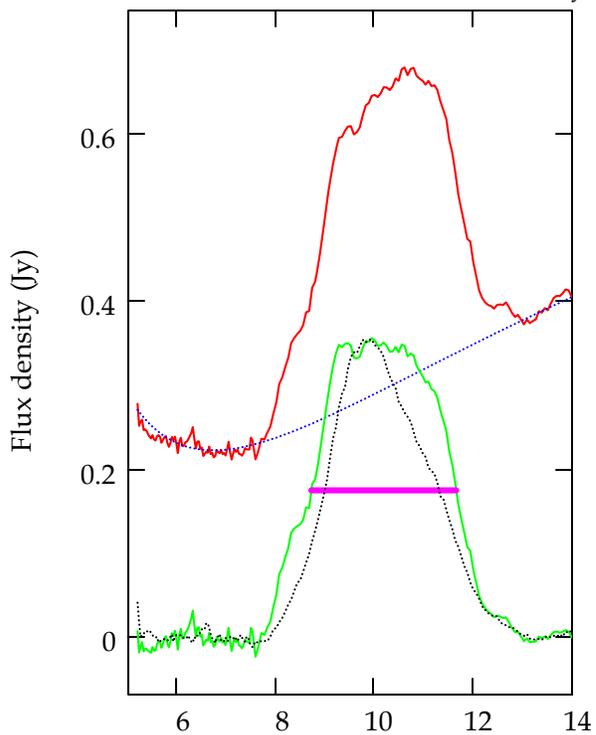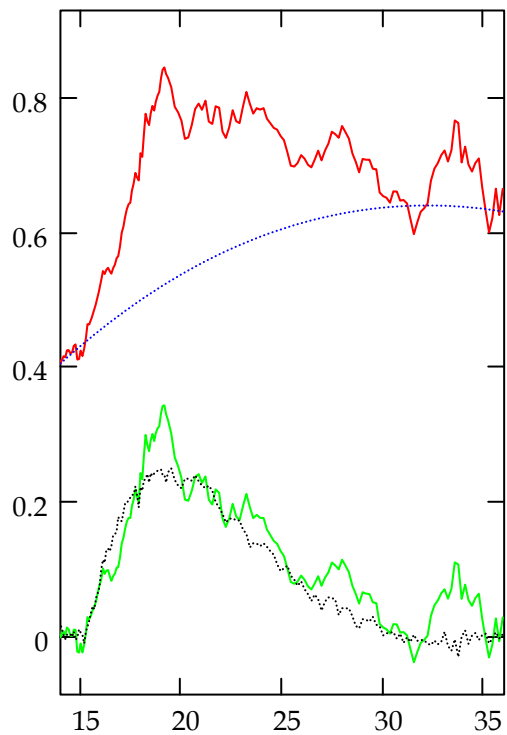

Names$_J$ = "F04297+2246B"

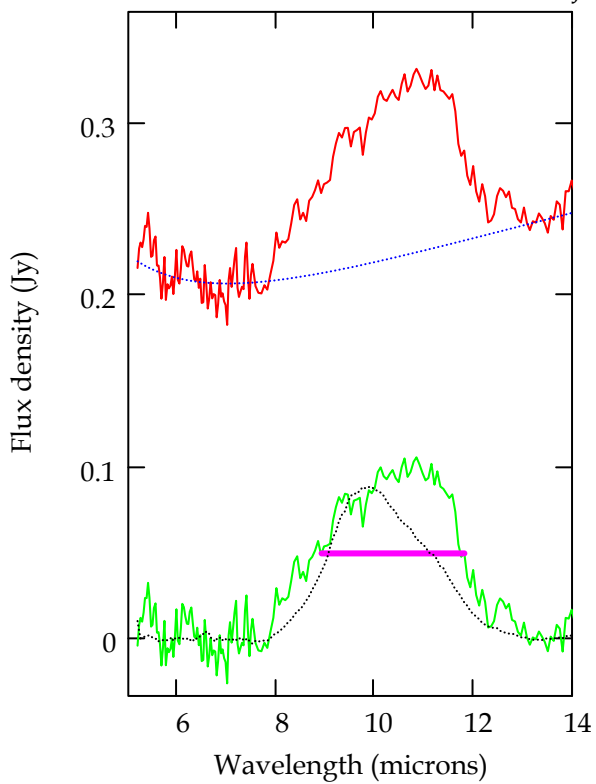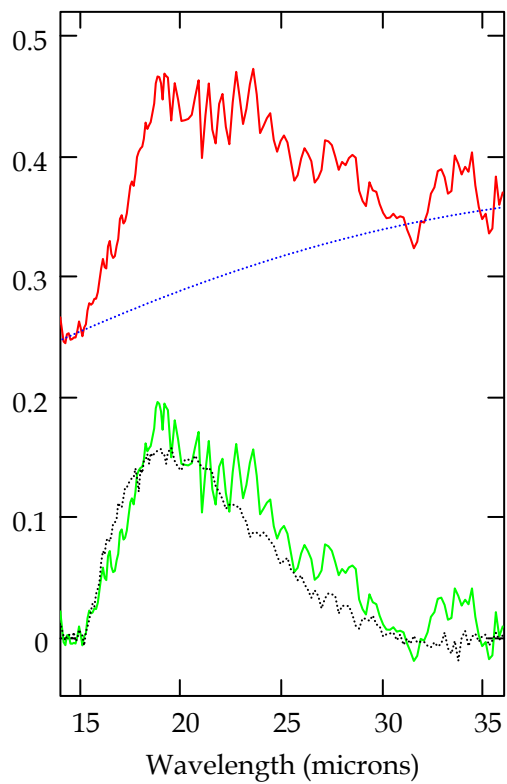

Wavelength (microns)

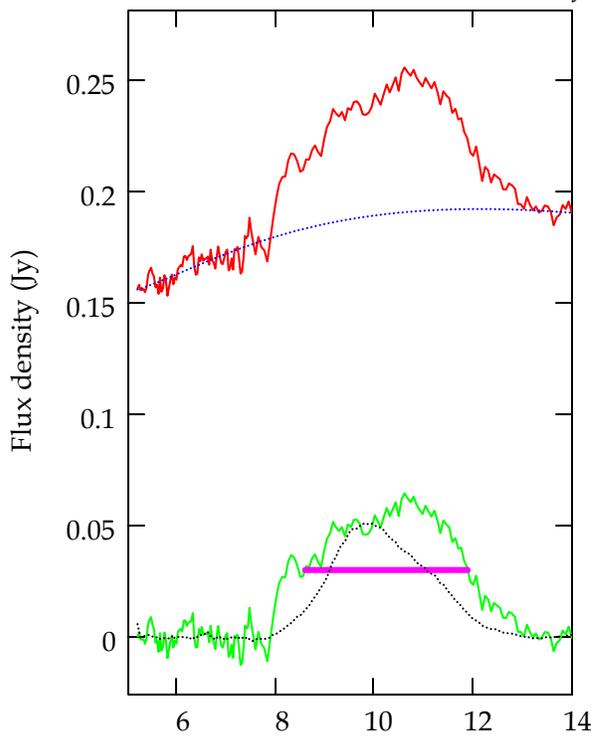
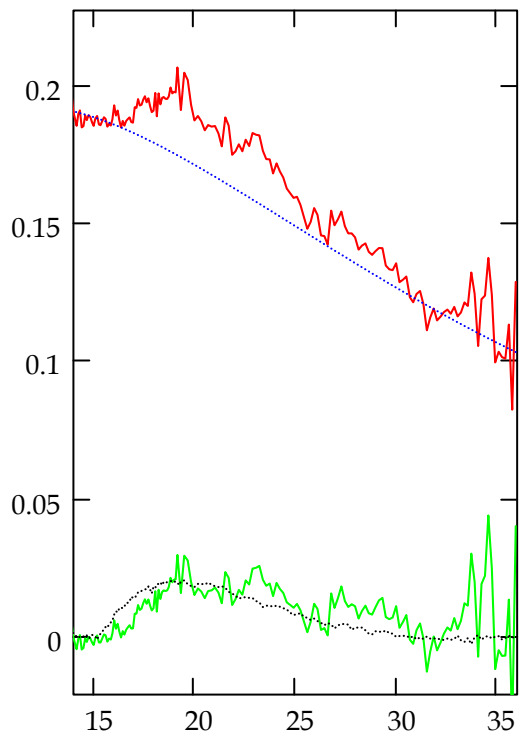

Names$_J$ = "F04570+2520"

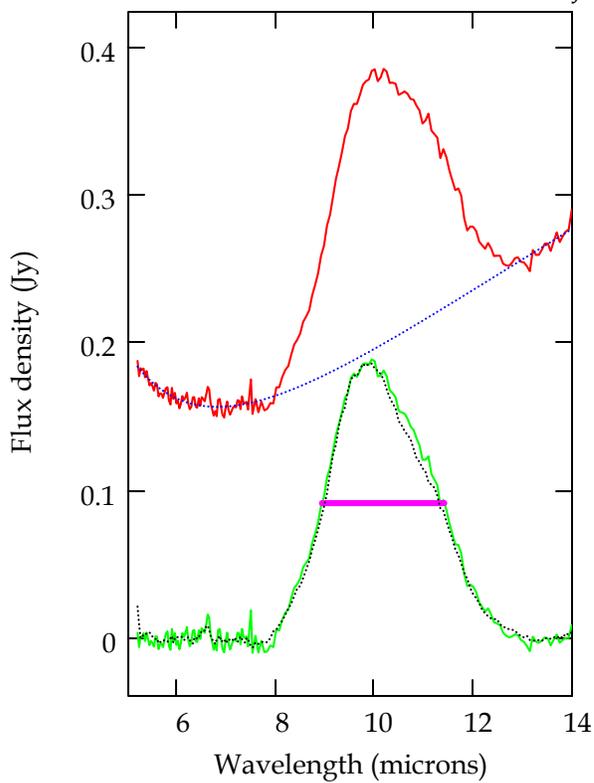
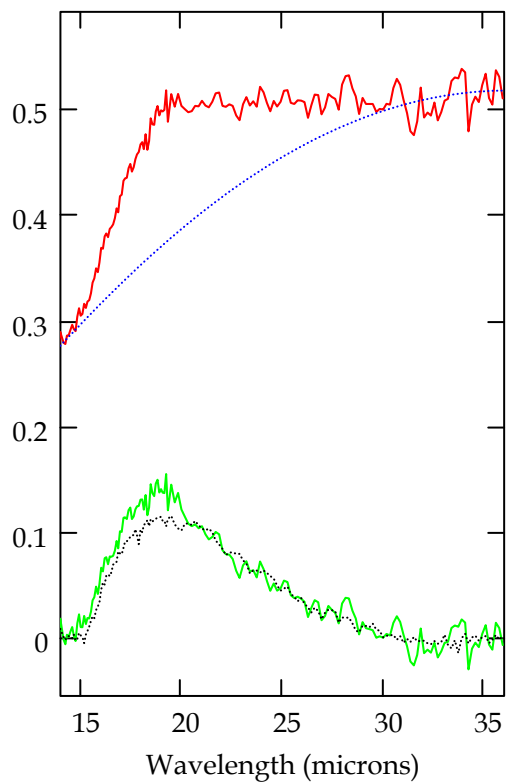

Names$_J$ = "FM Tau"

Names$_J$ = "FN Tau"

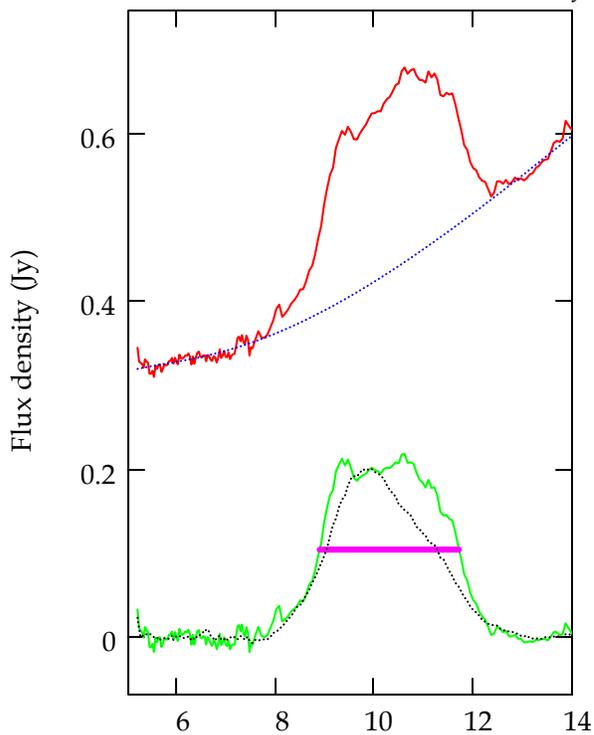
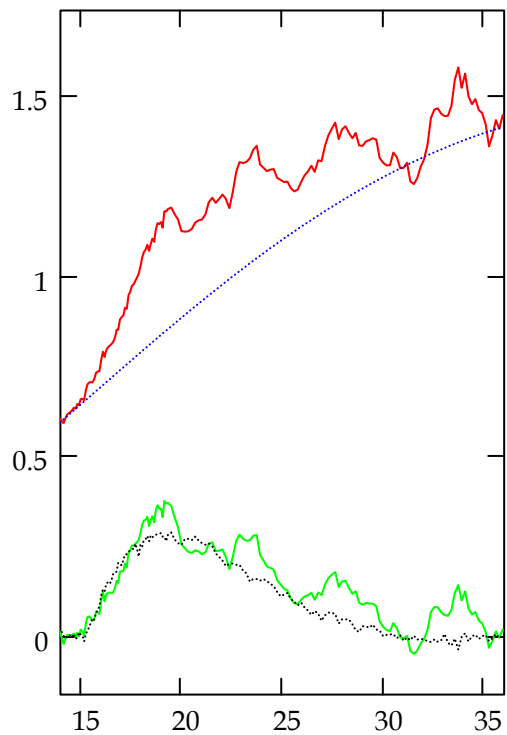

Names$_J$ = "FO Tau"

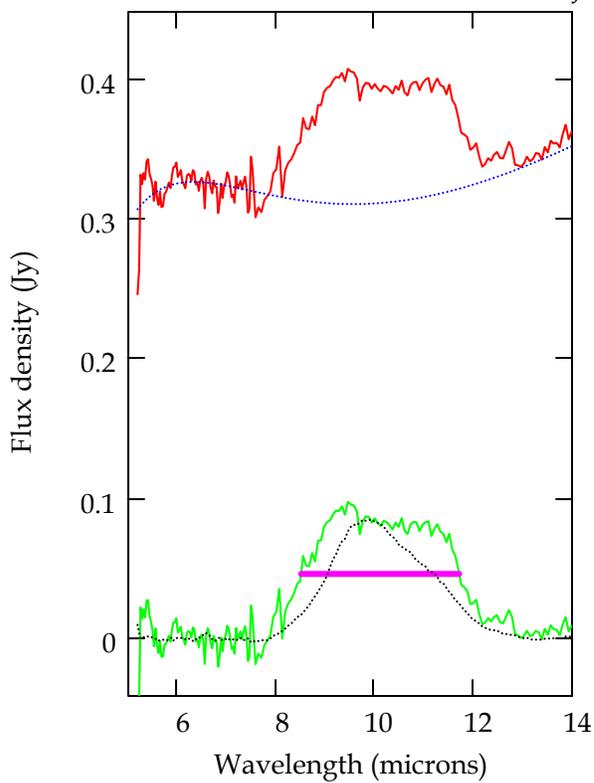
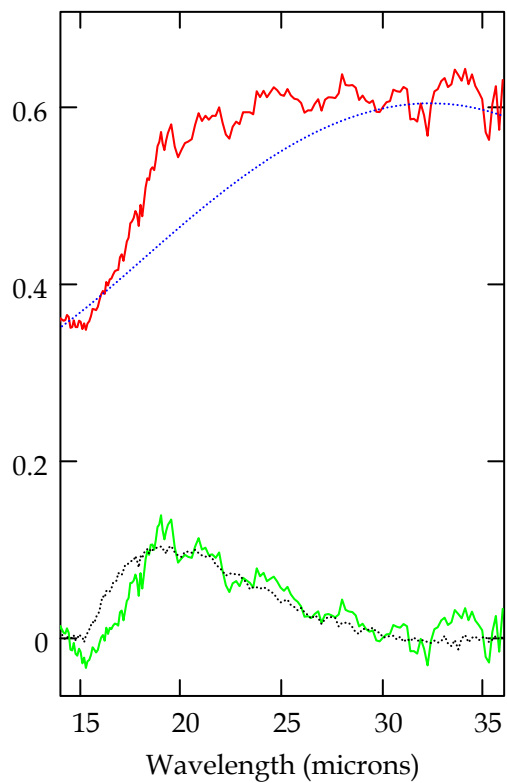

Flux density (Jy)

Wavelength (microns)

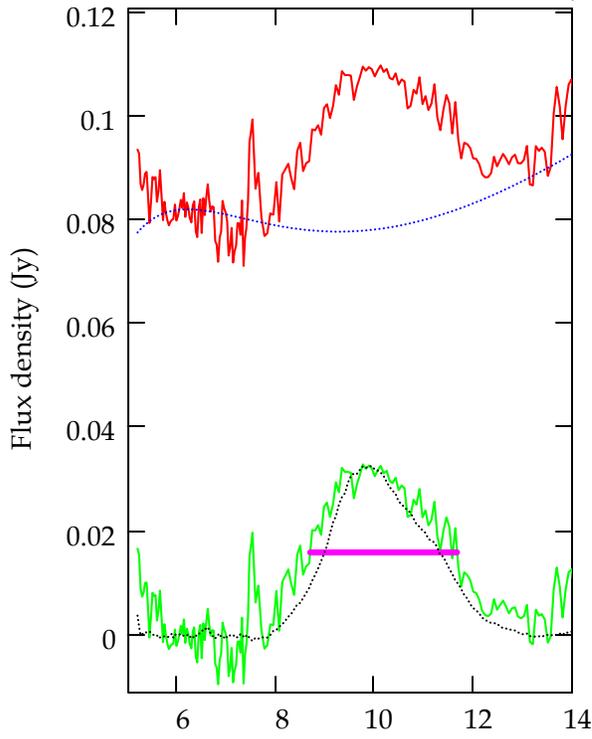
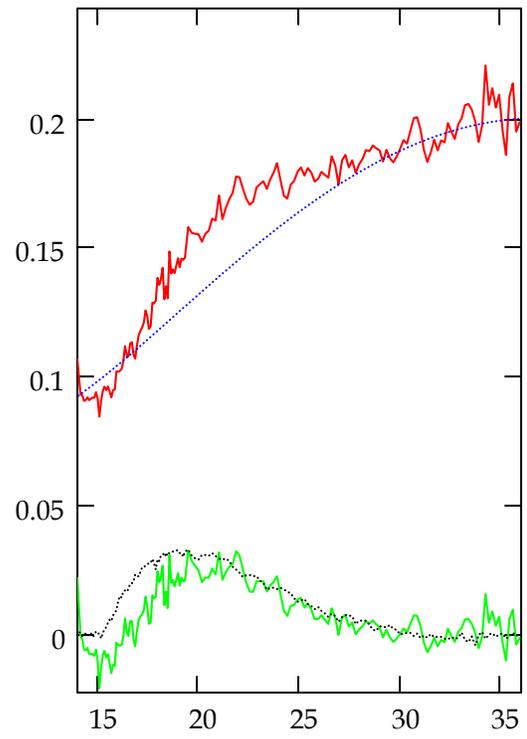

Names$_J$ = "FP Tau"

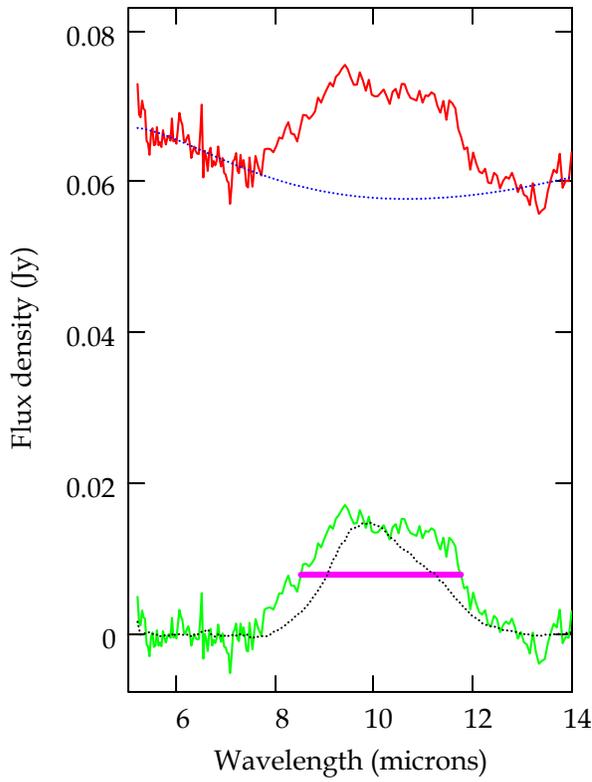
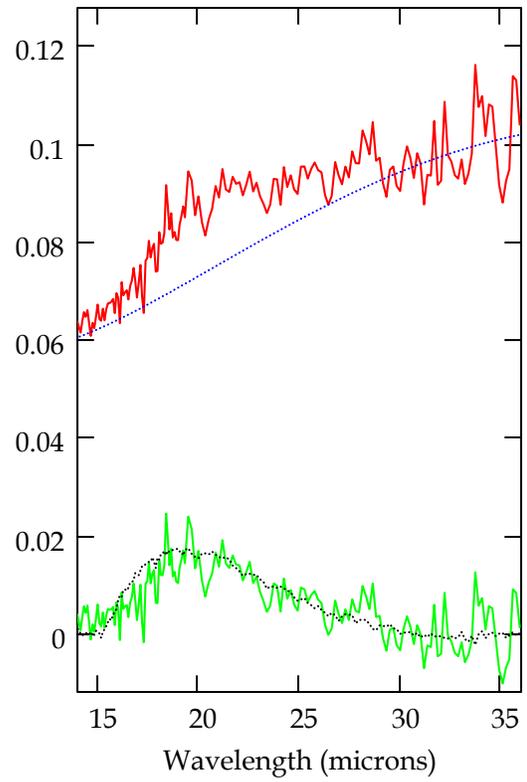

Names$_J$ = "FQ Tau"

Wavelength (microns)

Flux density (Jy)

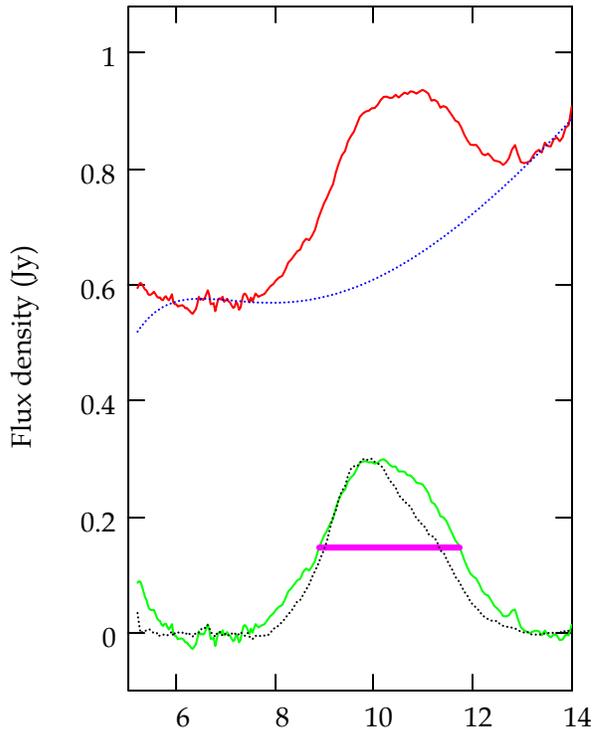
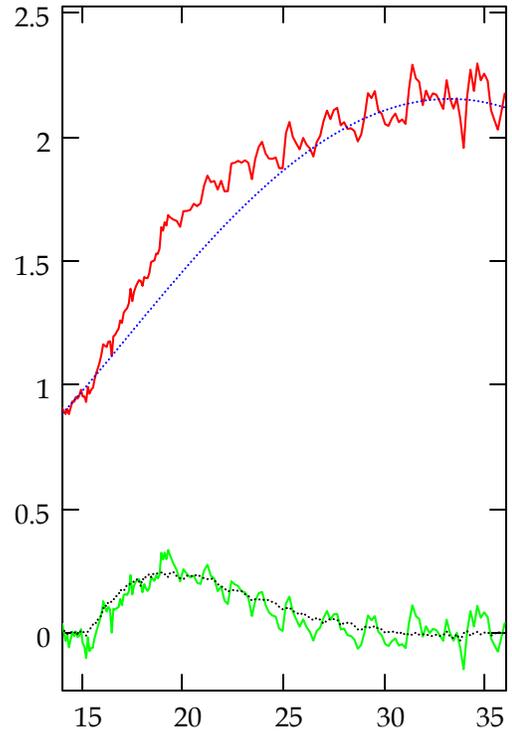
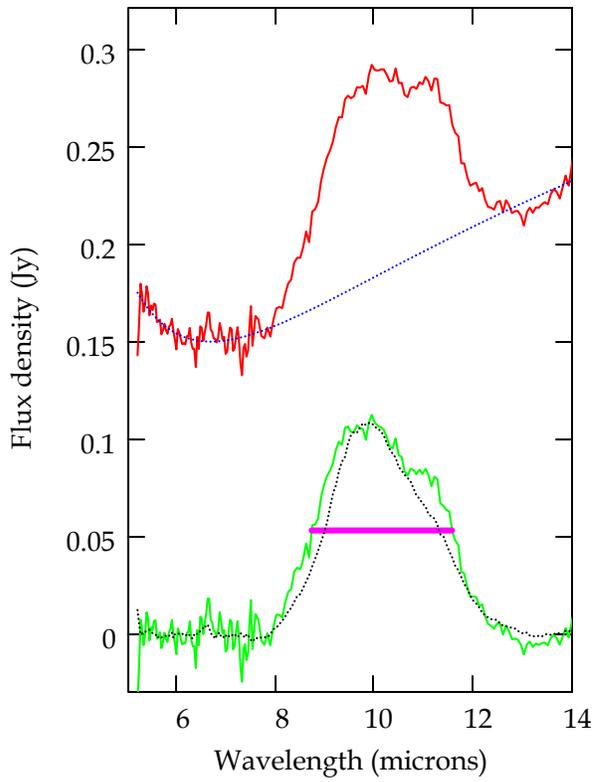
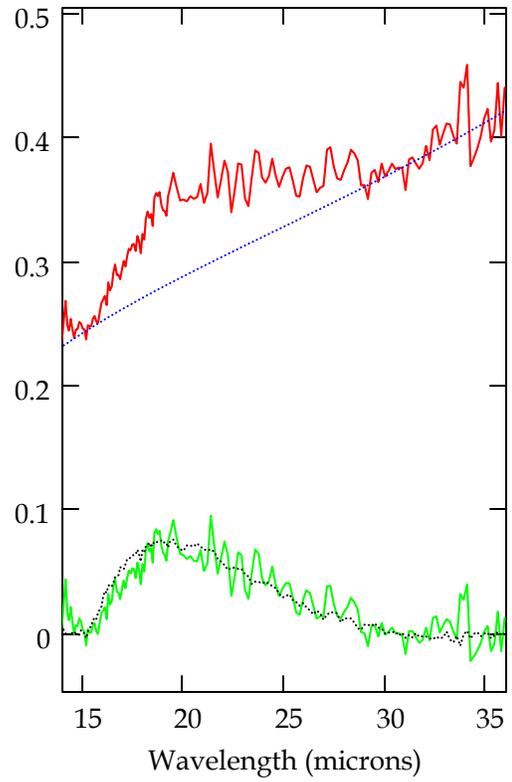

Names$_J$ = "FV Tau"

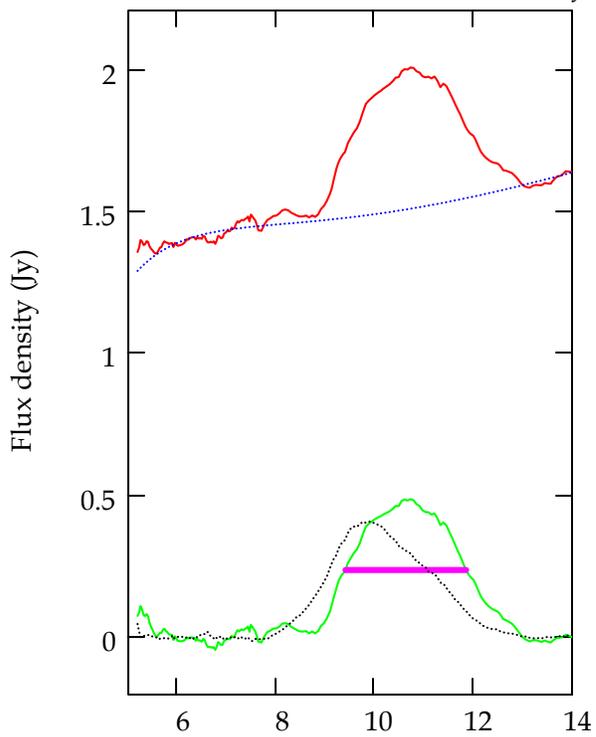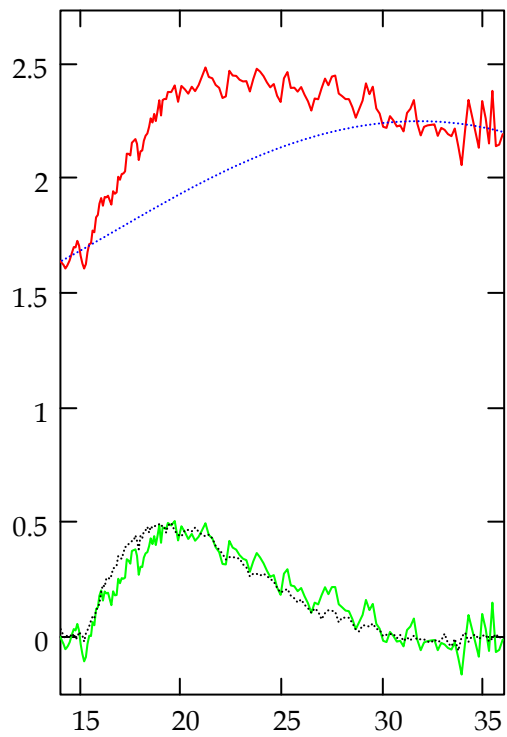

Names$_J$ = "FX Tau"

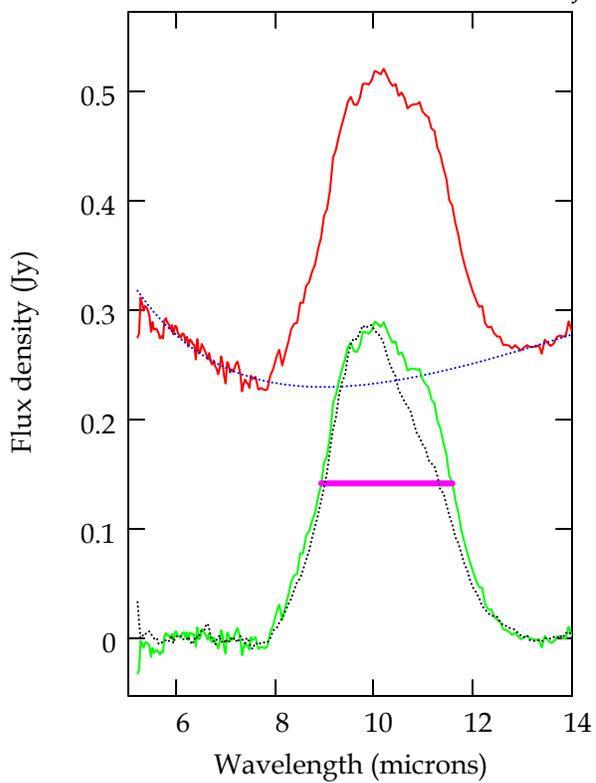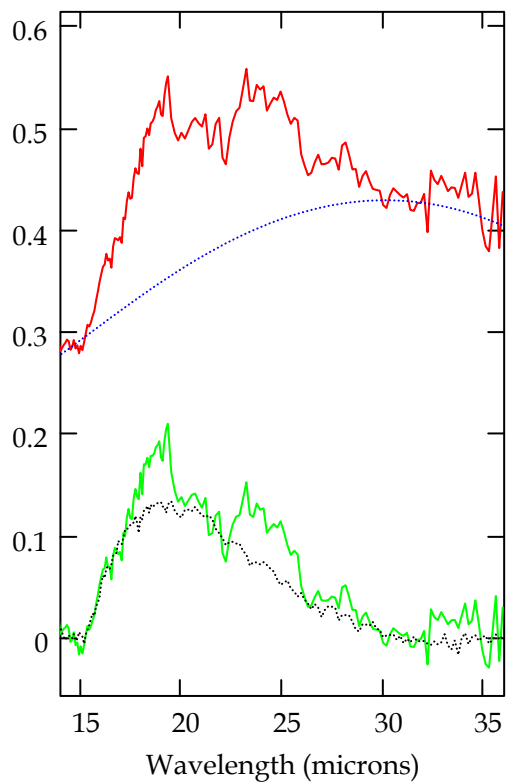

Wavelength (microns)

Names$_J$ = "FZ Tau"

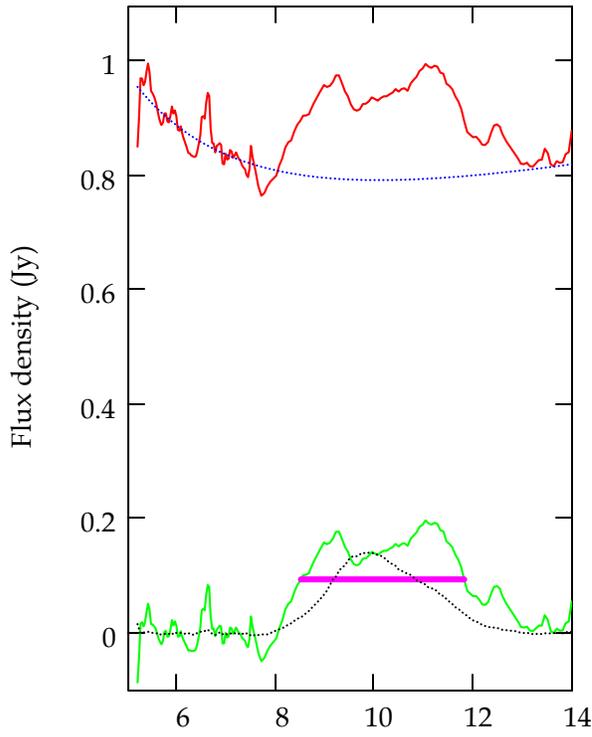
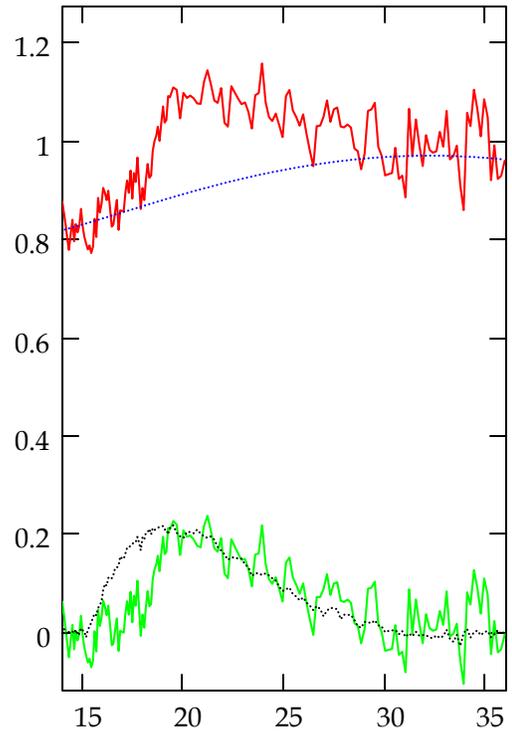

Names$_J$ = "GG Tau A"

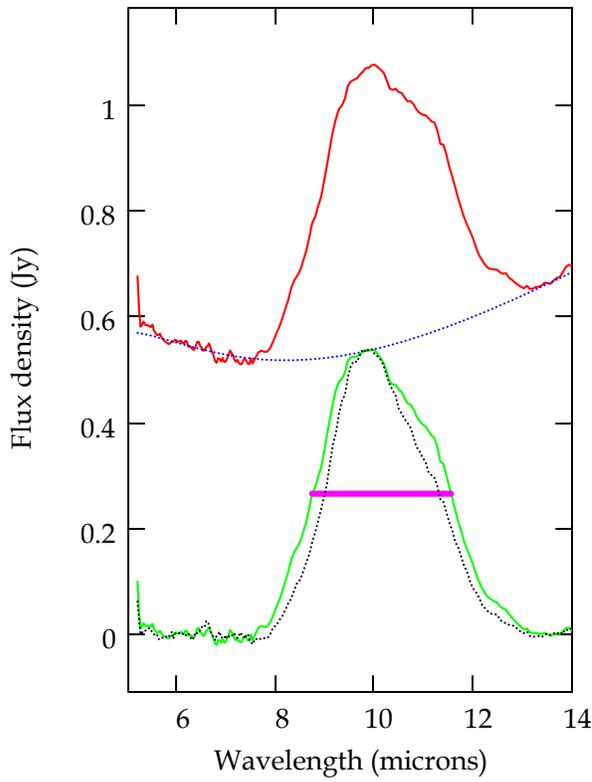
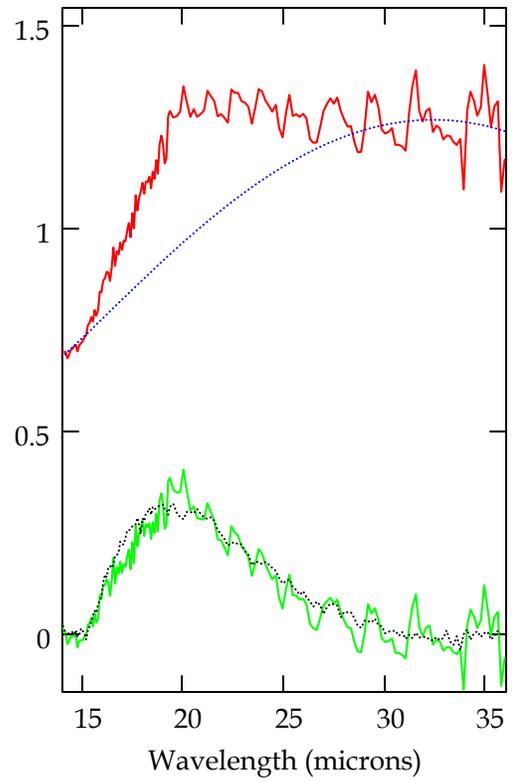

Wavelength (microns)

Names$_J$ = "GH Tau"

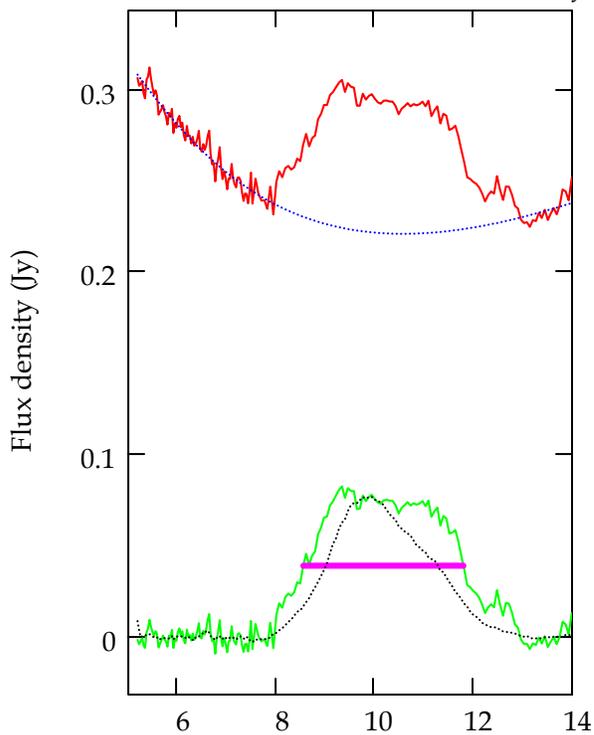
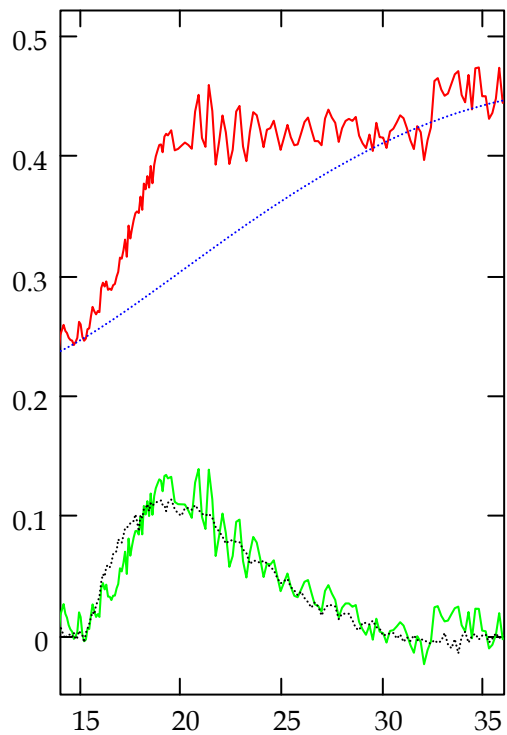

Names$_J$ = "GI Tau"

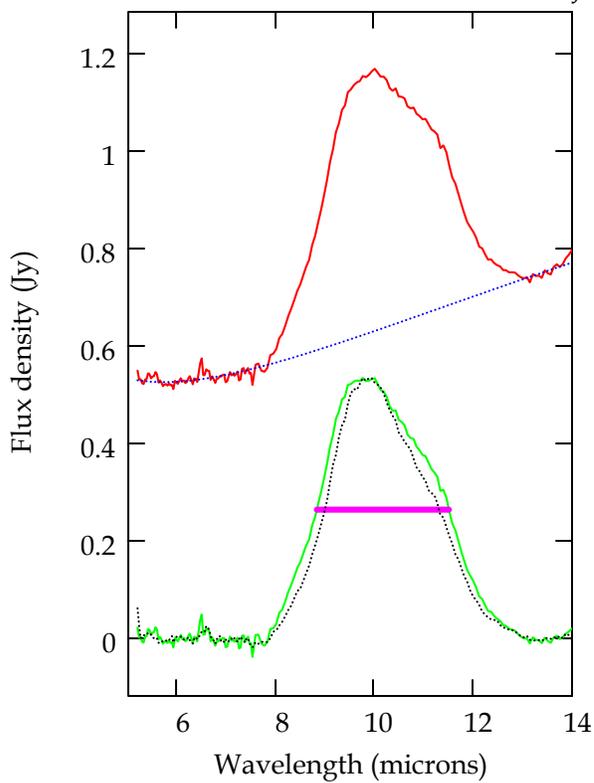
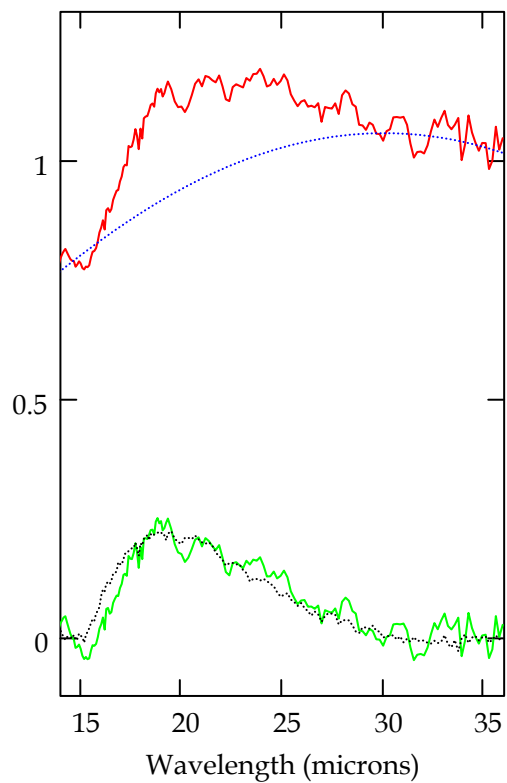

Wavelength (microns)

Names$_J$ = "GK Tau"

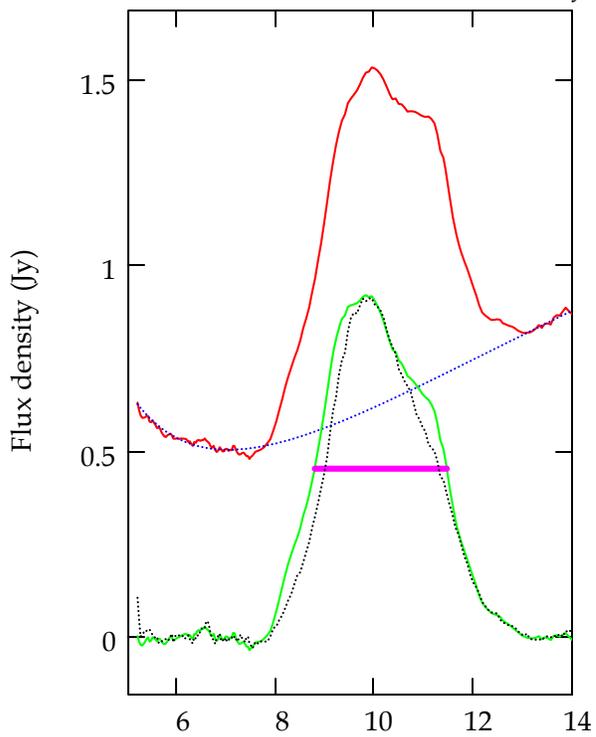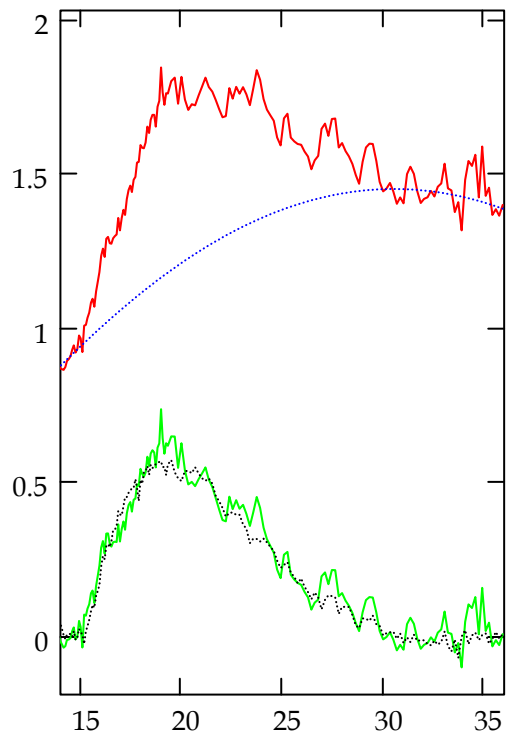

Names$_J$ = "GM Aur"

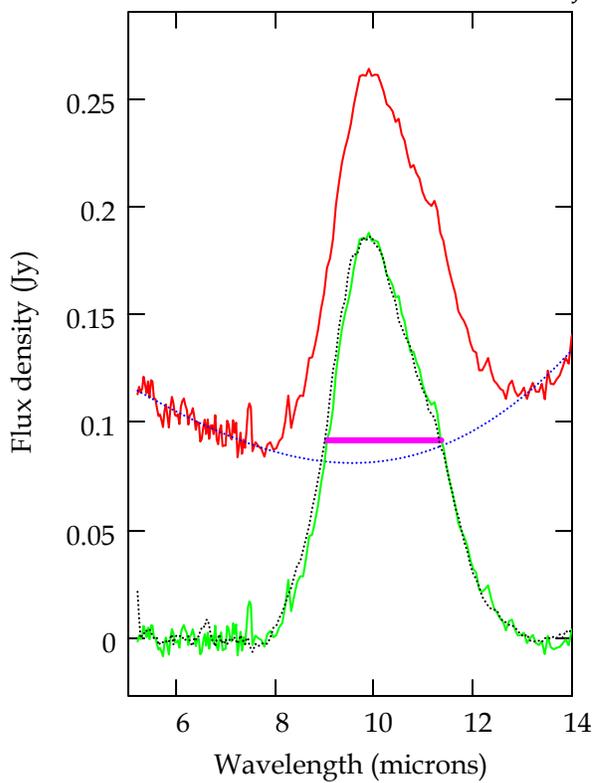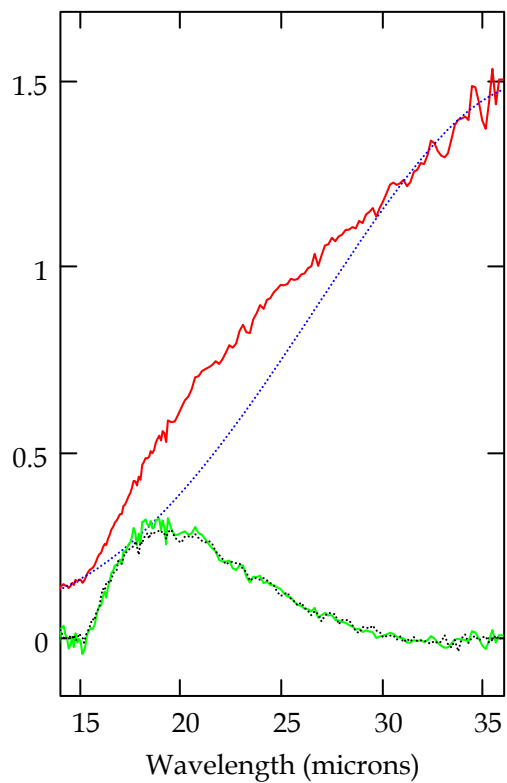

Names$_J$ = "GN Tau"

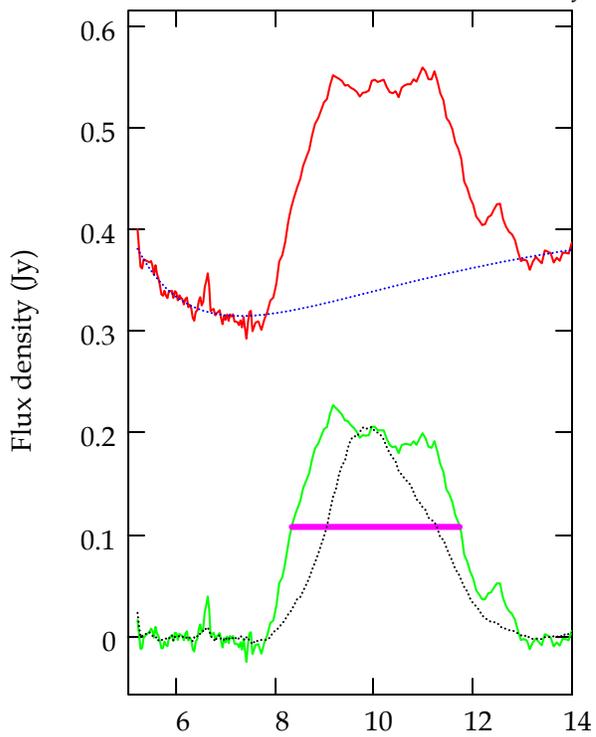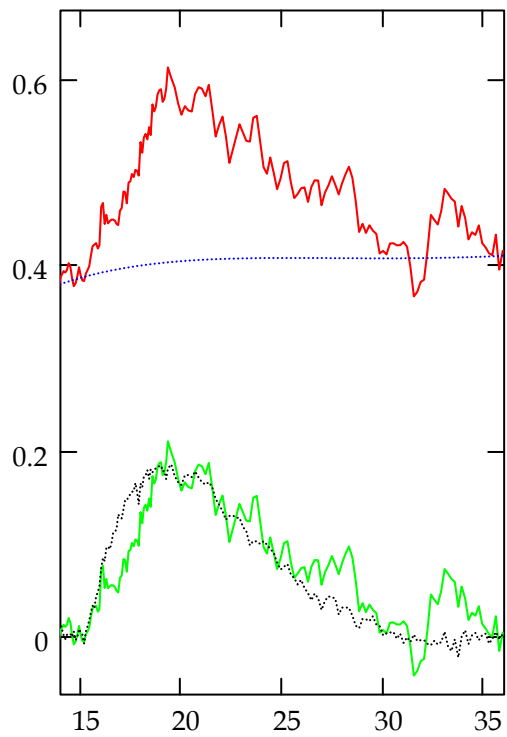

Names$_J$ = "GO Tau"

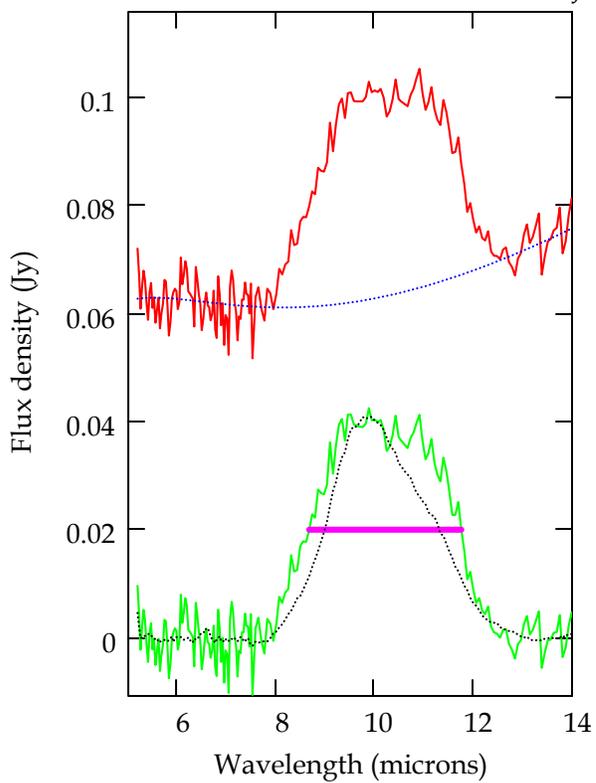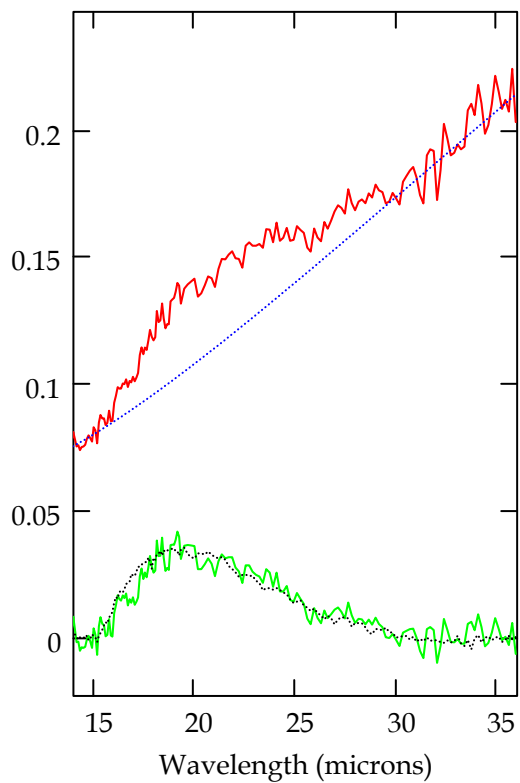

Flux density (Jy)

Wavelength (microns)

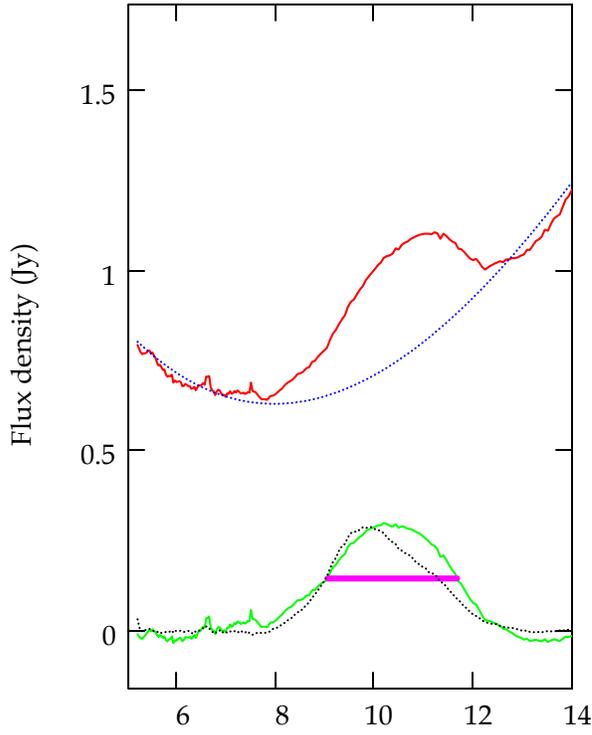
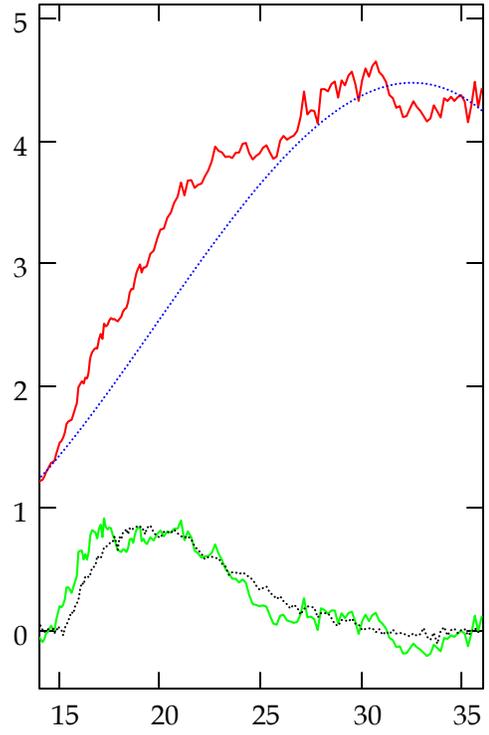
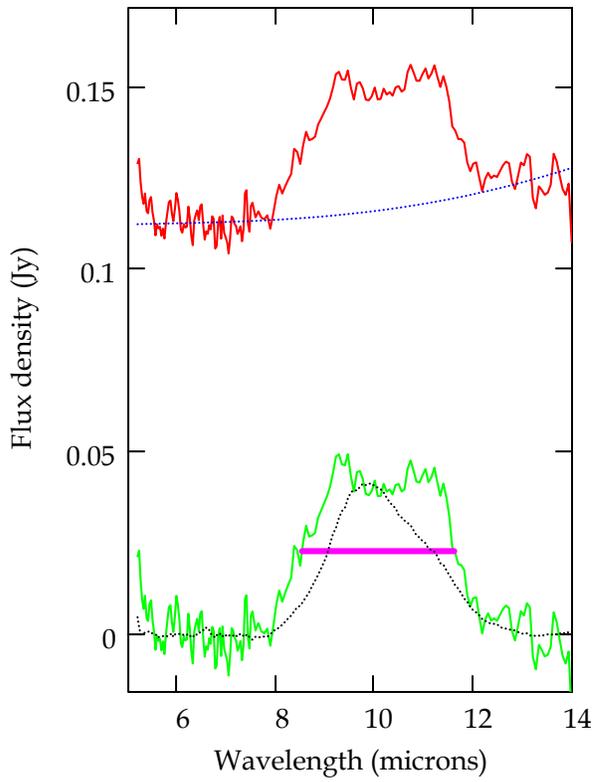
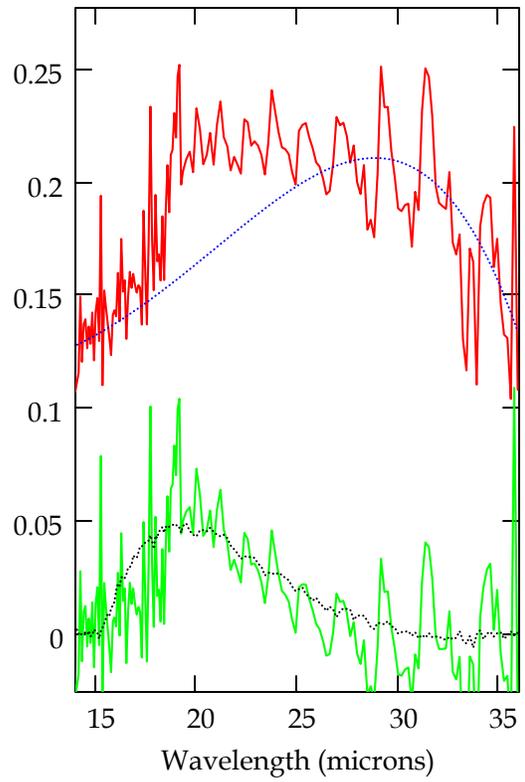

Names$_J$ = "Haro 6-37"

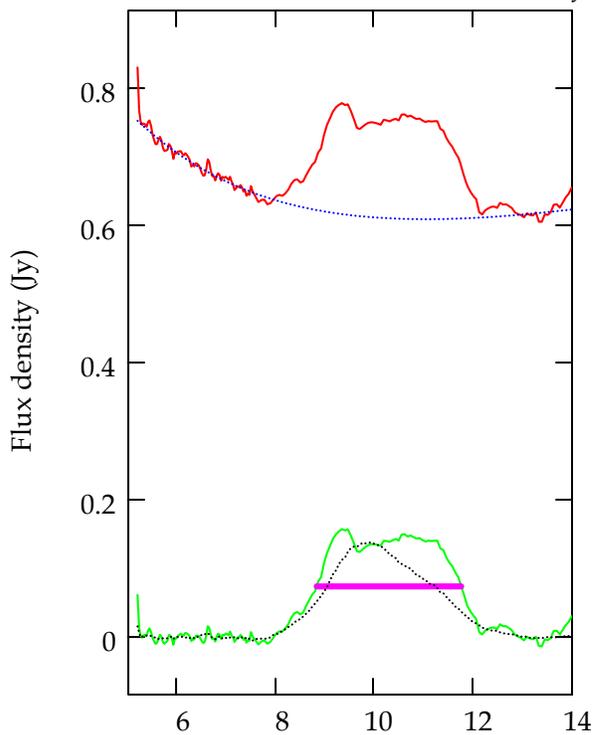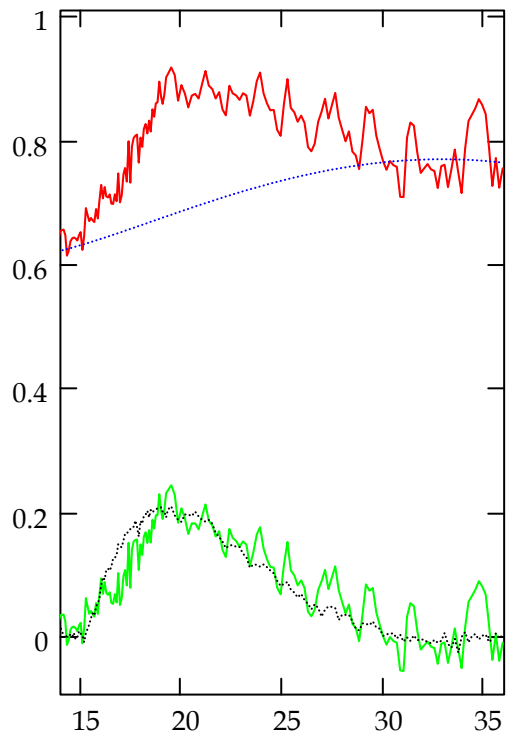

Names$_J$ = "HK Tau"

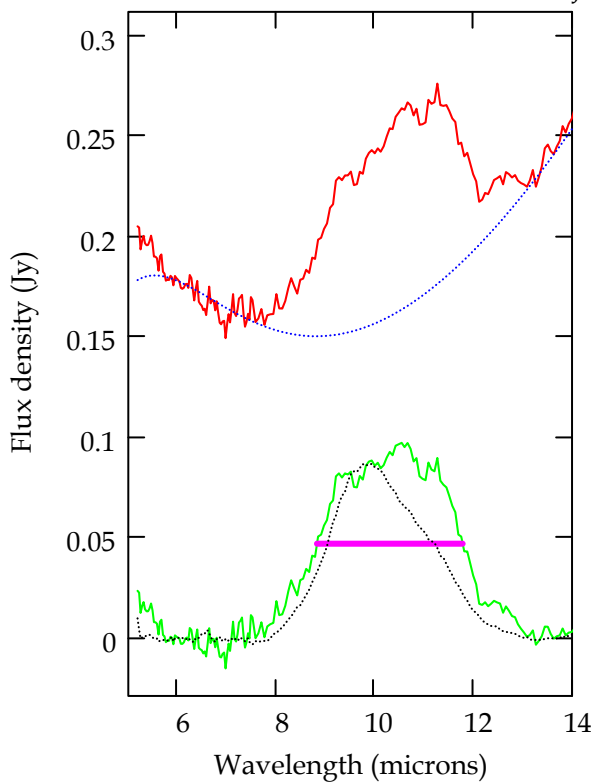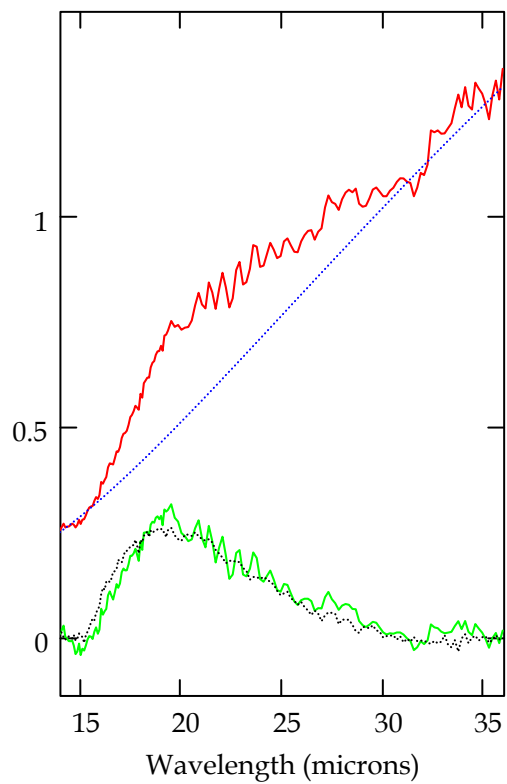

Names$_J$ = "HN Tau"

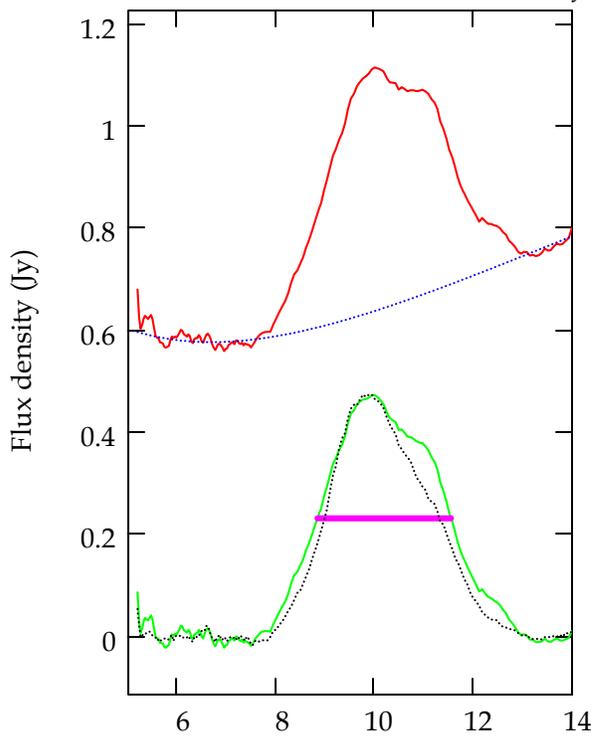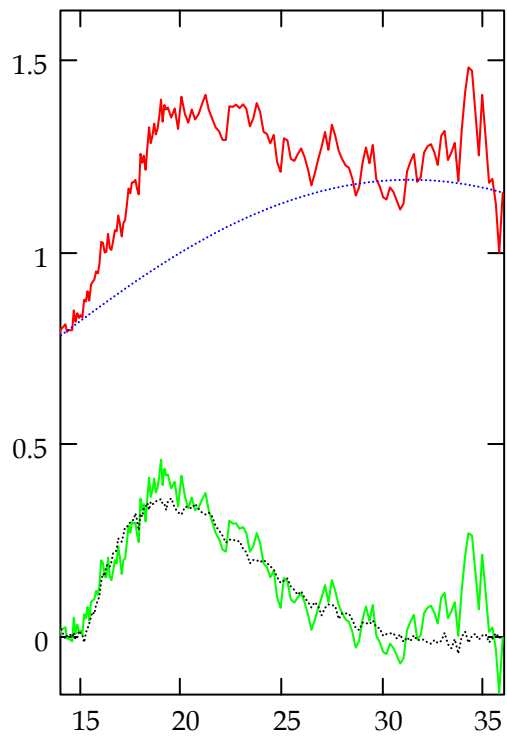

Names$_J$ = "HO Tau"

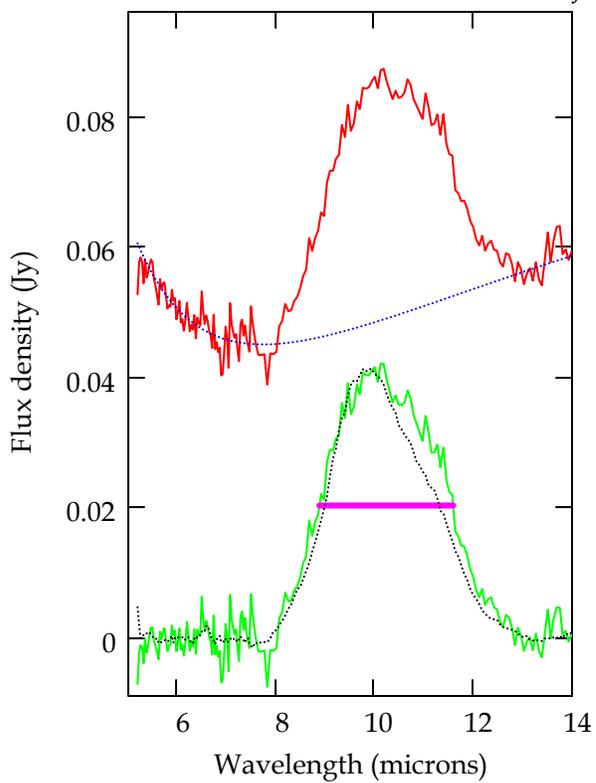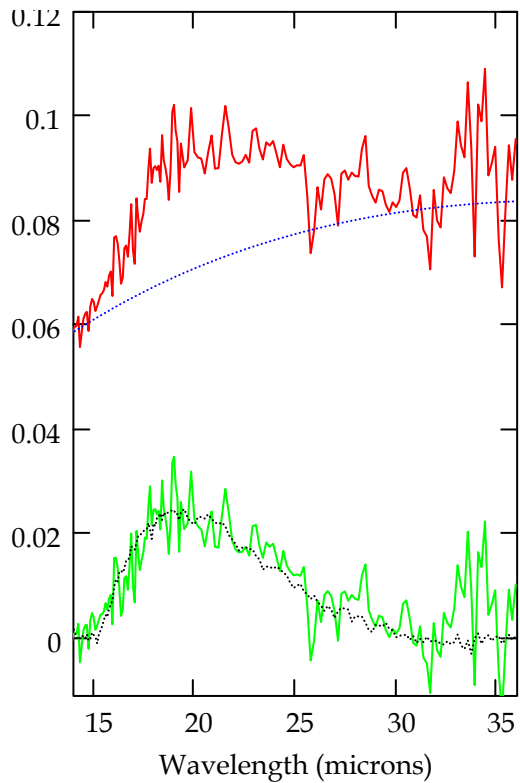

Wavelength (microns)

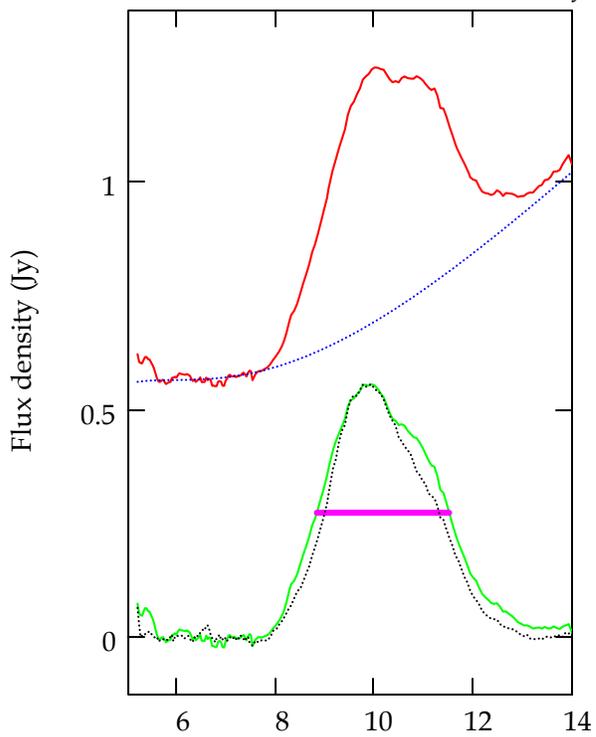
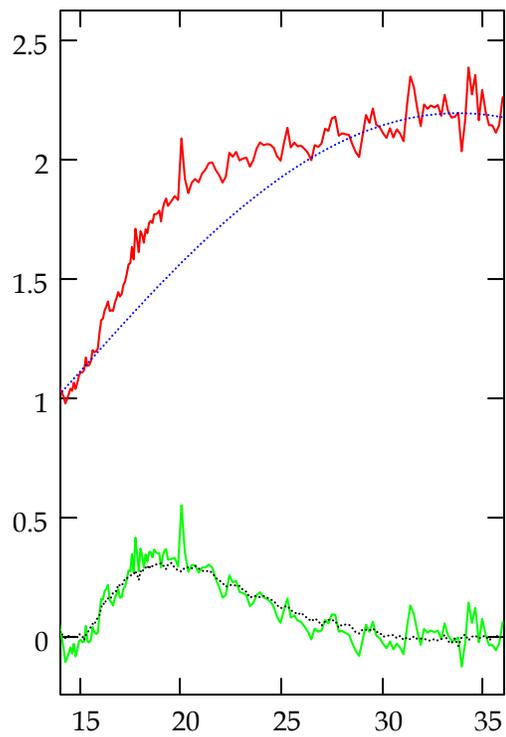

Names$_J$ = "HP Tau"

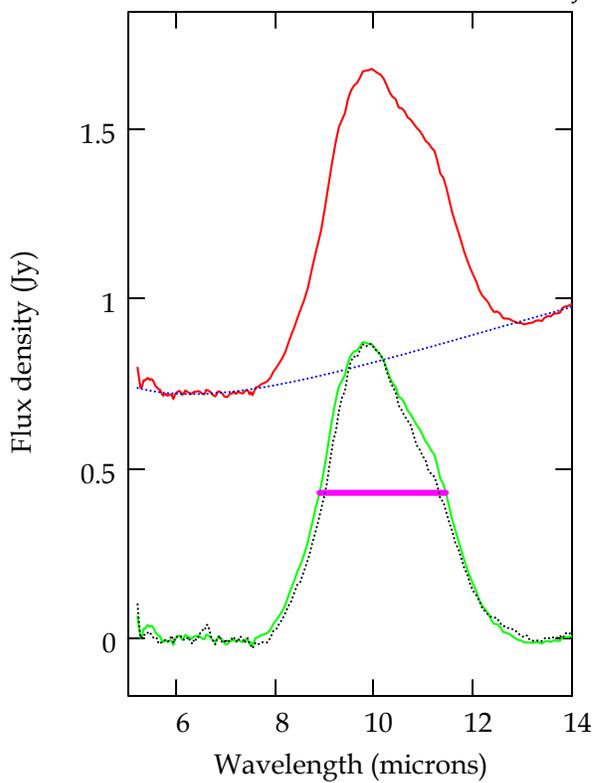
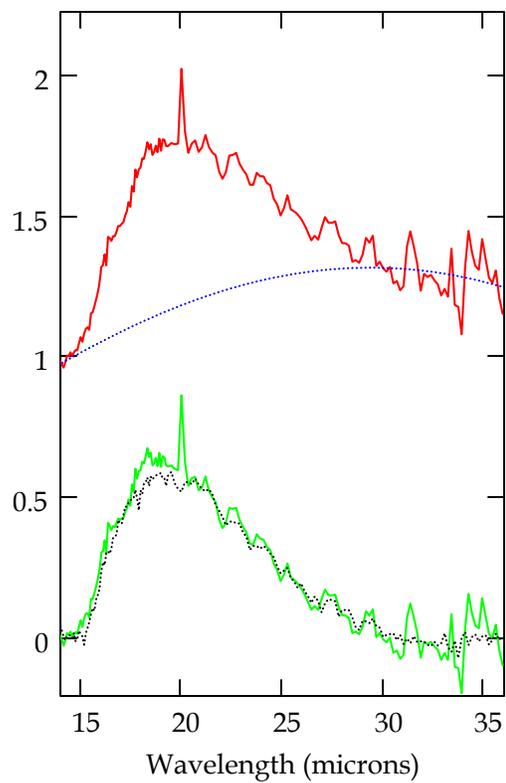

Names$_J$ = "HQ Tau"

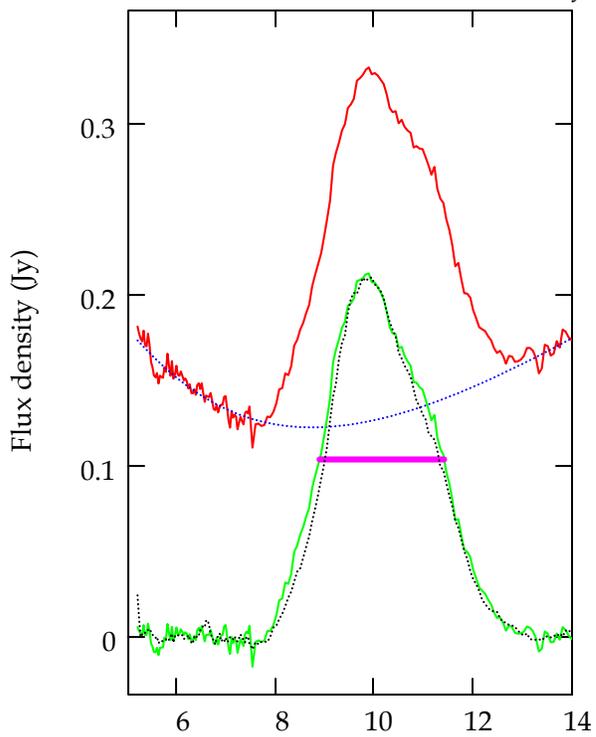
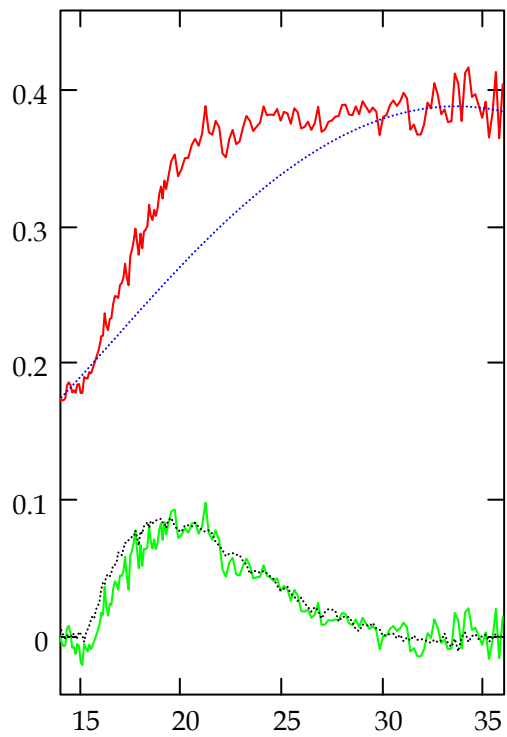

Names$_J$ = "IP Tau"

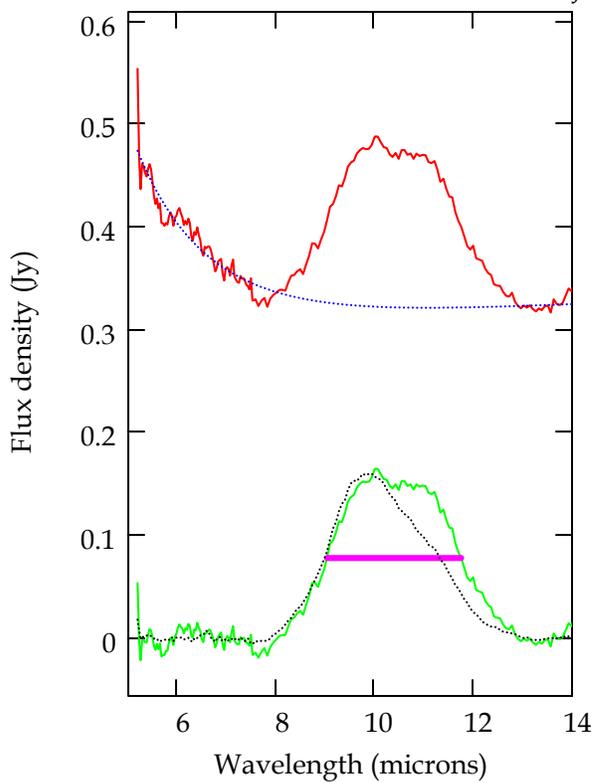
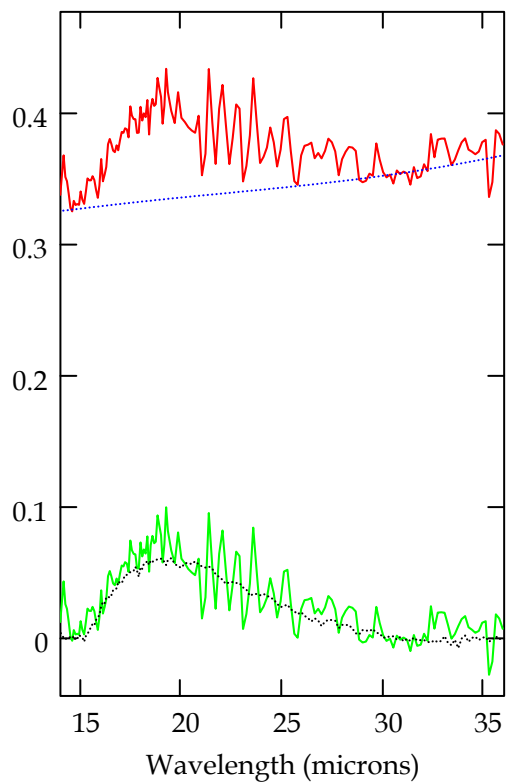

Names$_J$ = "IQ Tau"

Wavelength (microns)

Flux density (Jy)

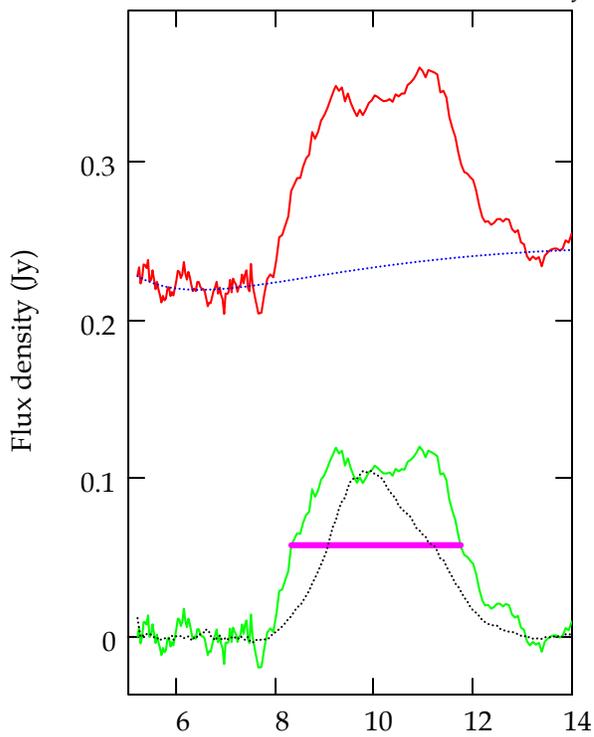
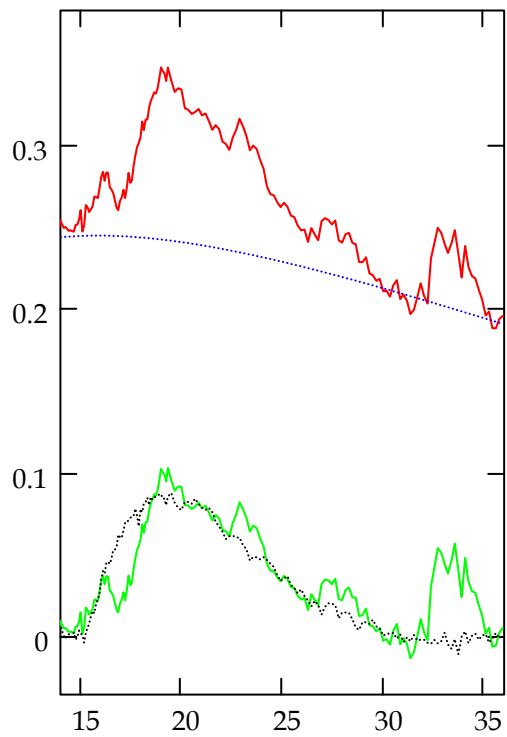
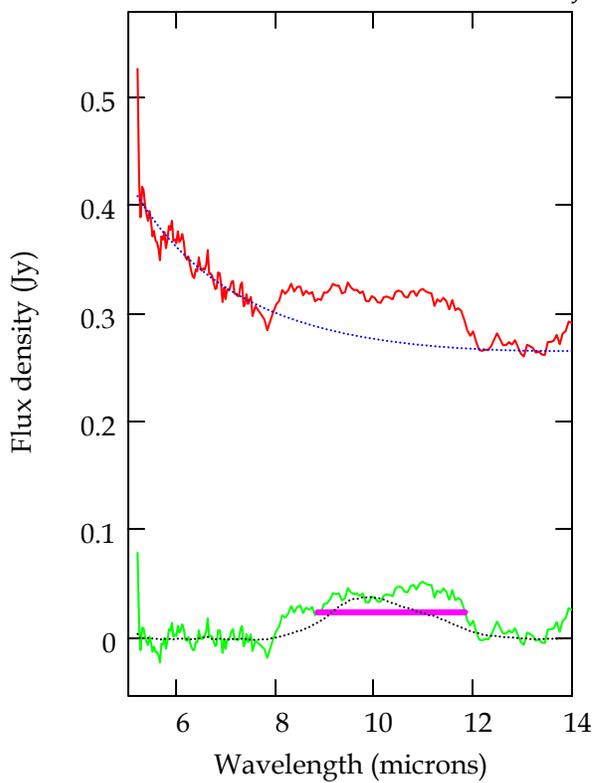
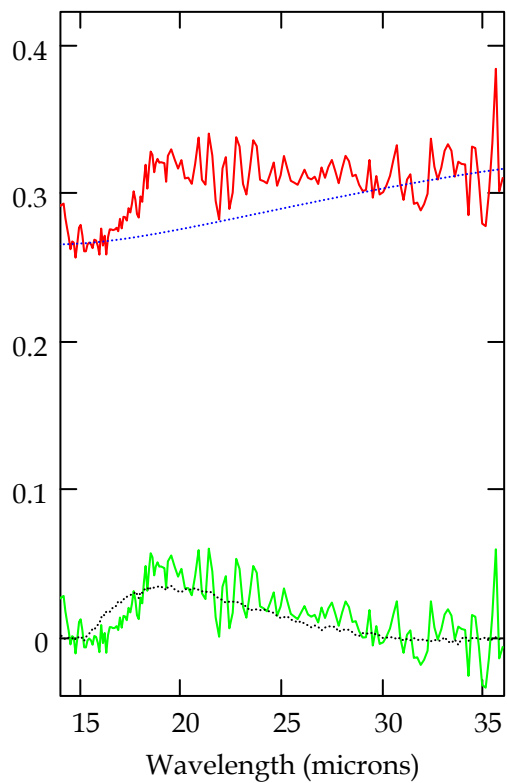

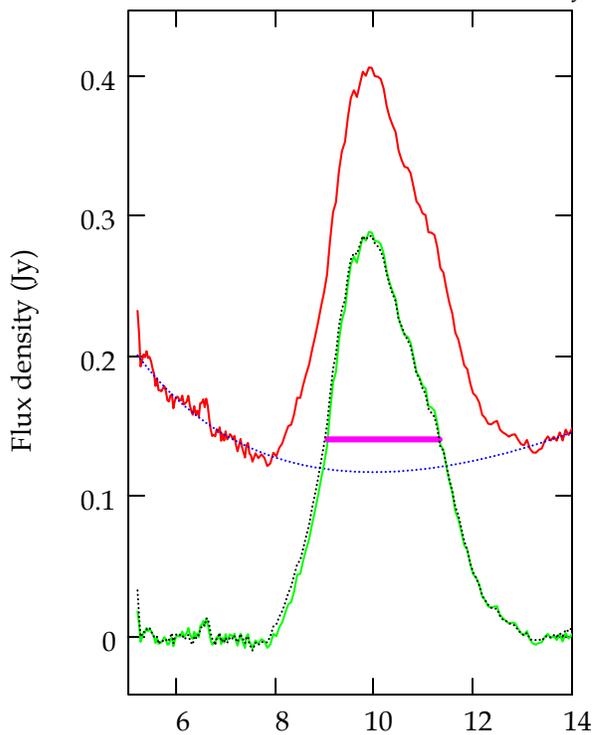
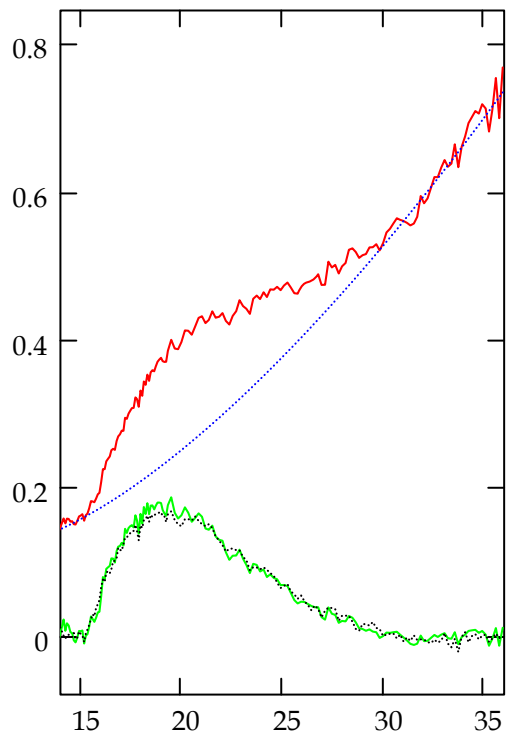

Names$_J$ = "LkCa 15"

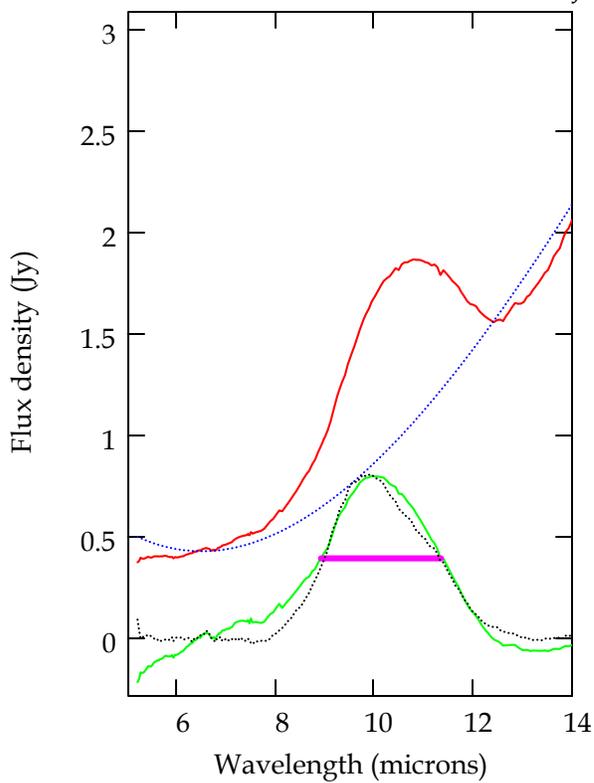
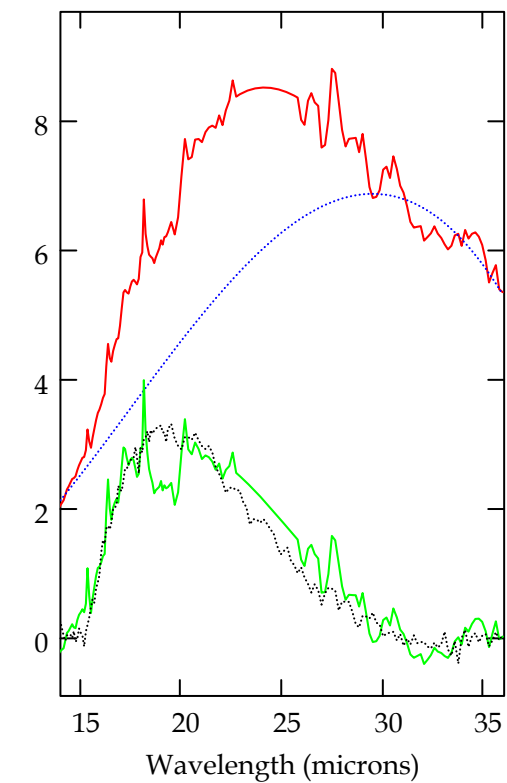

Names$_J$ = "MHO-3"

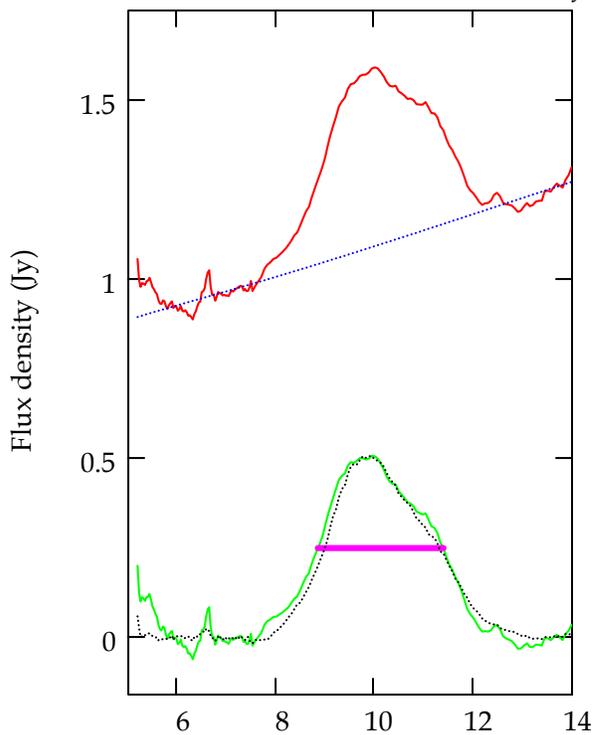
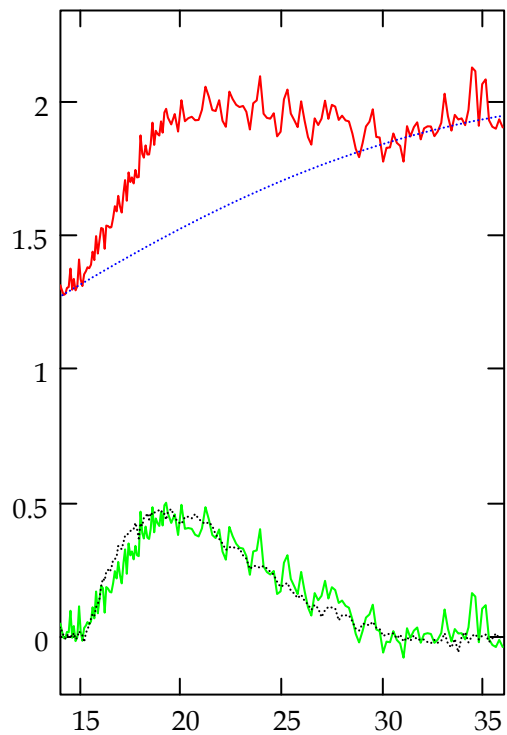

Names$_J$ = "RW Aur A"

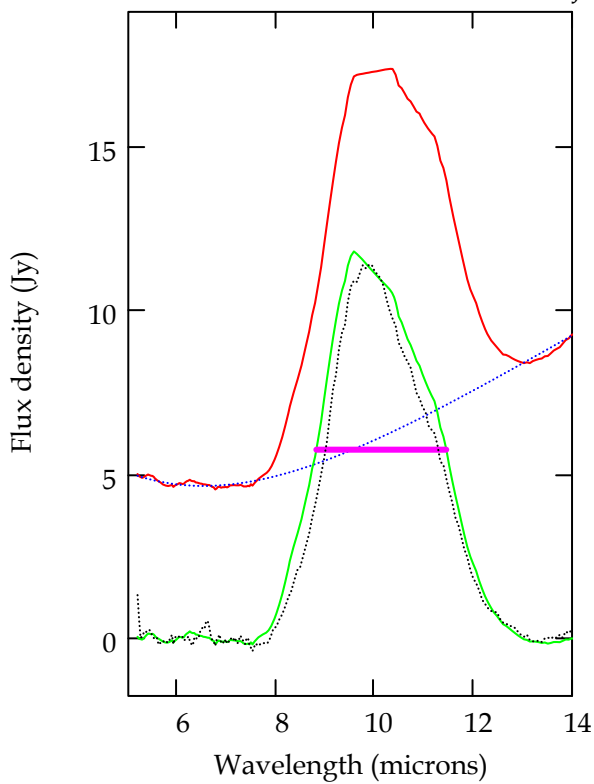
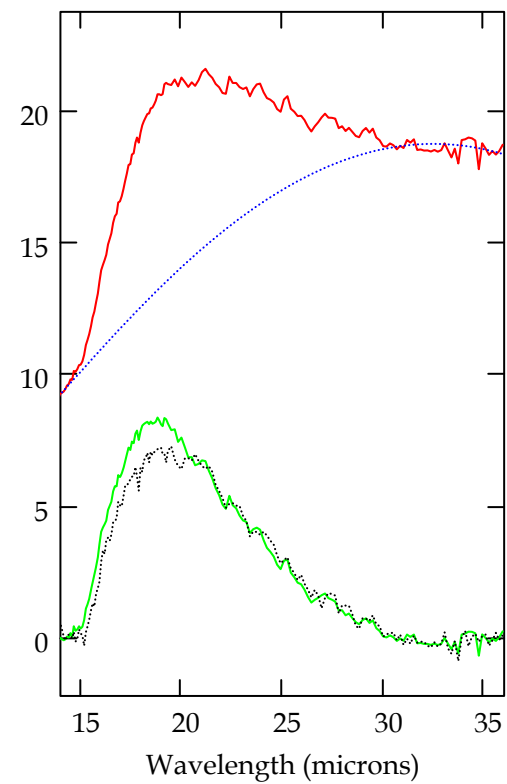

Names$_J$ = "RY Tau"

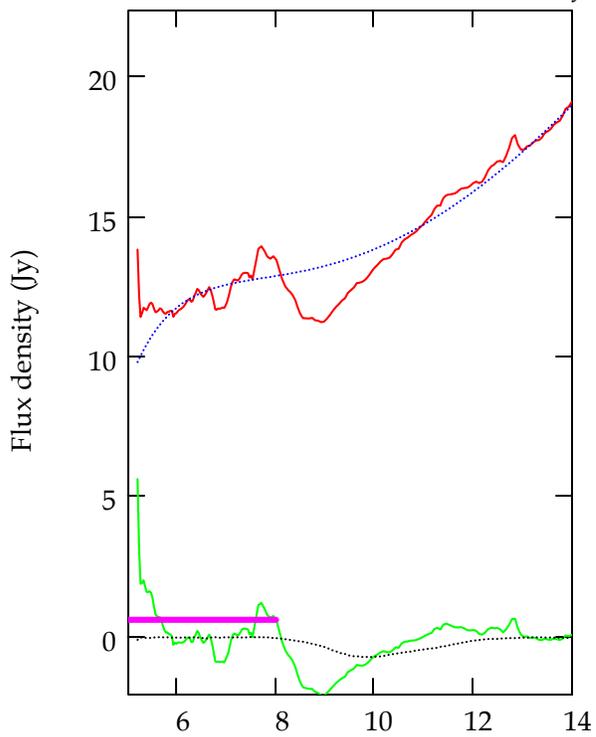
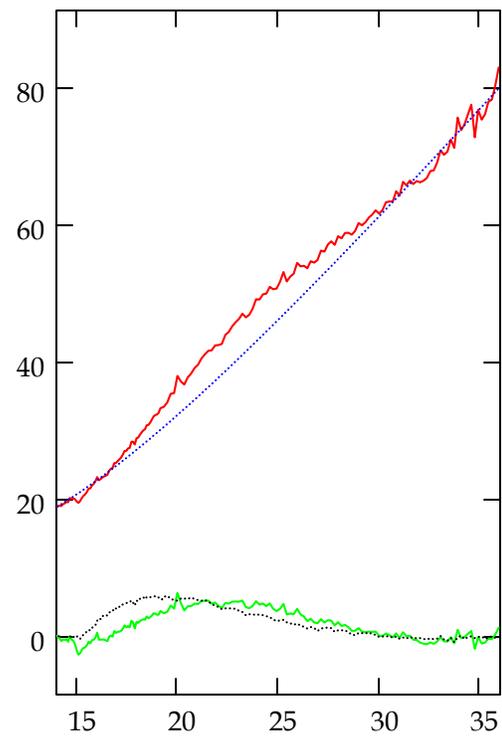

Names$_J$ = "T Tau "

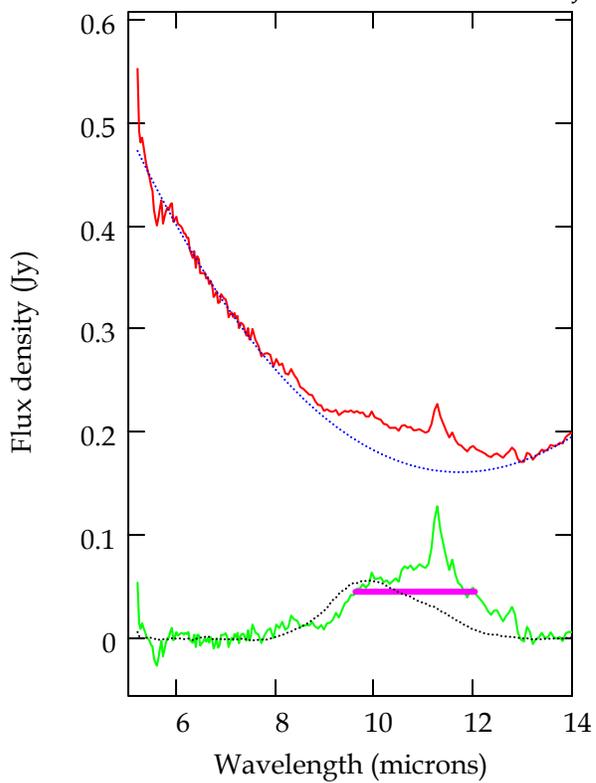
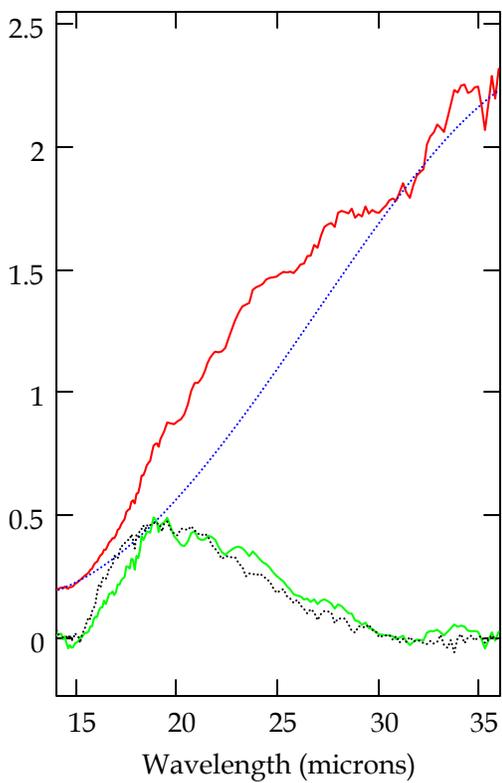

Names$_J$ = "UX Tau A"

Flux density (Jy)

Wavelength (microns)

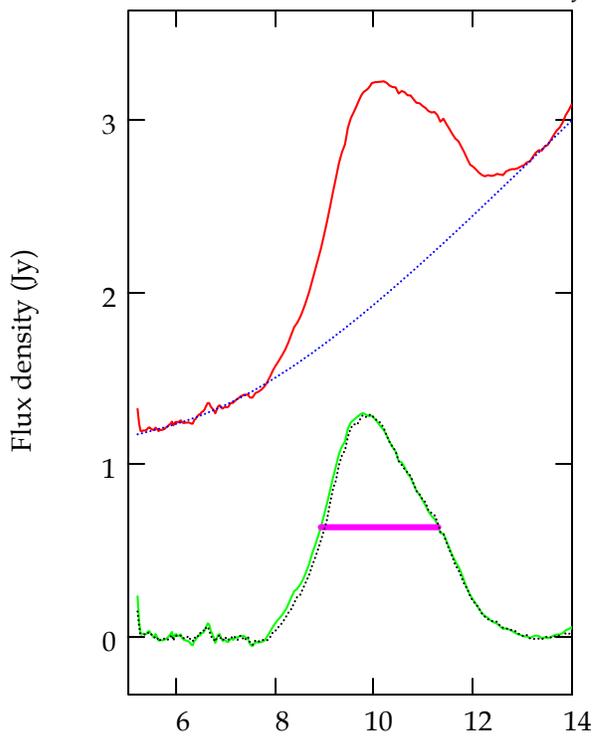
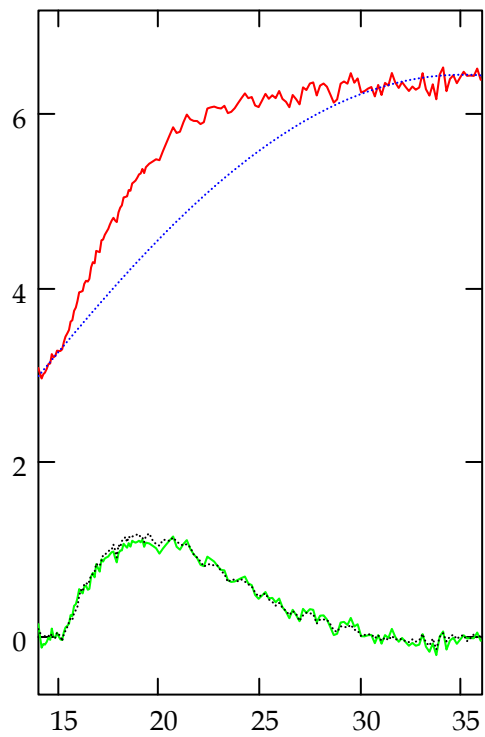
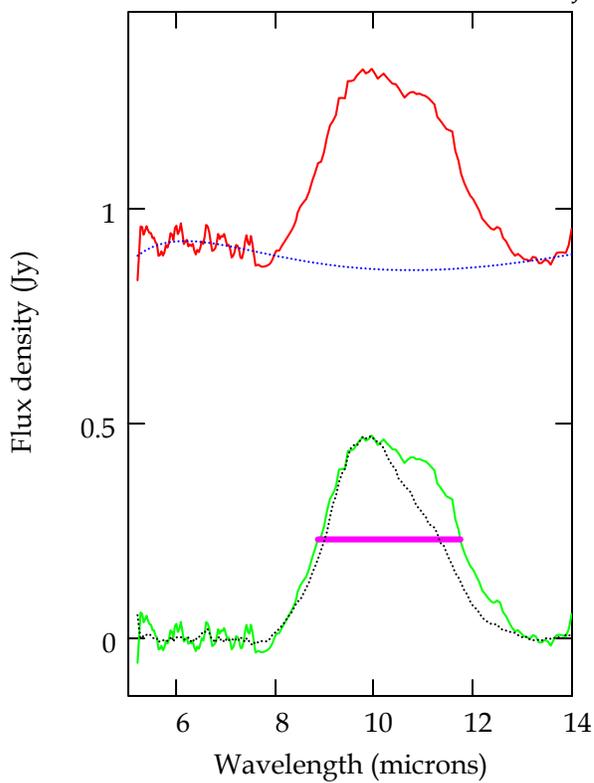
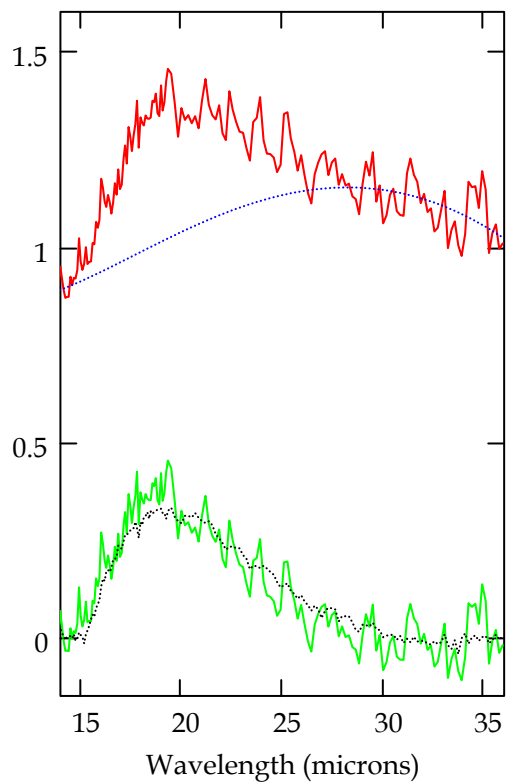

Names$_J$ = "UY Aur"

Names$_J$ = "UZ Tau/e"

Flux density (Jy)

Wavelength (microns)

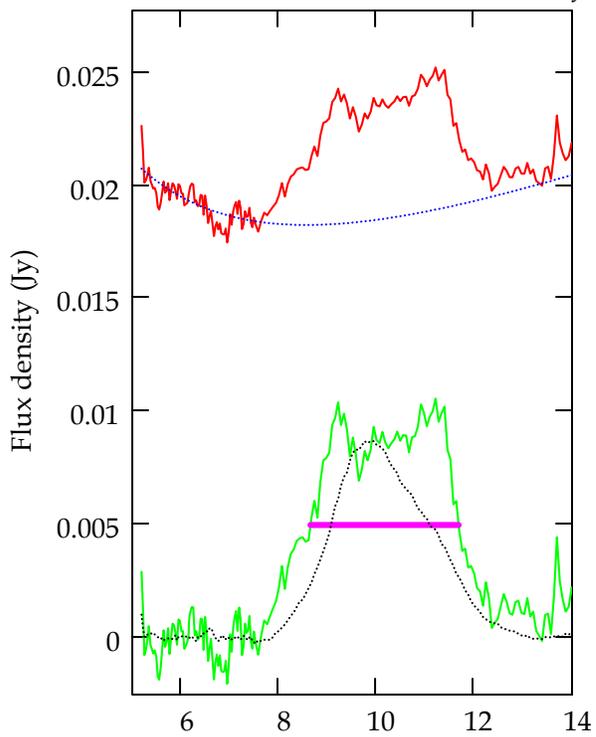
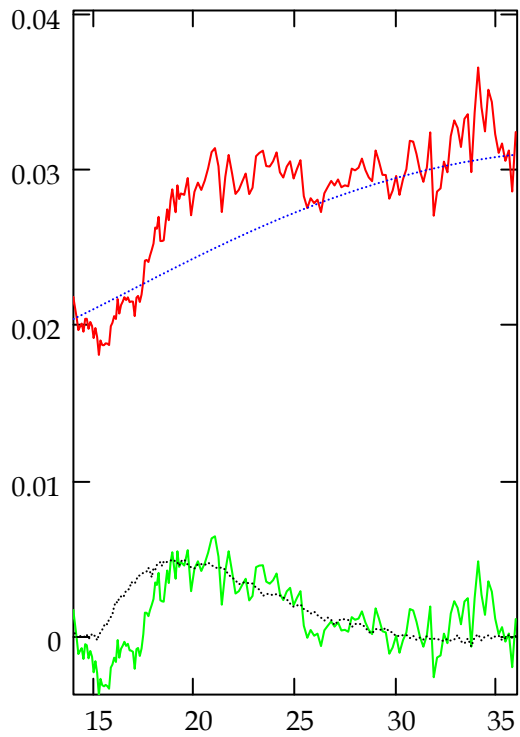

Names$_J$ = "V410 Anon 13"

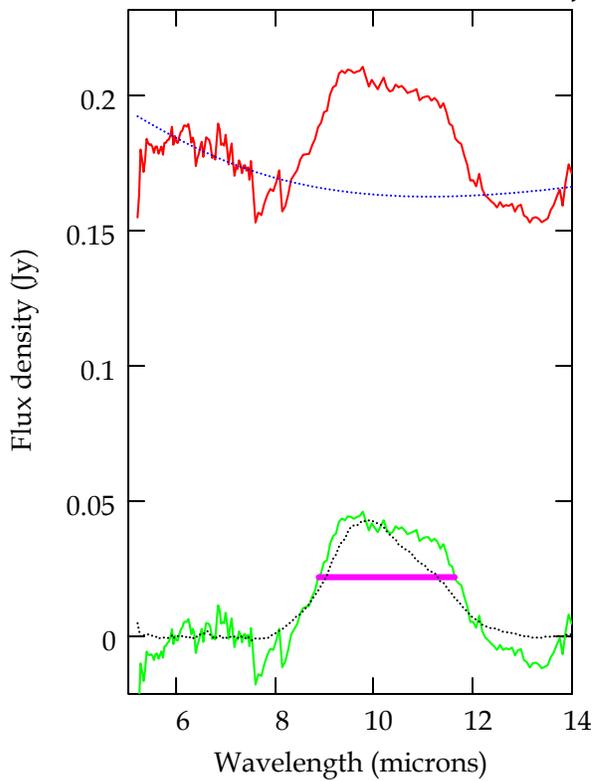
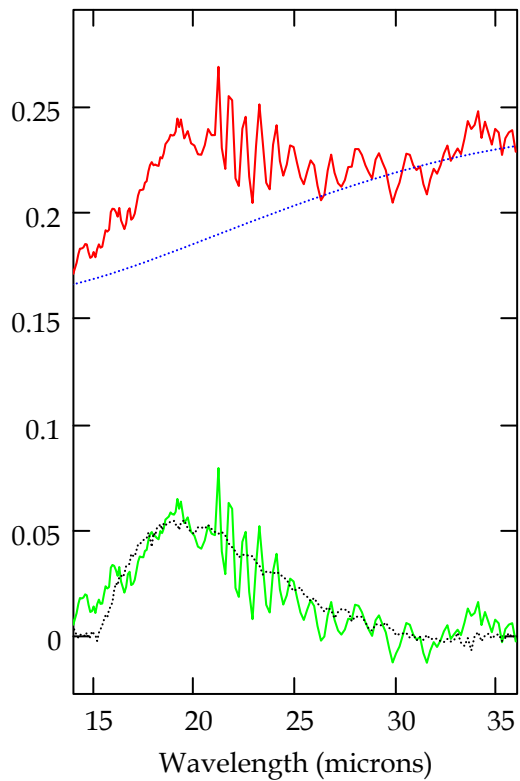

Names$_J$ = "V710 Tau"

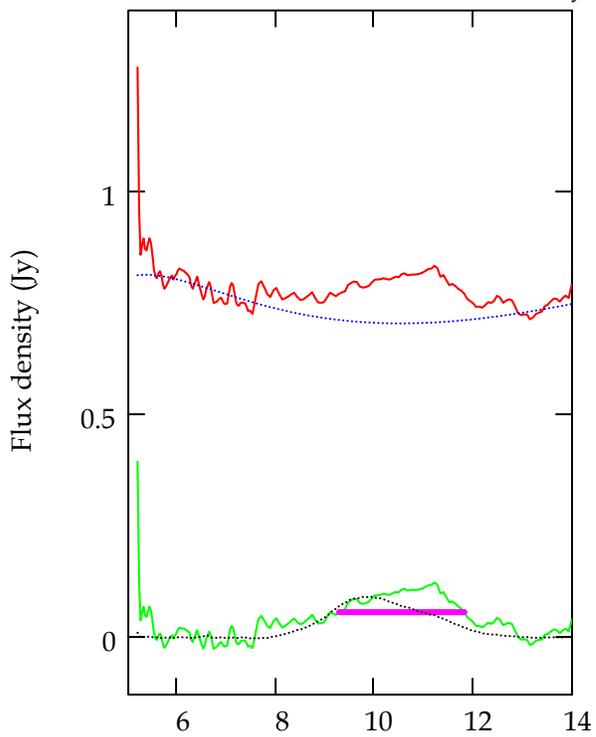
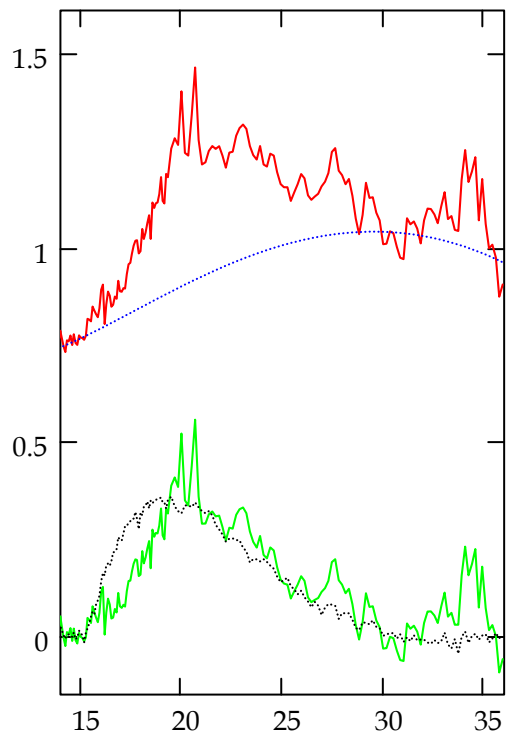
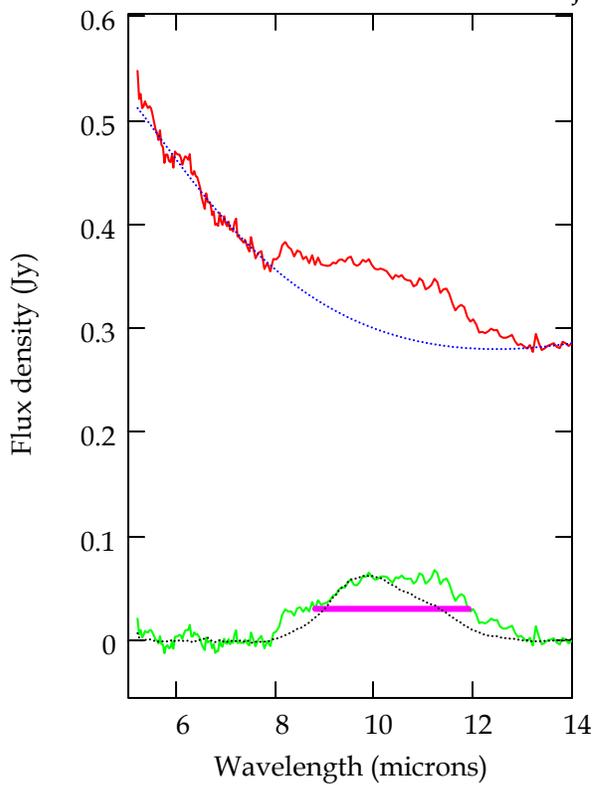
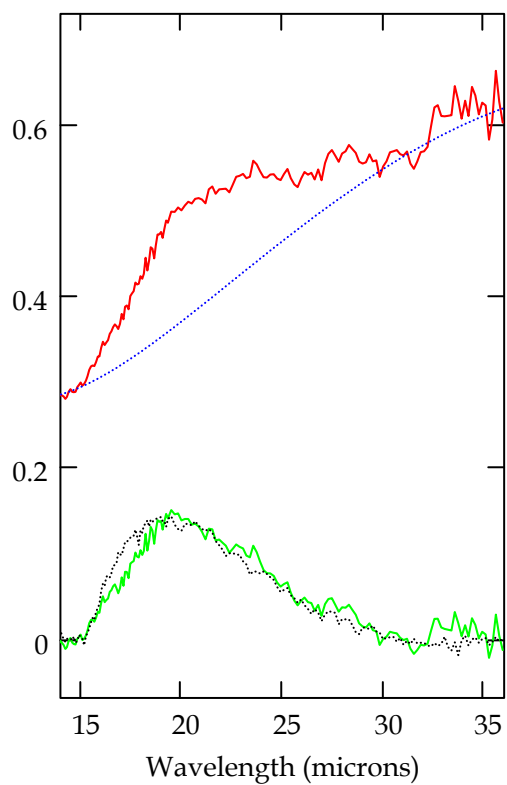

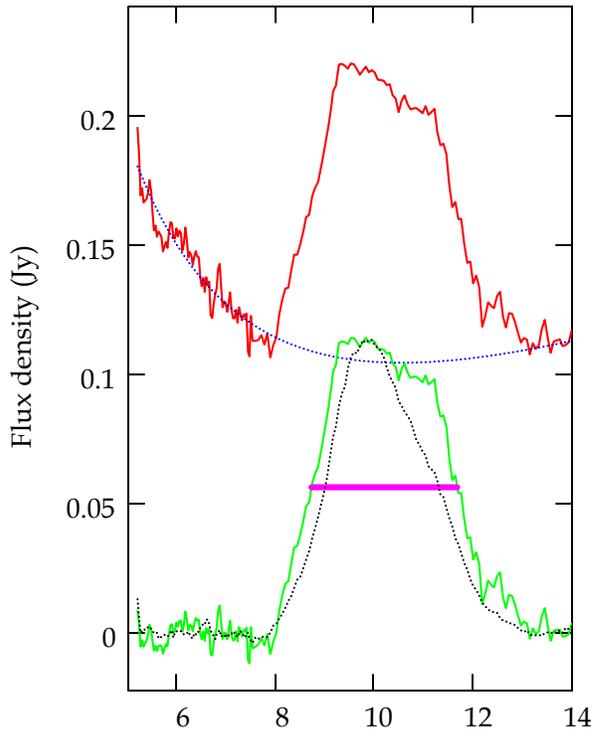
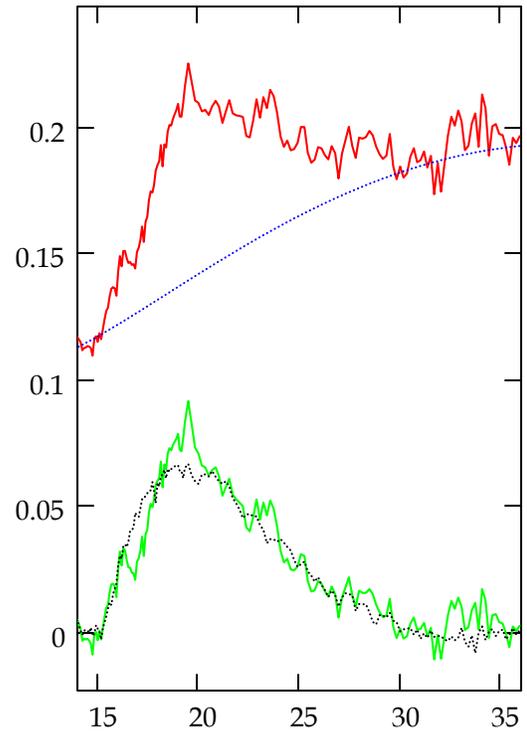
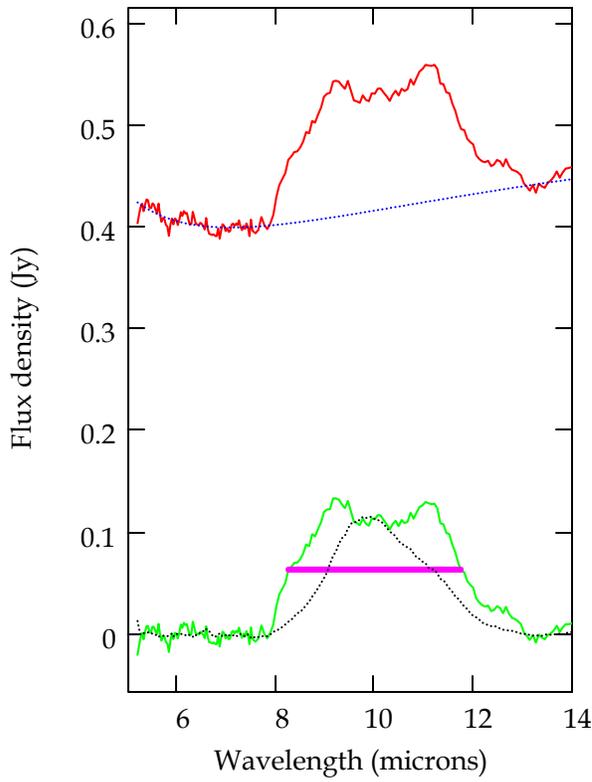
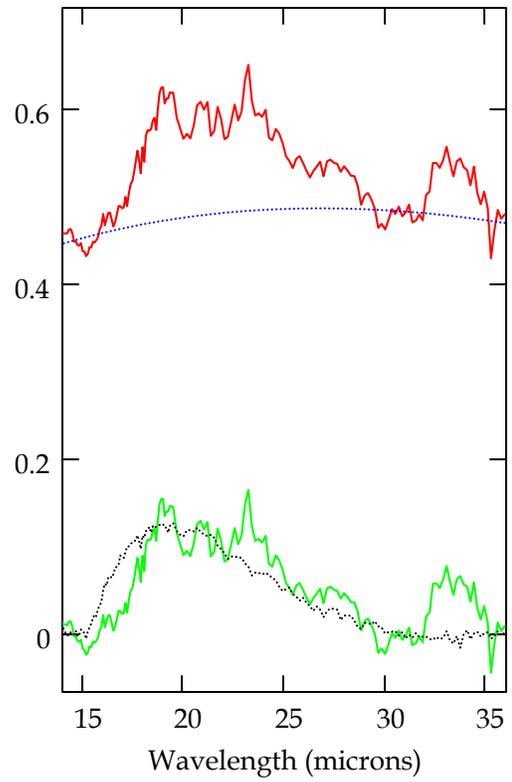

Names$_J$ = "V836 Tau"

Names$_J$ = "V955 Tau"

Flux density (Jy)

Wavelength (microns)

Names$_J$ = "VY Tau"

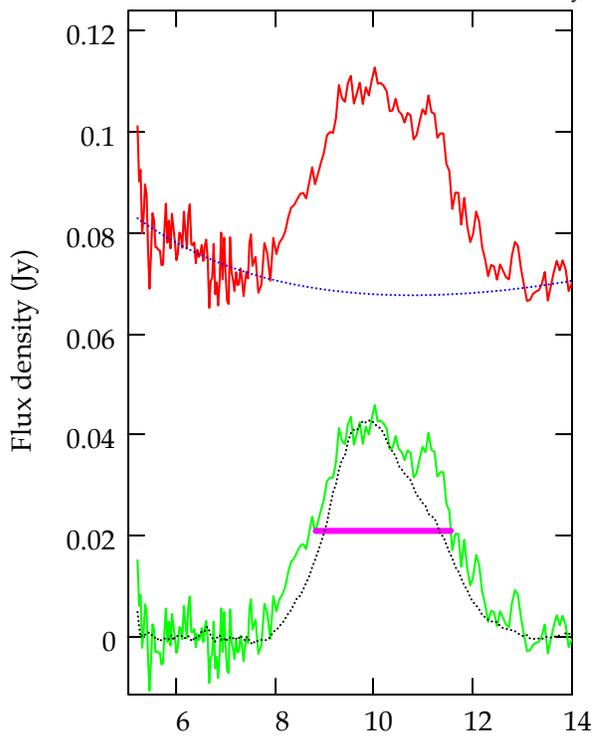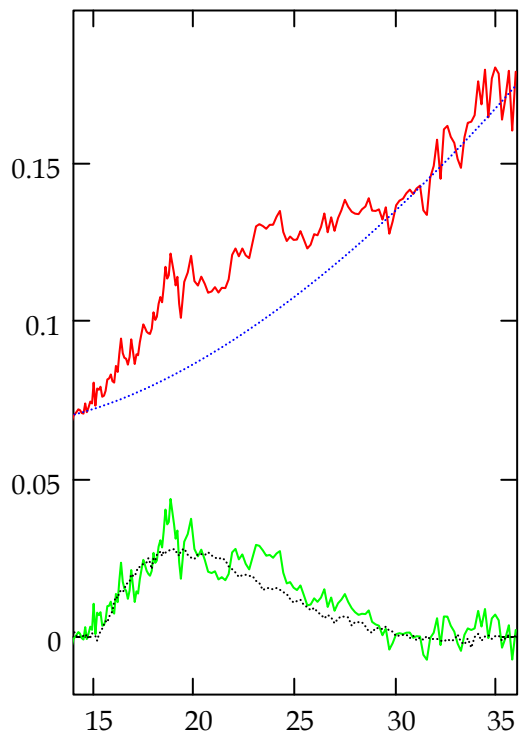

Names$_J$ = "XZ Tau"

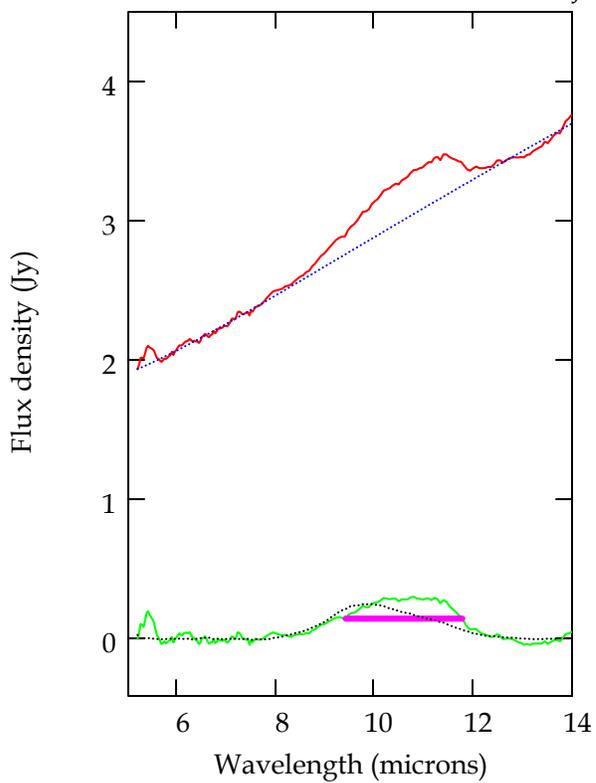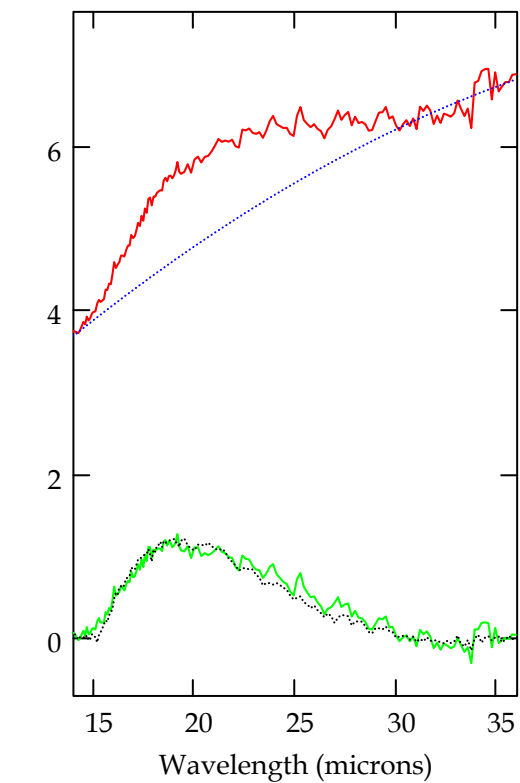

Wavelength (microns)

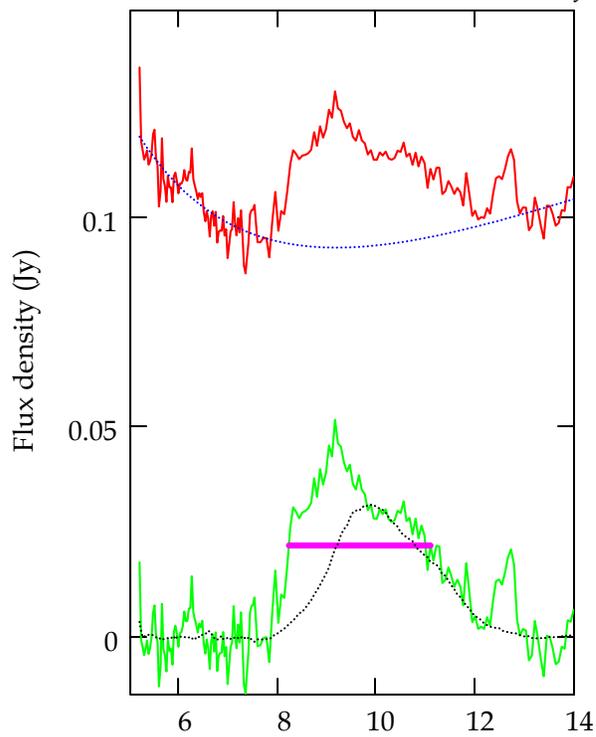
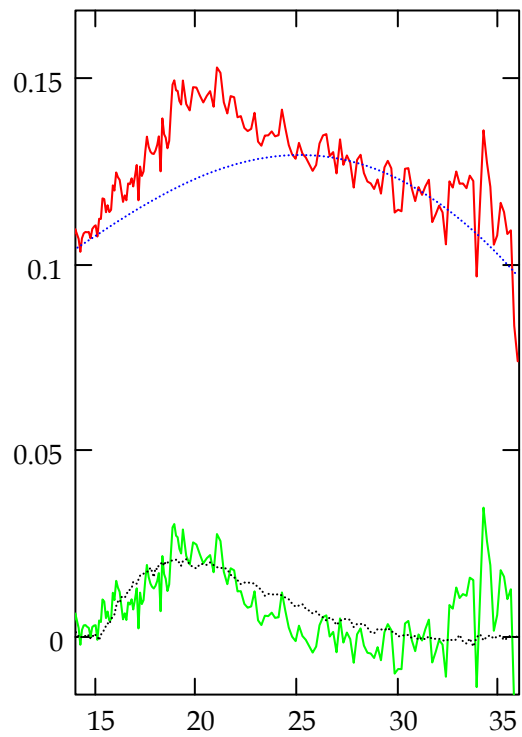
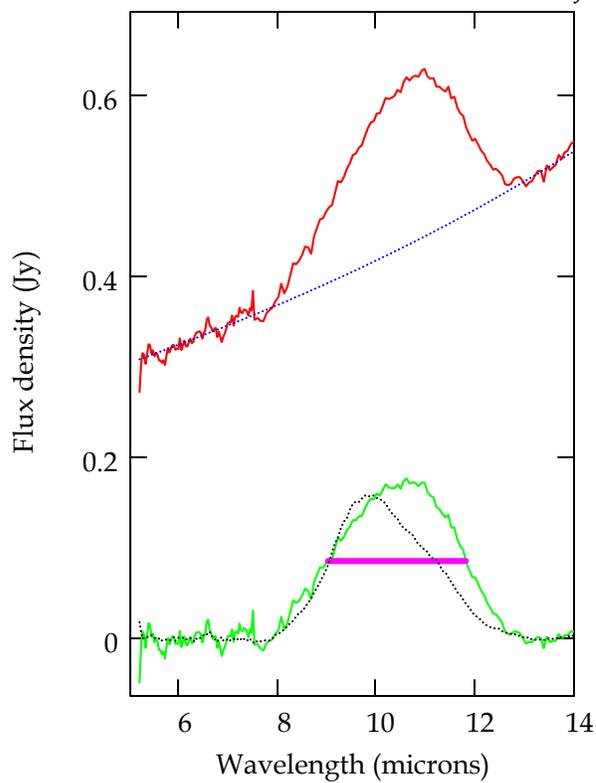
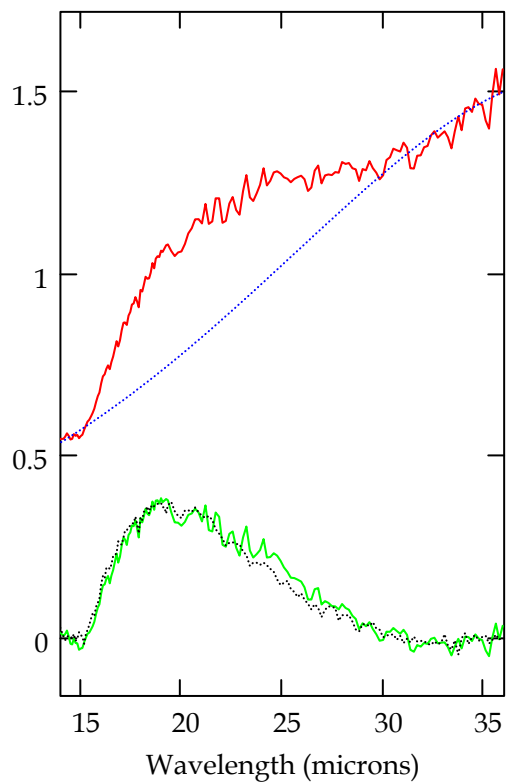

Names$_J$ = "ZZ Tau"

Names$_J$ = "ZZ Tau IRS"

Wavelength (microns)

Wavelength (microns)

Flux density (Jy)